\documentclass[10pt]{article}

\usepackage[top=1in, bottom=1.25in, left=1.25in, right=1.25in]{geometry}

\usepackage{setspace}
\usepackage{algorithm}

\usepackage{algpseudocode}

\usepackage{amssymb}

\usepackage{amsthm}

\usepackage{pifont}

\usepackage{amsmath}

\usepackage{adjustbox}

\usepackage{bm}

\usepackage{bbm}

\usepackage{natbib}

\usepackage{graphicx}

\usepackage{booktabs}

\usepackage{blkarray}

\usepackage{titlesec}  

\usepackage{authblk}

\usepackage[labelfont=bf]{caption}

\titleformat*{\section}{\sffamily\large}
\titleformat*{\subsection}{\sffamily\normalsize}

\newcommand{\mat}[1]{\bm{#1}}
\newcommand{\vect}[1]{\bm{#1}}

\newcommand{\T}{\mathsf{T}}

\newtheorem{lemma}{Lemma}
\newtheorem{theorem}{Theorem}
\newtheorem{definition}{Definition}
\newtheorem{assumption}{Assumption}
\newtheorem*{theorem*}{Theorem}

\title{Statistical properties of sketching algorithms}

\author[1,2]{Daniel Ahfock}
\author[1]{William J. Astle}
\author[1]{Sylvia Richardson}
\affil[1]{MRC Biostatistics Unit, University of Cambridge}
\affil[2]{School of Mathematics and Physics, University of Queensland}
\date{}                     
\begin{document}
\maketitle

\begin{abstract}
Sketching is a probabilistic data compression technique that has been largely developed in the computer science community. Numerical operations on big datasets can be intolerably slow; sketching algorithms address this issue by generating a smaller surrogate dataset. Typically, inference proceeds on the compressed dataset. Sketching algorithms generally use random projections to compress the original dataset and this stochastic generation process makes them amenable to statistical analysis. We argue that the sketched data can be modelled as a random sample, thus placing this family of data compression methods firmly within an inferential framework. In particular, we focus on the Gaussian, Hadamard and Clarkson-Woodruff sketches, and their use in single pass sketching algorithms for linear regression with huge $n$. We explore the statistical properties of sketched regression algorithms and derive new distributional results for a large class of sketched estimators. A key result is a conditional central limit theorem for data oblivious sketches. An important finding is that the best choice of sketching algorithm in terms of mean square error is related to the signal to noise ratio in the source dataset. Finally, we demonstrate the theory and the limits of its applicability on two real datasets.
\end{abstract}

\section{Introduction}
\label{sec:introduction}
Sketching is a general probabilistic data compression technique designed for Big Data applications \citep{cormode_sketch_2011}. Even routine calculations can be prohibitively computationally expensive on massive datasets. Computation time can be reduced to an acceptable level by allowing for some approximation error in the results. Sketching algorithms relax the computational task by generating a compressed version of the original dataset which then serves as a surrogate for calculations. The compressed dataset is referred to as a sketch, as it acts as a compact representation of the full dataset.  Sketching algorithms use a randomised compression stage which makes them interesting from a statistical viewpoint. Sketching algorithms for linear regression  have attracted significant attention in the numerical linear algebra and theoretical computer science communities \citep{woodruff_sketching_2014,mahoney_randomized_2011}. In this paper we investigate the statistical properties of sketched regression algorithms, a perspective which has received little attention up to now.

To describe sketched regression in more detail, we first assume the data consists of a $n$-length response vector $\vect{y}$ and a $n \times p $ matrix of covariates, $\mat{X}$ which is of full rank. It is assumed throughout that $n > p$.  The objective is to find the optimal least squares coefficients. Given sufficient computational resources, these could be computed exactly as 
\begin{align*}
\vect{\beta}_{F} = (\mat{X}^{\T}\mat{X})^{-1}\mat{X}^{\T}\vect{y}.
\end{align*}
The subscript $F$ is used to indicate the connection to the full dataset. Only two quantities are needed in order to determine $\vect{\beta}_{F}$, the Gram matrix $\mat{X}^{\T}\mat{X}$, and the marginal associations $\mat{X}^{\T}\vect{y}$. Calculation of $\mat{X}^{\T}\mat{X}$ requires $O(np^2)$ operations while computation of $\mat{X}^{\T}\vect{y}$ needs only $O(np)$ calculations. There are two broad methods for sketched regression, complete sketching and partial sketching. Complete sketching is based on approximating both $\mat{X}^{\T}\mat{X}$ and $\mat{X}^{\T}\vect{y}$, whereas partial sketching only approximates the Gram matrix. \citet{drineas_sampling_2006} establish many important results for complete sketching, and \citet{dhillon_new_2013} and \citet{pilanci_iterative_2016} give foundational results for partial sketching.

Sketching algorithms use random linear mappings to reduce the size of the dataset from $n$ to $k$ observations. The random linear mapping can be represented as a $k \times n$ sketching matrix $\mat{S}$. Complete sketching generates a $k$-length sketched response vector $\widetilde{\vect{y}}$ and a $k \times p$ matrix of sketched predictors $\widetilde{\mat{X}}$. The sketched data are computed through the linear mappings $\widetilde{\vect{y}} = \mat{S}\vect{y}$ and $\widetilde{\mat{X}} = \mat{S}\mat{X}$. Partial sketching only generates a $k \times p$ matrix of sketched covariates $\widetilde{\mat{X}}$. We again use the random mapping $ \widetilde{\mat{X}}= \mat{S}\mat{X}$. 

The complete sketching estimator, $\vect{\beta}_{S}$, is defined as the least squares coefficients using the sketched responses and predictors, 
\begin{align}
\vect{\beta}_{S} = (\widetilde{\mat{X}}^{\T}\widetilde{\mat{X}})^{-1}\widetilde{\mat{X}}^{\T}\widetilde{\vect{y}}.
\label{eq:beta_S}
\end{align}
The partial sketching estimator, $\vect{\beta}_{P}$,  is defined as
\begin{align}
\vect{\beta}_{P} = (\widetilde{\mat{X}}^{\T}\widetilde{\mat{X}})^{-1}{\mat{X}}^{\T}{\vect{y}}.
\label{eq:beta_P}
\end{align}
The key difference between \eqref{eq:beta_S} and \eqref{eq:beta_P} is that the partial sketched estimator $\vect{\beta}_{P}$ is constructed using the exact marginal associations $\mat{X}^{\T}\vect{y}$. Given the sketched data, computation of $\vect{\beta}_{S}$ or $\vect{\beta}_{P}$ requires only $O(kp^2)$ operations, compared with the $O(np^2)$ required for $\vect{\beta}_{F}$. 

The estimand within a sketching algorithm is the optimal coefficient vector $\vect{\beta}_{F}$. Sketching algorithms have the property that given a fixed $k$,  the approximation error $\lVert \vect{\beta}_{S} - \vect{\beta}_{F} \rVert_{2}$  or $\lVert \vect{\beta}_{P} - \vect{\beta}_{F} \rVert_{2}$ remains probabilistically bounded even as $n \to \infty$. Designing estimators for approximate computation with such properties is very difficult, and is  a common goal in the development of techniques for Big Data \citep{bardenet_note_2015, phillips_coresets_2016}. The favourable scaling properties of sketching algorithms are a critical factor in making them stand apart from simple subsampling approaches, where it can be difficult to establish universal worst case bounds for large $n$ \citep{drineas_sampling_2006, ma_statistical_2015}. The fact that sketching algorithms provide finite $k$ guarantees for arbitrarily large $n$ is a major reason they have received so much attention in the computer science community.

There is a large literature concerned with designing appropriate distributions for the random sketching matrix $\mat{S}$. Our focus is on data-oblivious random projections, where the distribution of the sketching matrix is not a function of the source data $[\vect{y}, \mat{X}]$. An example is the Gaussian sketch, where each element is independently distributed as a $N(0, 1/k)$ variate. We also consider the Hadamard sketch and the Clarkson-Woodruff sketch, random projections that exploit structure and sparsity for computational efficiency. 

Most existing results on the accuracy of sketching are universal worst case bounds \citep{woodruff_sketching_2014, mahoney_structural_2016}. This is typical for randomised algorithms, however a more detailed error analysis can provide important insights \citep{halko_finding_2011}. We investigate the statistical properties of $\vect{\beta}_{P}$ and $\vect{\beta}_{S}$ when using data oblivious sketches. An important finding is that the signal to noise ratio in the source dataset strongly influences the relative efficiency of complete to partial sketching. The statistical analysis also allows the construction of exact confidence intervals for the Gaussian sketch, and asymptotic confidence intervals for other  random projections, paving the way for their wider use in the statistical community interested in Big Data methods.  

We start by reviewing the existing literature on sketching algorithms before investigating the statistical properties in more detail. At its core, sketched regression is a randomised algorithm for approximate computation of $\vect{\beta}_{F}$. Repeated application of the sketching algorithm on the same dataset will produce different results. The first stage in our analysis is to establish the distributional properties of the sketched estimators with the source dataset fixed. This gives a clear statistical picture of the behaviour of the randomised algorithm. An important result is a conditional central limit theorem for the sketched dataset that connects the Hadamard and Clarkson-Woodruff projections to the Gaussian sketch. The regularity conditions have a intuitive interpretation in terms of the geometry of the source dataset. Our conditional analysis of the randomised algorithms is then extended to cover situations where sketching is used for approximate statistical inference. Given a statistical model for the response $\vect{y}=\mat{X}\vect{\beta}_{0} + \vect{\epsilon}$, for a vector of population parameters $\vect{\beta}_{0}$, and error terms $\vect{\epsilon}$, we can determine properties of 
$\vect{\beta}_{P}$ and $\vect{\beta}_{S}$ by integrating over the conditional distributions of the sketched estimators that take $\vect{y}$ as fixed. 

\section{Background and related work}
\subsection{Preliminaries}
Before proceeding, it is worth mentioning alternatives to sketching, in particular iterative methods for calculating the least squares coefficients $\vect{\beta}_{F}$. These include coordinate descent or stochastic gradient methods. Iterative methods are guaranteed to converge to $\vect{\beta}_{F}$ under very mild conditions. These iterative techniques assume that the entire dataset can stored in memory in a single location, or require regular communication if the full dataset is distributed across multiple sites. Sketching algorithms are not burdened by these memory and communication costs, with the drawback of no convergence guarantees to $\vect{\beta}_{F}$. Connections to iterative methods are postponed until the discussion, the focus for now is on the single pass estimators $\vect{\beta}_{S}$ and $\vect{\beta}_{P}$. 

The purpose of this section is to review the existing theoretical framework for sketching algorithms. Sketching algorithms are largely motivated through worst case guarantees. We recap how these bounds can be developed before studying the statistical properties of the sketched estimators. 

It will be helpful to define a number of quantities related to the full dataset before moving on. Let 
$TSS_{F}=\vect{y}^{\T}\vect{y},
RSS_{F}=\lVert\vect{y}-\mat{X}\vect{\beta}_{F}\rVert_{2}^{2}, 
{MSS}_{F} = \lVert \mat{X}\vect{\beta}_{F}\rVert_{2}^{2}$ and
$R^2_{F} = MSS_{F}/TSS_{F}$. These terms summarise the goodness of fit of the model. The total, residual and model sum of squares are given by $TSS_{F}$, $RSS_{F}$ and $MSS_{F}$ respectively, with $TSS_{F}=MSS_{F}+RSS_{F}$. The proportion of variance explained by the model is given by $R^2_{F}$. These values will be important in characterising the behaviour of $\vect{\beta}_{S}$ and $\vect{\beta}_{P}$. 

\subsection{Worst case bounds}
\label{subsec:worst_case}
A key concept in the construction of sketching algorithms is the notion of an $\epsilon$-subspace embedding \citep{woodruff_sketching_2014, meng_low-distortion_2013, yang_implementing_2015}. 
\begin{definition}{$\epsilon$-subspace embedding. }
\label{defn:subspace_embedding_matrices}\newline
For a given $n \times d$ matrix $\mat{A}$, we call a $k \times n $ matrix $\mat{S}$ an $\epsilon$-subspace embedding for $\mat{A}$, if for all vectors $\vect{z} \in \mathbb{R}^{d}$
\begin{align*}
(1-\epsilon) \lVert \mat{A}\vect{z} \rVert_{2}^{2} \le  \lVert \mat{S} \mat{A}\vect{z} \rVert_{2}^{2} \le (1+\epsilon)\lVert \mat{A}\vect{z}\rVert_{2}^{2}.
\end{align*}
\end{definition}
Speaking broadly, an $\epsilon$-subspace preserves the linear structure of the columns of the original dataset up to some multiplicative $(1\pm \epsilon)$ factor. In particular, if $\epsilon$ is small, the linear mapping $\mat{S}$ approximately preserves the covariance structure of the source dataset.  Most theoretical arguments for sketching algorithms are predicated on the idea that the sketching matrix $\mat{S}$ is an $\epsilon$-subspace embedding for the source dataset. The general notion is that it is possible to use a linear mapping $\mat{S}$ that reduces the sample size from $n$ to $k$ whilst preserving much of the linear information in the full dataset.  

The issue of how to generate $\epsilon$-subspace embeddings is deferred until section \ref{subsec:data_oblivious}, the present focus will be on the utility of $\epsilon$-subspace embeddings for linear regression problems. For now, assume that we have some method for generating $\epsilon$-subspace embeddings for the source data matrix $\mat{A}$. It will be convenient to refer to  $\widetilde{\mat{A}}=\mat{S}\mat{A}$ as an $\epsilon$-subspace embedding of $\mat{A}$ if $\mat{S}$ is an $\epsilon$-subspace embedding for $\mat{A}$. As regression is the focus from this point forward, we will define the source data matrix as $\mat{A}=[\vect{y}, \mat{X}]$, the sketched data matrix as $\widetilde{\mat{A}}=[\widetilde{\vect{y}}, \widetilde{\mat{X}}]$ and set $d=p+1$. 

The complete sketched estimator $\vect{\beta}_{S}$ is given by the least squares coefficients using the sketched responses $\widetilde{\vect{y}}$ and the sketched predictors $\widetilde{\mat{X}}$,
\begin{align*}
\vect{\beta}_{S} &= \underset{\vect{\beta} \in \mathbb{R}^p}{\text{argmin }} \lVert \widetilde{\vect{y}} -\widetilde{\mat{X}}\vect{\beta}\rVert_{2}^{2}.
\end{align*}
An $\epsilon$-subspace embedding is useful as it relates the sketched optimisation problem to the full dataset optimisation problem. If $\widetilde{\mat{A}} = [\widetilde{\vect{y}}, \widetilde{\mat{X}}]$ is an $\epsilon$-subspace embedding of $\mat{A}=[\vect{y}, \mat{X}]$, it must hold that for  all $\vect{\beta} \in \mathbb{R}^{p}$,
\begin{align*}
(1-\epsilon) \lVert \vect{y}-\mat{X}{\vect{\beta}}\rVert_{2}^{2} \le \lVert \widetilde{\vect{y}}-\widetilde{\mat{X}}{\vect{\beta}} \rVert_{2}^{2} \le (1+\epsilon)\lVert \vect{y}-\mat{X}{\vect{\beta}}\rVert_{2}^{2}. 
\end{align*}
If $\epsilon$ is small, minimising the sum of squared residuals on the sketched dataset is similar to minimising the sum of squared residuals on the full dataset. If this is the case, it can be expected that $\vect{\beta}_{S}$ will be close to $\vect{\beta}_{F}$. It is possible to establish the concrete bounds, that if $\widetilde{\mat{A}}$ is an $\epsilon$-subspace embedding of $\widetilde{\mat{A}}$  \citep{sarlos_improved_2006},
\begin{align}
\lVert \vect{\beta}_{S}- \vect{\beta}_{F}\rVert_{2}^{2} &\le  \dfrac{\epsilon^2}{\sigma_{\text{min}}^2(\mat{X})}RSS_{F}, \label{eq:beta_s_epsilon_bound}
\end{align}
where $\sigma_{\text{min}}(\mat{X})$ represents the smallest singular value of the design matrix $\mat{X}$. A very similar argument can be used to motivate the partial sketched estimator $\vect{\beta}_{P}$. Existing bounds for the partial sketch focus on the prediction error $\lVert \mat{X}\vect{\beta}_{P} - \mat{X}\vect{\beta}_{F} \rVert_{2}^{2}$ \citep{becker_robust_2015, pilanci_iterative_2016}.  To make a direct comparison to \eqref{eq:beta_s_epsilon_bound} we establish a bound on the coefficient error
\begin{theorem}
\label{thm:partial_sketching_worst_case}
Suppose that $\widetilde{\mat{X}}$ is an $\epsilon$-subspace embedding of $\mat{X}$ with $\epsilon < 0.5$. Then the following bound holds, 
\begin{align}
\lVert \vect{\beta}_{P}- \vect{\beta}_{F}\rVert_{2}^{2} &\le \dfrac{4\epsilon^2}{\sigma_{\mathrm{min}}^2(\mat{X})}MSS_{F}. \label{eq:beta_p_epsilon_bound}
\end{align}
\end{theorem}
For proof see the supplementary material. The mild requirement that $\epsilon < 0.5$ is imposed so that the bound matches the functional form of the complete sketching bound \eqref{eq:beta_s_epsilon_bound}. Comparing the partial sketching bound to \eqref{eq:beta_s_epsilon_bound}, we see that the tightness of the bound is controlled by the model sum of squares as opposed to the residual sum of squares. The sensitivity of partial sketching to the model sum of squares as opposed to the residual sum of squares has been noted in previous on partial sketching \citep{dhillon_new_2013, pilanci_iterative_2016, becker_robust_2015}. This suggests that the signal to noise ratio in the source dataset will be important when selecting which sketched estimator to use. A naive conclusion is that complete sketching is preferred when $RSS_{F} < 4MSS_{F}$, or equivalently $R^{2}_{F} > 0.25 $. Such a result is hardly prescriptive, as the worst case bound is not necessarily indicative of expected performance. A second point of interest is that if the $k \times n$ matrix $\mat{S}$ is an $\epsilon$-subspace embedding for $\mat{A}=[\vect{y}, \mat{X}]$, it is also an $\epsilon$-subspace embedding for $\mat{X}$. This suggests that it is reasonable to compute both $\vect{\beta}_{P}$ and $\vect{\beta}_{S}$ from a single sketch, although it is not clear how to combine the estimators into a single point estimator. These issues will be explored in more depth by examining the statistical properties of both complete and partial sketching. Before moving on to the statistical analysis we review some of the existing methods for generating $\epsilon$-subspace embeddings. 

There are two general categories of distributions for the random matrix $\mat{S}$, data aware random projections and data oblivious random projections. A data aware random projection uses information in the source data $\vect{y}$, $\mat{X}$ to generate $\mat{S}$. In contrast, a data oblivious random projection can be sampled without knowledge of $\vect{y}$ or $\mat{X}$. Data aware random projections are closely connected to finite population sampling methods in the statistics literature, and this is discussed in more detail in Section \ref{subsec:data_aware}. Data oblivious random projections are more closely related to dimension reduction techniques such as multidimensional scaling.  Our main focus is on data oblivious random projections. Data oblivious projections are designed to produce $\epsilon$-subspace embeddings for an arbitrary source data matrix with high probability.
\subsection{Data oblivious sketches}
\label{subsec:data_oblivious}
The Gaussian sketch was one of the first projections proposed for sketched regression \citep{sarlos_improved_2006}. Recall that a Gaussian sketch is formed by  independently sampling each element of $\mat{S}$ from a $N(0, 1/k)$ distribution. The drawback of the Gaussian sketch is that computation of the sketched data is quite demanding, taking $O(ndk)$ operations. As such, there has been work on designing more computationally efficient random projections. The Hadamard sketch and the Clarkson-Woodruff sketch are two examples of more efficient methods for generating $\epsilon$-subspace embeddings.  

The Hadamard sketch is a structured random matrix \citep{ailon_fast_2009}. The sketching matrix is formed as $\mat{S} = \Phi\mat{H}\mat{D}/\surd{k}$, where $\Phi$ is a $k \times n$ matrix and $\mat{H}$ and $\mat{D}$ are both $n \times n$ matrices. The fixed matrix $\mat{H}$ is a Hadamard matrix of order $n$. A Hadamard matrix is a square matrix with elements that are either $+1$ or $-1$ and orthogonal rows. Hadamard matrices do not exist for all integers $n$, the source dataset can be padded with zeroes so that a conformable Hadamard matrix is available. The random matrix $\mat{D}$ is a diagonal matrix where each nonzero element is an independent Rademacher random variable. The random matrix $\Phi$ subsamples $k$  rows of $\mat{H}$ with replacement. The structure of the Hadamard sketch allows for fast matrix multiplication, reducing calculation of the sketched dataset to $O(nd \log k)$ operations. 

The Clarkson-Woodruff sketch is a sparse random matrix \citep{clarkson_low_2013}. The  projection can be represented as the product of two independent random matrices, $\mat{S} = \mat{\Gamma}\mat{D}$, where $\mat{\Gamma}$ is a random $k \times n$ matrix and $\mat{D}$ is a random $n \times n$ matrix. The matrix $\mat{\Gamma}$ is formed by choosing one element in each column independently and setting the entry to $+1$. The matrix $\mat{D}$ is a diagonal matrix where each nonzero element is an independent Rademacher random variable. This results in a sparse $\mat{S}$, where there is only one nonzero entry per column. The sparsity of the Clarkson-Woodruff sketch speeds up matrix multiplication, dropping the complexity of  generating the sketched dataset to $O(nd)$.

\begin{figure}
\centering
\includegraphics[width=0.8\textwidth]{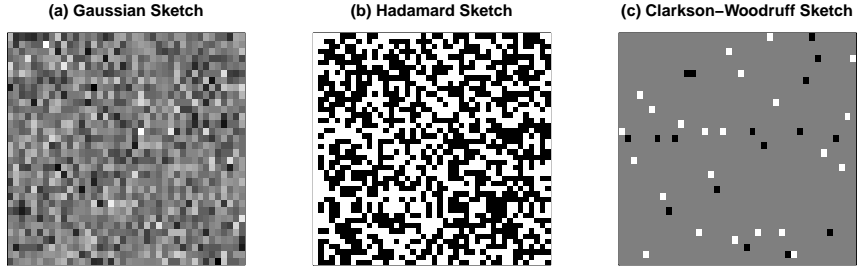}
\caption{Sampled sketching matrices $\mat{S}$ for $k=32, n=36$. Elements in the sketching matrix are coloured based on the value. One and negative one are coloured as black and white respectively. Intermediate values are in shades of grey. }
\label{fig:sketching_diagram}
\end{figure}

Figure \ref{fig:sketching_diagram} shows examples of the three sketches for $k=32, n=36$. The sketches are discussed in more detail in the supplementary material.



\begin{table}
\centering
\begin{tabular}{l r r r}
Algorithm & Sketching time  & Required sketch size $k$ \\
\hline
Gaussian sketch & $O(ndk) $  & $O [\lbrace d+\log(1/\delta)\rbrace/\epsilon^2] $\\
Hadamard sketch & $O(nd \log k)$ & $O [ (\surd{d}+\surd{\log n})^2\lbrace \log (d/\delta)\rbrace ]/ \epsilon^2]$\\
Clarkson-Woodruff Sketch & $O(nd)$& $O\lbrace d^2/(\delta\epsilon^2)\rbrace$ \\
\end{tabular}
\caption{Properties of different data oblivious random projections \citep{woodruff_sketching_2014}.The third column refers to the necessary sketch size $k$ to obtain an $\epsilon$-subspace embedding for an arbitrary $n \times d$ source dataset with at least probability $(1-\delta)$. }
\label{tab:run_time}
\end{table}

Data oblivious sketches are designed to give an $\epsilon$-subspace embedding for an arbitrary source dataset with at least probability $(1-\delta)$. Sketching algorithms are appealing for large $n$ problems as the required $k$ to attain the $(\delta, \epsilon)$ bound is independent of $n$ for the Gaussian and Clarkson-Woodruff sketches, and very weakly dependent on $n$ for the Hadamard sketch. Table \ref{tab:run_time} summarises existing results on the necessary $k$ to attain the $(\epsilon, \delta)$ bound.  Probabilistic worst case bounds for sketched regression are formed by noting that if a sketch produces an $\epsilon$-subspace embedding with probability at least $(1-\delta)$, then the bounds in Section \ref{subsec:worst_case} must hold with probability at least $(1-\delta)$. \citet{woodruff_sketching_2014} gives an excellent survey of work in this area. 

\subsection{Data aware sketches}
\label{subsec:data_aware}
As mentioned, data aware random projections can also be used to generate $\epsilon$-subspace embeddings. Data aware sketching is closely related to finite population subsampling methods \citep{ma_leveraging_2015}, in particular classic Hansen-Hurwitz estimators \citep{hansen_sampling_1943}. Suppose we sample $k$ observations from the original dataset with replacement using observation sampling weights $\pi_{1}, \ldots, \pi_{n}$. Let the data aware sketching matrix be constructed as $\mat{S}=\mat{R}\mat{W}$. The $k \times n$ matrix $\mat{W}$ subsamples $k$ rows of the source dataset with replacement. Each row of $\mat{W}$ contains a single nonzero entry. Element $W_{ij}$ is equal to one if the $j$th original observation is sampled in the $i$th sampling round for $i=1, ldots, k$ and $j \in \left\lbrace 1, \ldots, n \right\rbrace$. The $k \times k$ diagonal matrix $R$ rescales the subsampled rows. The $i$th diagonal element of $R$ is set to $1/(k\pi_{j})^{1/2}$ if $W_{ij}$ is equal to one, that is if row $j$ in the source dataset is subsampled by the $i$th row of the subsampling matrix $\mat{W}$. Using a data aware sketch, the sketched dataset is defined as
\begin{align*}
\widetilde{\mat{A}} &= \mat{S}\mat{A} \\
&= \mat{R}\mat{\Phi}\mat{A}.
\end{align*}
The sketched dataset has the property that $\mat{E}_{S}(\widetilde{\mat{A}}^{\T}\widetilde{\mat{A}} \mid \mat{A}^{\T}\mat{A}) = \mat{A}^{\T}\mat{A}$. The subsampling and rescaling can be interpreted as a Hansen-Hurwitz estimator of the full dataset sufficient statistics $\mat{A}^{\T}\mat{A}$.

Data aware sketching algorithms use the leverage scores of the observations to define the sampling weights $\pi_{1}, \ldots, \pi_{n}$ \citep{mahoney_randomized_2011, woodruff_sketching_2014}. Let the singular value decomposition of the source data matrix $\mat{A}$ be given by $\mat{A}=\mat{U}\mat{D}\mat{V}^{\T}$, where $\mat{U}$ is the $n \times d$ matrix of left singular vectors, $\mat{D}$ is a $d \times d$ matrix with the singular values of $\mat{A}$ on the diagonal, and $\mat{V}$ is the $d \times d$ matrix of right singular vectors. Let $\vect{u}_{i}^{\T}$ give the $i$th row in $\mat{U}$. The leverage score for observation $i$ is defined as $\lVert \vect{u}_{i} \rVert_{2}^{2}$.

Suppose the original dataset is centred, so each column of $\mat{A}$ has mean zero. The leverage scores then have a particularly intuitive interpretation in terms of the principal components decomposition of the source dataset. The row vector $\vect{u}_{i}^{T}\mat{D}$ gives the coordinates of observation $i$ on the principal component axes. The elements of the vector $\vect{u}_{i}$ give the coordinates of observation $i$ in a scaled system where the variance along each principal coordinate axis is set to be one. The leverage score $\lVert \vect{u}_{i} \rVert_{2}^{2}$ gives a measure of the distance from the origin in the principal coordinate system. This geometric perspective will also be of use when analysing data oblivious random projections.

Data oblivious random projections operate in a different manner to data aware random projections, as the sketched dataset is not a rescaled subset of the original instances. Data oblivious random projections generate a pseudo-dataset of $k$ observations using the source dataset as a component in the generative process. In Section \ref{sec:asymptotics} we will show that the leverage scores have an important role in describing the asymptotic behaviour of data oblivious random projections. We first establish some exact distributional results for the estimators $\vect{\beta}_{S}$ and $\vect{\beta}_{P}$ under the Gaussian sketch in Section \ref{sec:gaussian_complete}. In Section \ref{sec:asymptotics} we establish corresponding asymptotic results for the Hadamard and Clarkson-Woodruff projections under regularity conditions on the statistical leverage scores.

\section{Gaussian sketching}
\subsection{Complete sketching}
\label{sec:gaussian_complete}
The Gaussian sketch is mathematically tractable, and it is possible to establish a number of exact finite sample results regarding the performance of the sketched estimators. In this section we will develop the distribution of $\vect{\beta}_{S}$ when using a Gaussian sketch. As mentioned previously, all results treat $\vect{y}$ and $\mat{X}$ as fixed. The variability in $\vect{\beta}_{S}$ is solely due to the use of the random sketching matrix $\mat{S}$. Let $(\widetilde{y}_{j}, \widetilde{\vect{x}}_{j}^{\T})^{\T}$ refer to the $j$th row in the sketched data matrix $\widetilde{\mat{A}} = [\widetilde{\vect{y}}, \widetilde{\mat{X}}]$ for $j=1, \ldots, k$. Similarly, let $\vect{s}_{j}^{\T}$ denote the $j$th row in the sketching matrix $\mat{S}$. The sketched dataset consists of $k$ random units $(\widetilde{y}_{j}, \widetilde{\vect{x}}_{j}^{\T})$, $(j=1, \ldots, k)$.  The $j$th sketched response is given by $\widetilde{y}_{j} = \vect{s}_{j}^{\T}\vect{y}$, and the $j$th sketched predictor is calculated as $\widetilde{\vect{x}}_{j}=\vect{s}_{j}^{\T}\mat{X}$  $(j=1, \ldots, k)$. The $k$ sketched instances are independently distributed, because rows of the sketching matrix are independent. 

We take an indirect route to find the distribution of $\vect{\beta}_{S}$, by focusing on the distribution of the sketched data $\widetilde{\mat{A}}=[\widetilde{\vect{y}}, \widetilde{\mat{X}}]$ conditional on the original dataset $\mat{A}=[\vect{y}, \mat{X}]$. 
The initial step is to decompose the joint distribution on the sketched responses and predictors as the product of a marginal and conditional distribution. Specifically,
\begin{align*}
p(\widetilde{\vect{y}}, \widetilde{\mat{X}} \mid \vect{y}, \vect{X}) = p(\vect{\widetilde{y}}\mid\widetilde{\mat{X}}, \vect{y}, \vect{X})p(\widetilde{\mat{X}} \vert \vect{y}, \vect{X}).
\end{align*}
It can be shown that $p(\vect{\widetilde{y}}\mid\widetilde{\mat{X}}, \vect{y}, \vect{X})p(\widetilde{\mat{X}} \vert \vect{y}, \vect{X})$ has the structure of a hierarchical Gaussian linear model. 
We first show that the sketched dataset has a multivariate normal distribution, conditional on the source dataset. This follows as the sketched dataset can be expressed as a linear combination of Gaussian random variables. Specifically, row $j$ in the sketched dataset is $(\widetilde{y}_{j}, \widetilde{\vect{x}}_{j}^{\T})=\vect{s}_{j}^{\T}\mat{A}$. The random vector $(\widetilde{y}_{j}, \widetilde{\vect{x}}_{j}^{\T})^{\T}$ is given by the linear combination
\begin{align*}
 \begin{bmatrix}
\widetilde{y}_{j} \\
\widetilde{\vect{x}}_{j}
\end{bmatrix}  &=
\mat{A}^{\T}\vect{s}_{j}. 
\end{align*}
Conditional on $\mat{A}=[\vect{y}, \mat{X}]$,  $\mat{A}^{\T}\vect{s}_{j}$ is a linear combination of independent Gaussians as $\vect{s}_{j} \sim N(\vect{0}, \mat{I}_{d}/k)$.  As affine transformations of Gaussians are also multivariate normal,  $(\widetilde{y}_{j}, \widetilde{\vect{x}}_{j}^{\T})$ must then be jointly  normally distributed, conditional on the source data $\mat{A}=[\vect{y}, \mat{X}]$. It is easily shown that the joint distribution of the sketched responses and predictors is then
\begin{align*}
\quad 
\left. \begin{bmatrix}
\widetilde{y}_{j} \\
\widetilde{\vect{x}}_{j}
\end{bmatrix} \right\vert \vect{y}, \mat{X} &\sim N\left(\begin{bmatrix}
0 \\
\vect{0}
\end{bmatrix},   \dfrac{1}{k}\begin{bmatrix}
\vect{y}^{\T}\vect{y} & \ \vect{y}^{\T}\mat{X} \\
\mat{X}^{\T}\vect{y} &  \ \mat{X}^{\T}\mat{X}
\end{bmatrix}\right), \quad (j=1, \ldots, k). 
\end{align*}
Standard results on the multivariate normal distribution give that the conditional distribution of $\widetilde{y}_{j}$ given $\widetilde{\vect{x}}_{j}$ is also normal. A routine calculation shows that the conditional mean is related to $\vect{\beta}_{F}$, that is $
E_{S}(\widetilde{y}_{j} \mid \widetilde{\vect{x}}_{j}, \vect{y}, \mat{X}) = \widetilde{\vect{x}}_{j}^{\T}\vect{\beta}_{F}$.  The subscript $S$ is used on the expectation operator to emphasise that only random quantity is the sketching matrix. The conditional variance is related to the prediction error on the source dataset $RSS_{F}$,
\begin{align*}
\text{var}_{S}\left(\widetilde{y}_{j}\mid \widetilde{\vect{x}}_{j}, \vect{y}, \mat{X}\right) &=\dfrac{1}{k}\lbrace \vect{y}^{\T}\vect{y} - \mat{y}^{\T}\mat{X}(\mat{X}^{\T}\mat{X})^{-1}\mat{X}^{\T}\vect{y} \rbrace  \\
&= \dfrac{1}{k}RSS_{F}. 
\end{align*}
The subscript $S$ is again used to recognise that the source of the variance is the random sketching matrix, the source dataset is fixed. The step in the second line follows from sum of squares partitions in linear models \citep[Chapter 3]{searle_linear_1997}. Therefore, the conditional  distribution of $\widetilde{y}_{j}$ given the sketched predictors $\widetilde{\vect{x}}_{j}$ and the source dataset $[\vect{y}, \mat{X}]$ is
\begin{align*}
\qquad \qquad \widetilde{y}_{j} \mid \widetilde{\vect{x}}_{j}, \vect{y}, \mat{X} \sim N\left(\widetilde{\vect{x}}_{j}^{\T}\vect{\beta}_{F}, \dfrac{RSS_{F}}{k}\right) \quad  (j=1, \ldots, k). 
\end{align*}
This is the exact form of a standard Gaussian linear model. The distribution $p(\widetilde{\mat{X}}\mid \vect{y}, \mat{X})$ is easily obtained as the marginal distribution of $\widetilde{\vect{x}}_{j}$ is also multivariate normal,
\begin{align*}
\widetilde{\vect{x}}_{j} \sim N\left( \vect{0}, \mat{X}^{\T}\mat{X}/k\right), \quad (j=1,\ldots,k). 
\end{align*}
The sketching process can be described using the following hierarchical model,
\begin{align*}
\widetilde{\vect{y}} \mid \widetilde{\mat{X}}, \vect{y}, \mat{X}  & \sim N\left(\widetilde{\vect{X}}\vect{\beta}_{F}, \dfrac{RSS_{F}}{k}\mat{I}_{k}\right), \\
\widetilde{\mat{X}} \mid  \vect{y}, \mat{X}  & \sim MN\left(\vect{0}, \mat{I}_{k}, \dfrac{1}{k}\mat{X}^{\T}\mat{X}\right). 
\end{align*}
A Gaussian sketch effectively simulates a series of observations from a Gaussian linear model parametrised in terms of $\vect{\beta}_{F}$ and $RSS_{F}$, where the design matrix has a matrix normal distribution. We now turn to the distribution of $\vect{\beta}_{S}$. The distribution of $\vect{\beta}_{S}$ conditional on the sketched predictors follows immediately from standard results on linear models \citep[Chapter 3]{searle_linear_1997}.
\begin{align}
\vect{\beta}_{S} \mid \widetilde{\mat{X}}, \vect{y}, \mat{X} &\sim N\left(\vect{\beta}_{F}, \dfrac{RSS_{F}}{k}(\widetilde{\mat{X}}^{\T}\widetilde{\mat{X}})^{-1}\right). \label{eq:beta_S_condtional}
\end{align}
To obtain the marginal distribution of $\vect{\beta}_{S}$ it is necessary to integrate over the random sketched design matrix $\widetilde{\mat{X}}$. From properties of the normal distribution \citep{eaton_2007_multivariate}, it is possible to show $(\widetilde{\mat{X}}^{\T}\widetilde{\mat{X}}) \mid \vect{y}, \mat{X} \sim \text{Wishart}(k, \mat{X}^{\T}\mat{X}/k)$. As such,  
\begin{align*}
(\widetilde{\mat{X}}^{\T}\widetilde{\mat{X}})^{-1} \mid \vect{y}, \mat{X} &\sim \text{Inverse-Wishart}\left(k, k(\mat{X}^{\T}\mat{X})^{-1} \right).
\end{align*}
As seen in equation \eqref{eq:beta_S_condtional}, $\vect{\beta}_{S}$ is normally distributed when conditioned on the random Inverse-Wishart matrix $(\widetilde{\mat{X}}^{\T}\widetilde{\mat{X}})^{-1}$. The marginal distribution of $\vect{\beta}_{S}$ can then be described using the Normal Inverse-Wishart distribution \citep[p.73]{gelman_bayesian_2014}. The following theorem characterises the distribution of $\vect{\beta}_{S}$ under the Gaussian sketch.
\begin{theorem}
\label{thm:gaussian_exact_distribution}
Suppose $\vect{\beta}_{S}$ is computed using a Gaussian sketch and $k > p +1$. The conditional distribution of $\vect{\beta}_{S}$ is
\begin{align*}
(i)  \vect{\beta}_{S} \vert \widetilde{\mat{X}}, \vect{y}, \mat{X}  & \sim N \left(\vect{\beta}_{F}, \dfrac{RSS_{F}}{k}\left(\widetilde{\mat{X}}^{\T}\widetilde{\mat{X}}\right)^{-1} \right).
\end{align*} 
 The marginal distribution of $\vect{\beta}_{S}$ is
\begin{align*}
(ii)  \vect{\beta}_{S} \vert \mat{y}, \vect{X} \sim \textnormal{Student}\left(\vect{\beta}_{F}, \dfrac{RSS_{F}}{k-p+1}\left({\mat{X}}^{\T}{\mat{X}}\right)^{-1},  \ k-p+1\right).
\end{align*}
\end{theorem}
For proof see the supplementary material. 

An immediate application of result $(i)$ is the ability to generate exact confidence intervals for the elements of $\vect{\beta}_{S}$, methodology that does not appear to be present in the existing literature. Let $\vect{\beta}_{S}^{(i)}$ give the $i$th element of $\vect{\beta}_{S}$ and let $\vect{\beta}_{F}^{(i)}$ give the $i$th element of $\vect{\beta}_{F}$. Let $RSS_{S}$ denote the sketched residual sum of squares, $RSS_{S} = \lVert \widetilde{\vect{y}} -\widetilde{\mat{X}}\vect{\beta}_{S}\rVert_{2}^{2}$. To construct a $100(1-\alpha)\%$ confidence interval, let $w_{ii} = (\widetilde{\mat{X}}^{\T}\widetilde{\mat{X}})^{-1}_{ii}$, and $t_{\text{crit}}$ denote the $100(1-\alpha/2)$th percentile of the \textit{t}-distribution with $k-p$ degrees of freedom. Then from standard results on Gaussian linear models \citep{searle_linear_1997},
\begin{align}
\vect{\beta}_{S}^{(i)} &\pm t_{\text{crit}} \times (w_{ii}RSS_{S}/(k-p))^{1/2} \label{eq:gaussian_exact_ci}
\end{align}
gives an exact $100(1-\alpha)\%$ confidence interval for $\vect{\beta}_{F}^{(i)}$. Again assuming that $k > p+1$, it should be noted that the variance of $\vect{\beta}_{S}$, 
\begin{align}
\text{var}(\vect{\beta}_{S} \mid \vect{y}, \mat{X}) &= \dfrac{RSS_{F}}{(k-p+1)}(\mat{X}^{\T}\mat{X})^{-1} \label{eq:beta_s_variance}
\end{align}
is not dependent on the compression ratio $k/n$. Although $RSS_{F}$ can be expected to grow linearly with $n$, this will generally be counterbalanced by $(\mat{X}^{\T}\mat{X})^{-1}$ decreasing linearly with $n$. The distribution of the approximation error  $\lVert \vect{\beta}_{S} - \vect{\beta}_{F} \rVert_{2}$ will largely be controlled by the target dimension $k$. This speaks to the defining characteristic of sketching algorithms, that given a fixed $k$, the stochastic approximation error does not necessarily increase with size of the original dataset $n$. 
\subsection{Partial sketching}
\label{subsec:partial_sketching}
Partial sketching was first proposed by \cite{dhillon_new_2013} using uniform subsampling, and later studied for general sketches by \cite{pilanci_iterative_2016}. Existing results on partial sketching highlight that the model sum of squares influences the approximation error of the partial sketched estimator $\vect{\beta}_{P}$. It is simple to see that the variance of the partial sketched estimator will not be a function of the residual sum of squares. From the normal equations it holds that
$
\mat{X}^{\T}\vect{y} = \mat{X}^{\T}\mat{X}\vect{\beta}_{F}
$. 
Using this property, we see that conditional on $\vect{y}, \mat{X}$, the variance of the random linear combination $\vect{\beta}_{P} =(\mat{X}^{\T}\mat{S}^{\T}\mat{S}\mat{X})^{-1}\mat{X}^{\T}\vect{y}= (\mat{X}^{\T}\mat{S}^{\T}\mat{S}\mat{X})^{-1}\mat{X}^{\T}\mat{X}\vect{\beta}_{F}$ will be a function of the covariates $\mat{X}$ and the fitted values $\mat{X}\vect{\beta}_{F}$. The residual vector has no influence on the variance of the partial sketching estimator, and as such the variance of $\vect{\beta}_{P}$ will not be related to the residual sum of squares.  This suggests that when the noise level is high, partial sketching may become preferable to complete sketching. This idea has been touched on in the existing literature, but specific guidelines are lacking \citep{becker_robust_2015, dhillon_new_2013}. A statistical analysis can provide some insight into this issue.

The hierarchical model for complete sketching gave an intuitive statistical perspective on the mechanics of the algorithm. Partial sketching seems to lack a similar conceptual device. The least squares coefficients can be represented as the solution to the linear system of the equations $\mat{X}^{\T}\mat{X}\vect{b}=\mat{X}^{\T}\vect{y}$. Partial sketching simply returns the solution, $\vect{b}$, to the approximate linear system $\widetilde{\mat{X}}^{\T}\widetilde{\mat{X}}\vect{b}=\mat{X}^{\T}\vect{y}$. Lacking a convenient representation for the estimator, we must proceed in a more pedestrian manner. The mean square error of the estimator $\vect{\beta}_{P}$ can be determined using only mean and variance information, and this will be the goal for now. The key observation is that $
(\widetilde{\mat{X}}^{\T}\widetilde{\mat{X}})^{-1} \mid \vect{y}, \mat{X} \sim \text{InvWishart}\left(k, k(\mat{X}^{\T}\mat{X})^{-1} \right).$
Conditional on $\vect{y}, \mat{X}$, the estimator $\vect{\beta}_{P}=(\widetilde{\mat{X}}^{\T}\widetilde{\mat{X}})^{-1}\mat{X}^{\T}\vect{y}$ is a linear combination of the elements of an Inverse-Wishart random matrix. However, this is a non-standard distribution and it is difficult to directly express the distribution function of $\vect{\beta}_{P}$. Despite this, it is straightforward to determine the mean and variance of $\vect{\beta}_{P}$. From properties of the Inverse-Wishart distribution, it can be seen that the partial sketched estimator is biased, with mean 
\begin{align*}
E_{S}(\vect{\beta}_{P} \mid \vect{y}, \mat{X})
&= \dfrac{k}{(k-p-1)}\vect{\beta}_{F},
\end{align*}
where it is assumed that $k > p+3$. This motivates an alternative unbiased estimator
 \begin{align*}
\vect{\beta}_{P}^* 
&=   \dfrac{(k-p-1)}{k}(\widetilde{\mat{X}}^{\T}\widetilde{\mat{X}})^{-1}{\mat{X}}^{\T}{\vect{y}}.
\end{align*}
Determining the variance of $\vect{\beta}_{P}$ and the unbiased $\vect{\beta}_{P}^*$ is a more lengthy computation (see supplementary material). The variance of the biased estimator $\vect{\beta}_{P}$ is
\begin{align}
\text{var}(\vect{\beta}_{P}\mid \vect{y}, \mat{X})&= \dfrac{k^2}{(k-p)(k-p-1)(k-p-3)}\left(MSS_{F}(\mat{X}^{\T}\mat{X})^{-1} + \dfrac{k-p+1}{k-p-1} \vect{\beta}_{F}\vect{\beta}_{F}^{\T}\right). \label{eq:beta_p_variance}
\end{align}
The variance of the unbiased estimator $\vect{\beta}_{P}^{*}$ is
\begin{align}
\text{var}(\vect{\beta}_{P}^* \mid \vect{y}, \mat{X})&= \dfrac{(k-p-1)}{(k-p)(k-p-3)}\left(MSS_{F}(\mat{X}^{\T}\mat{X})^{-1} + \dfrac{k-p+1}{k-p-1}\vect{\beta}_{F}\vect{\beta}_{F}^{\T}\right). \label{eq:beta_p_star_variance}
\end{align}
The variances of $\vect{\beta}_{P}$ and $\vect{\beta}_{P}^{*}$ have a similar structure to the variance of $\vect{\beta}_{S}$. The main point of difference is that the variance of $\vect{\beta}_{S}$ depends on the residual sum of squares, whereas the variance of $\vect{\beta}_{P}$ and $\vect{\beta}_{P}^{*}$ depends on the model sum of squares. 

As mentioned the explicit form of the sampling distribution is hard to obtain, but by making a connection with method of moments estimation it is possible to establish asymptotic normality of both $\vect{\beta}_{P}$ and $\vect{\beta}_{P}^*$ as $k$ tends to infinity. This motivates the construction of approximate confidence intervals. As the exact variance is unknown we propose the following estimator 
\begin{align}
\text{var}(\vect{\beta}_{P}^* \mid \vect{y}, \mat{X})&\approx \dfrac{(k-p-1)}{(k-p)(k-p-3)}\left(\left(\dfrac{k-p-1}{k}\right)MSS_{S}(\widetilde{\mat{X}}^{\T}\widetilde{\mat{X}})^{-1} + \vect{\beta}_{P}^{*}\vect{\beta}_{P}^{*\T}\right). \label{eq:hat_beta_p_star_variance}
\end{align}
\subsection{Relative efficiency}
The relative efficacy of complete and partial sketching is also of interest. As the plug in estimator $\vect{\beta}_{P}$ has a higher mean square error than $\vect{\beta}_{P}^*$, it will not be considered in this section. The performance of the complete sketching estimator $\vect{\beta}_{S}$ and the unbiased partial sketched estimator $\vect{\beta}_{P}^*$ will be compared in terms of mean squared error. As both $\vect{\beta}_{F}$ and $\vect{\beta}_{P}^*$ are unbiased, the mean squared error can be computed using their respective covariance matrices, that is
\begin{align*}
E_{S} \left  (\lVert \vect{\beta}_{S} - \vect{\beta}_{F} \rVert_{2}^{2} \ \mid \ \vect{y}, \mat{X} \right)  &=  \text{tr}(\text{var}(\vect{\beta}_{S})), \\
E_{S} \left  (\lVert \vect{\beta}_{P}^{*} - \vect{\beta}_{F} \rVert_{2}^{2} \ \mid \ \vect{y}, \mat{X} \right) &=  \text{tr}(\text{var}(\vect{\beta}_{P}^*)).
\end{align*}
Comparing  \eqref{eq:beta_s_variance} and \eqref{eq:beta_p_star_variance}, the variance of $\vect{\beta}_{P}^{*}$ is dependent on $MSS_{F}$, whereas the variance of $\vect{\beta}_{S}$ is dependent on $RSS_{F}$. This suggests that the signal to noise ratio in the source dataset will be an influential factor in determining which estimator is more efficient. When $R^2_{F}$ is close to one complete sketching can be orders of magnitude more efficient than partial sketching, and when $R^2_{F}$ is close to zero, partial sketching can be orders of magnitude more efficient than complete sketching. 

\subsection{Combined estimator}
\label{subsec:combined}
So far we have assumed that an analyst much choose between one of the two methods. Obtaining both $\vect{\beta}_{P}^{*}$ and $\vect{\beta}_{S}$ from a single sketch is computationally cheap, and may be an attractive strategy. The most demanding operation with the sketched data is calculating $(\widetilde{\mat{X}}^{\T}\widetilde{\mat{X}})^{-1}$. Given this quantity it is economical to compute both $\vect{\beta}_{S}$ and $\vect{\beta}_{P}^{*}$. \cite{becker_robust_2015} mention they are presently investigating such a strategy, but do not give any details. Our motivation for a  combined estimator is driven by the fact even when using a single sketch $(\widetilde{\vect{y}}, \widetilde{\mat{X}})$, the two estimators are uncorrelated, that is
$\text{cov}\left( \vect{\beta}_{P}^*, \vect{\beta}_{S}\right) =\mat{0}
$. This is established by taking iterated expectations, and using the hierarchical model established in Section \ref{sec:gaussian_complete} (see supplementary material). A simple strategy is then to take a weighted combination of $\vect{\beta}_{S}$ and $\vect{\beta}_{P}^{*}$. A combined estimator $\vect{\beta}_{C}$ can be defined as
\begin{align*}
\vect{\beta}_{C} &= \phi \vect{\beta}_{S} + (1-\phi)\vect{\beta}_{P}^{*},
\end{align*} 
for some $0 \le \phi \le 1$. The value of $\phi$ that minimises the mean square error is 
\begin{align*}
\phi_{\text{opt}} &= \dfrac{\text{tr}(\text{var}(\vect{\beta}_{P}^{*}))}{\text{tr}(\text{var}(\vect{\beta}_{P}^{*})) + \text{tr}(\text{var}(\vect{\beta}_{S}))}. 
\end{align*}

Use of the weighted estimator is expected to be most beneficial when the signal to noise ratio is moderate, that is $R^{2}_{F} \approx 0.5$. When the signal to noise ratio is either very high or very low, there is little gain from using the weighted estimator as either the complete or partial estimator will dominate. 
\subsection{One-step correction}
As noted by a referee, the combined estimator is related to another strategy in the sketching literature for improving $\vect{\beta}_{S}$. \cite{dhillon_new_2013} and \cite{pilanci_iterative_2016} propose a refinement procedure using gradient information from the source dataset. The one-step corrected estimator, $\vect{\beta}_{H}$, is defined as
\begin{align}
\vect{\beta}_{H} &= \vect{\beta}_{S} + (\widetilde{\mat{X}}^{\T}\widetilde{\mat{X}})^{-1}\mat{X}^{\T}(\vect{y}-\mat{X}\vect{\beta}_{S}) \nonumber \\
&= (\mat{I}-(\widetilde{\mat{X}}^{\T}\widetilde{\mat{X}})^{-1}\mat{X}^{\T}\mat{X})\vect{\beta}_{S} + (\widetilde{\mat{X}}^{\T}\widetilde{\mat{X}})^{-1}\mat{X}^{\T}\vect{y}. \label{eq:beta_h_iterative}
\end{align}
The one-step estimator can be interpreted as a single step of the iterative Hessian sketch proposed by \cite{pilanci_iterative_2016}, initialised at $\vect{\beta}_{S}$. The optimal least square solution $\vect{\beta}_{F}$ satisfies $\mat{X}^{\T}(\vect{y}-\mat{X}\vect{\beta}_{F})=\vect{0}$ so
\begin{align}
\vect{\beta}_{F} &= \vect{\beta}_{F} + (\widetilde{\mat{X}}^{\T}\widetilde{\mat{X}})^{-1}\mat{X}^{\T}(\vect{y}-\mat{X}\vect{\beta}_{F}) \nonumber \\ 
&= (\mat{I}-(\widetilde{\mat{X}}^{\T}\widetilde{\mat{X}})^{-1}\mat{X}^{\T}\mat{X})\vect{\beta}_{F} + (\widetilde{\mat{X}}^{\T}\widetilde{\mat{X}})^{-1}\mat{X}^{\T}\vect{y}. \label{eq:beta_f_iterative}
\end{align}
Subtracting \eqref{eq:beta_f_iterative} from \eqref{eq:beta_h_iterative} gives the following expression for the error
\begin{align}
\vect{\beta}_{H}-\vect{\beta}_{F} &= (\mat{I}-(\widetilde{\mat{X}}^{\T}\widetilde{\mat{X}})^{-1}\mat{X}^{\T}\mat{X})(\vect{\beta}_{S}-\vect{\beta}_{F}).
\end{align}
The expected squared error is then
\begin{align*}
E_{S} (\lVert \vect{\beta}_{H}-\vect{\beta}_{F} \rVert_{2}^{2}) &= E_{S} \left\lbrace (\vect{\beta}_{S}-\vect{\beta}_{F})^{\T}(\mat{I}-(\widetilde{\mat{X}}^{\T}\widetilde{\mat{X}})^{-1}\mat{X}^{\T}\mat{X})^{\T}(\mat{I}-(\widetilde{\mat{X}}^{\T}\widetilde{\mat{X}})^{-1}\mat{X}^{\T}\mat{X}) (\vect{\beta}_{S}-\vect{\beta}_{F}) \right\rbrace.
\end{align*}
We can then take iterated expectations using the hierarchical model in Section \ref{sec:gaussian_complete}. The action of the sketch can be taken over $\widetilde{\mat{X}}$then over the conditional distribution $\widetilde{\vect{y}}$ given $\widetilde{\mat{X}}$. Theorem \ref{thm:gaussian_exact_distribution} (i) gives the distribution of $\vect{\beta}_{S}$ conditional on $\widetilde{\mat{X}}$. We thus have
\begin{align}
E_{S} (\lVert \vect{\beta}_{H}-\vect{\beta}_{F} \rVert_{2}^{2}) &= E_{\widetilde{\mat{X}}}\left[ E_{\widetilde{\vect{y}} \mid \widetilde{\mat{X}}} \left\lbrace(\vect{\beta}_{S}-\vect{\beta}_{F})^{\T}(\mat{I}-(\widetilde{\mat{X}}^{\T}\widetilde{\mat{X}})^{-1}\mat{X}^{\T}\mat{X})^{\T}(\mat{I}-(\widetilde{\mat{X}}^{\T}\widetilde{\mat{X}})^{-1}\mat{X}^{\T}\mat{X}) (\vect{\beta}_{S}-\vect{\beta}_{F}) \mid \widetilde{\mat{X}} \right\rbrace  \right]  \nonumber \\
&= E_{\widetilde{\mat{X}}}\left[ \text{tr}\left(\text{var}(\vect{\beta}_{S} \mid \widetilde{\mat{X}}) (\mat{I}-(\widetilde{\mat{X}}^{\T}\widetilde{\mat{X}})^{-1}\mat{X}^{\T}\mat{X})^{\T}(\mat{I}-(\widetilde{\mat{X}}^{\T}\widetilde{\mat{X}})^{-1}\mat{X}^{\T}\mat{X}\right)\right] \nonumber \\
&= E_{\widetilde{\mat{X}}}\left\lbrace \text{tr}\left(\dfrac{RSS_{F}}{k}(\widetilde{\mat{X}}^{\T}\widetilde{\mat{X}})^{-1} (\mat{I}-(\widetilde{\mat{X}}^{\T}\widetilde{\mat{X}})^{-1}\mat{X}^{\T}\mat{X})^{\T}(\mat{I}-(\widetilde{\mat{X}}^{\T}\widetilde{\mat{X}})^{-1}\mat{X}^{\T}\mat{X})\right) \right\rbrace. \label{eq:one_step_mse}
\end{align} 
The key term in \eqref{eq:one_step_mse} is the random matrix $(\widetilde{\mat{X}}^{\T}\widetilde{\mat{X}})^{-1}$. Now as $(\widetilde{\mat{X}}^{\T}\widetilde{\mat{X}})^{-1} \sim \text{Inverse-Wishart}(k, k(\mat{X}^{\T}\mat{X})^{-1})$, it is possible to evaluate the expectation in \eqref{eq:one_step_mse} using moments of the Inverse-Wishart distribution. The exact expression involves $E(\widetilde{\mat{X}}^{\T}\widetilde{\mat{X}})^{-1}$,  $E(\widetilde{\mat{X}}^{\T}\widetilde{\mat{X}})^{-2}$ and  $E(\widetilde{\mat{X}}^{\T}\widetilde{\mat{X}})^{-3}$. Formulae for the required moments are given in \cite{letac_invariant_2004}. The main conclusions are that the one-step estimator $\vect{\beta}_{H}$ can have a larger mean square error than $\vect{\beta}_{S}$ when the sketch size to variables ratio $k/p$ is close to one. As $k$ increases the one-step estimator becomes more efficient than both $\vect{\beta}_{S}$ and $\vect{\beta}_{C}$ with the optimal weight $\phi_{\text{opt}}$. The relative efficiency of $\vect{\beta}_{C}$ to $\vect{\beta}_{S}$ is at most two. The relative efficiency of $\vect{\beta}_{H}$ to $\vect{\beta}_{S}$ can be much larger, providing that $k/p$ is sufficiently large. The exact relationship is a function of $k$ and $p$. Direct comparisons between $\vect{\beta}_{H}$,  $\vect{\beta}_{S}$ and $\vect{\beta}_{P}^{*}$ are difficult, as the one-step estimator $\vect{\beta}_{H}$ requires gradient information $\mat{X}^{\T}(\vect{y}-\vect{X}\vect{\beta}_{S})$. Calculation of the gradient requires access to the full dataset. The single pass estimators $\vect{\beta}_{S}$ and $\vect{\beta}_{P}^{*}$ require only the sketched dataset and the summary statistic $\mat{X}^{\T}\vect{y}$. Additionally, the iterative correction can also be applied to $\vect{\beta}_{P}$ or $\vect{\beta}_{P}^{*}$. We are currently investigating the properties of iterative sketching algorithms in more detail using the asymptotic results developed in this paper. 
\section{Asymptotics}
\label{sec:asymptotics}
\subsection{Preliminaries}
Finite sample distributions of random projection estimators can be mathematically intractable, and as such asymptotic analysis can be a powerful tool \citep{li_very_2006, diaconis_1984_asymptotics}. It is a very difficult task to establish meaningful finite sample results for the Hadamard and Clarkson-Woodruff sketches, as they are discrete distributions over an enormous combinatorial space. The explicit finite sample distribution  of the sketched estimators can be written as a sum over all these possible combinations, but such a representation is not very informative. Instead, it is useful to study the large $n$ distribution of the estimators $\vect{\beta}_{S}$ and $\vect{\beta}_{P}$ to obtain an interpretable expression. 

As $\vect{\beta}_{F}$ is the estimand in sketching algorithms, this requires conditioning on the source data in the asymptotic analysis. To elaborate, let $\mat{A}_{(n)} = [\vect{y}_{(n)}, \mat{X}_{(n)}]$ represent the $n \times d$ source data matrix of full column rank. Any source data matrix $\mat{A}_{(n)}$ has a set of associated least squares coefficients, which will here be denoted $\vect{\beta}_{F}^{(n)}$. The overall goal is to determine the asymptotic form of the distributions $p(\vect{\beta}_{S} \mid \mat{A}_{(n)})$ and $p(\vect{\beta}_{P}^{*} \mid \mat{A}_{(n)})$ for some arbitrary large dataset $\mat{A}_{(n)}$. 

To take limits, we employ a fixed sequence of $n \times d$ datasets, all of rank $d$.  In the regression scenario this amounts to assuming that $\mat{X}_{(n)}$ is of full column rank and that $\vect{y}_{(n)}$ is not a perfect linear combination of the columns of $\mat{X}_{(n)}$ for all $n$. Conditioning on $\mat{A}_{(n)}$ is effectively the same as treating the full dataset as an arbitrary sequence of constants $A_{ij}$ for $i=1, \ldots, n$, $j=1, \ldots, d$. This is analogous to large sample results for regression models where the design matrix is treated as arbitrary set of constants, and the random variables of interest are the error terms, for example see \citet[Section 2.5]{van_der_vaart_asymptotic_1998}. Here the source dataset is treated as a sequence of constants and the random variables of interest are the elements of the sketching matrix. 

The asymptotic analysis is carried out in two stages. The initial step is to establish asymptotic normality of the sketched dataset. The regularity condition for the central limit theorem highlights the influential role of the leverage scores of the observations in the source dataset. This is then followed by an analysis of the limiting distribution of $\vect{\beta}_{S}$, and $\vect{\beta}_{P}^{*}$. There is some related work by \cite{ma_statistical_2015} who develop Taylor series approximations for the bias and variance of data aware sketched regression estimators, where the asymptotic expansion is taken in the sketch size $k$. Our work is different as we study data  oblivious random projections and build our asymptotic results from  a conditional central limit theorem for the sketched data matrix. The conditional central limit theorem is established for fixed $k$ and $d$, taking the number of source observations to $n$ to infinity. 

\subsection{Sketching central limit theorem}
A central limit theorem for sparse sketching matrices with independent entries is given in \cite{li_very_2006}.  The Clarkson-Woodruff sketch and the Hadamard sketch have dependent entries, and as such we use a different method of proof. Under some regularity conditions the Hadamard and Clarkson-Woodruff sketches produce sketched data that asymptotically has the same matrix normal distribution as under the Gaussian sketch. Using a Gaussian sketch, conditional on $\mat{A}$, 
\begin{align}
\widetilde{\mat{A}} \sim {MN}(\mat{0}, \mat{I}_{k}, \mat{A}^{\T}\mat{A}/k). \label{eq:gaussian_sketch_matrix_normal}
\end{align}
Each row is statistically independent, and marginally normally distributed with covariance matrix $\mat{A}^{\T}\mat{A}/k$. Although asymptotic normality may not be particularly surprising seeing as the sketched data are linear combinations of random vectors, the proof is not immediate due to the dependence in the Hadamard and Clarkson-Woodruff sketches. The difficulties caused by the dependence are most easily illustrated for the Clarkson-Woodruff sketch.

\begin{algorithm}[!h]
	\caption{Clarkson-Woodruff sketch} \label{alg:CW_sketch}
	\begin{tabbing}
		\enspace $\widetilde{\mat{A}} \gets \mat{0}$ \qquad {Initialise sketched dataset as $k \times d$ matrix of zeroes}\\
		\enspace For $i=1$ to $i=n$ \\
		\qquad Sample $z \sim \text{Uniform}(1, \ldots, k)$ \qquad {Sample random index} \\
		\qquad Sample $r \sim \text{Uniform}(-1, +1)$ \qquad {Sample random sign} \\\
		\qquad $\widetilde{\mat{A}}_{z} \gets   r\times \mat{A}_{i} + \widetilde{\mat{A}}_{z}$ \qquad {Multiply by $r$ and add to row $z$ in sketch} \\
		\enspace Output $\widetilde{\mat{A}}$ \qquad{Output sketched dataset}
	\end{tabbing}
\end{algorithm}

The behaviour of the Clarkson-Woodruff sketch can be represented as a many to less mapping. Each row in the source dataset is assigned to a single row in the sketched dataset. The Clarkson-Woodruff sketch has an alternative streaming construction that highlights this property, given in Algorithm \ref{alg:CW_sketch}. As each row in the source dataset  only contributes to a single row in the sketched dataset, it might be expected that this results in some statistical dependence amongst the rows of the sketched dataset. Additionally, although it seems each row in the sketched dataset will be marginally normally distributed, it is not clear if joint asymptotic normality over all rows will hold.  Similar conundrums arise when examining the Hadamard sketch in detail. 

The $k \times d$ random matrix $\widetilde{\mat{A}}$ is the output of a stochastic process governed by the fixed  $n \times d$ source dataset $\mat{A}_{(n)}$ and the distribution of the random $k \times n$ sketching matrix $\mat{S}$. The sketched dataset is a  linear combination of random vectors, the number of which increases with $n$. As such, we can expect $\widetilde{\mat{A}}$ to demonstrate some stable limiting behaviour as $n$ grows larger. Under an assumption on the limiting leverage scores of the source data matrix, we can establish a central limit theorem for the sketched dataset. Recall the singular value decomposition of the source dataset $\mat{A}_{(n)} = \mat{U}_{(n)}\mat{D}_{(n)}\mat{V}^{\T}_{(n)}$. The leverage score of observation $i$ in the source dataset is defined as $\lVert \vect{u}_{(n)i} \rVert_{2}^{2}$ where $\vect{u}_{(n)i}^{\T}$ gives row $i$ in $\mat{U}_{(n)}$. The leverage scores of the observations in the source data matrix have been identified an important structural property of sketching algorithms \citep{mahoney_structural_2016}. Assumption  \ref{assump:leverage} highlights their role in establishing asymptotic normality of the sketched data matrix. 

\begin{assumption}
	\label{assump:leverage}
  Let the singular value decomposition of the $n \times d$ source dataset be given by $\mat{A}_{(n)}=\mat{U}_{(n)}\mat{D}_{(n)}\mat{V}_{(n)}^{\T}$. Let $\vect{u}_{(n)i}^{\T}$ give the $i$th row in $\mat{U}_{(n)}$. Assume that the maximum leverage score tends to zero, that is
	\begin{align*}
	\lim_{n \to \infty} \underset{i=1, \ldots, n}{\text{max}} \lVert \vect{u}_{(n)i} \rVert_{2}^{2} = 0. 
	\end{align*}
\end{assumption}
Theorem \ref{thm:sketching_clt} gives the sketching central limit theorem. 
\begin{theorem}
	\label{thm:sketching_clt}
	Consider a fixed sequence of arbitrary $n \times d$ data matrices $\mat{A}_{(n)}$, where $d$ is fixed. Let $\mat{A}_{(n)}=\mat{U}_{(n)}\mat{D}_{(n)}\mat{V}_{(n)}^{\T}$ represent the singular value decomposition of $\mat{A}_{(n)}$. Let $\mat{S}$ be a $k \times n$ Hadamard or Clarkson-Woodruff sketching matrix where $k$ is also fixed. Suppose that Assumption 1 on the maximum leverage score is satisfied. Then as $n$ tends to infinity with $k$ and $d$ fixed, we have the following convergence in distribution
	\begin{align*}
[\widetilde{\mat{A}}\mat{V}_{(n)}\mat{D}_{(n)}^{-1} \ \mid \ \mat{A}_{(n)}] \to \emph{MN}(\mat{0}, \mat{I}_{k}, \mat{I}_{d}/k).
	\end{align*}
\end{theorem}

The proof is given in the supplementary material. Heuristically, for large $n$ we expect the matrix normal result \eqref{eq:gaussian_sketch_matrix_normal} to approximately hold for the Hadamard and Clarkson-Woodruff sketches.  The significance of Assumption 1 is perhaps best explained by making a connection to a version of the Lindeberg-Feller theorem for triangular arrays of uniformly bounded random variables.
 
\begin{theorem}[\citealp{billingsley_1995_probability}, Chapter 5]
\label{thm:bounded_clt}
For each $n \in \mathbb{N}$, let $Z_{n1}, Z_{n2}, \ldots, Z_{nr_{n}}$ be a sequence of independent random variables with $E(Z_{ni})=0$ and $\textnormal{var}(Z_{ni})=\sigma^2_{ni}$ for $i=1, \ldots, r_{n}$. Let $s_{n}^{2}=\sum_{i=1}^{r_n}\sigma^2_{ni}$ and assume that $r_{n} \to \infty$ as $n \to \infty$. Suppose that we can form a sequence of upper bounds $(K_{n})_{n \in \mathbb{N}}$ such that for each $n$,
\begin{align*}
|Z_{ni}|  \le K_{n} \textnormal{ almost surely for $i=1, \ldots, r_n$}.    
\end{align*}
Then if $K_{n}/s_{n} \to 0$ as $n \to \infty$ we have the following convergence in distribution
\begin{align*}
    \dfrac{1}{s_{n}}\sum_{i=1}^{r_{n}}Z_{ni} \to N(0,1)
\end{align*}
\end{theorem} 
 
In Theorem \ref{thm:bounded_clt}, the condition that $K_{n}/s_{n} \to 0$ ensures that no random variable in a particular row of the array has too much pull over the sum $\sum_{i=1}^{r_n}Z_{ni}$. A triangular array of random variables satisfying the conditions in Theorem \ref{thm:bounded_clt} is often said to be uniformly asymptotically negligible, in that no single term has undue influence over the random sum. We can make an analogy to the leverage score condition in the sketching central limit theorem (Theorem \ref{thm:sketching_clt}). The sum of the statistical leverage scores is always equal to the rank of the source dataset. As we have assumed that each dataset in the sequence is of rank $d$, we have that $\sum_{i=1}^{n}\lVert \vect{u}_{(n)i} \rVert_{2}^{2}=d$ for all $n$. As $n$ grows we need the maximum contribution from a single term in the sum to tend to zero. The limiting leverage scores must satisfy an asymptotic negligibility condition, so that each individual observation provides a vanishingly small contribution to the total sum of the leverage scores. 

As mentioned in the discussion of data aware sketching, the leverage scores have a particularly intuitive interpretation in terms of the principal components decomposition of the source dataset. The row vector $\vect{u}_{(n)i}^{\T}\mat{D}_{(n)}$ gives the coordinates of observation $i$ on the principal component axes. The elements of the vector $\vect{u}_{(n)i}$ give the coordinates of observation $i$ in a scaled system where the variance along each principal coordinate axis is set to be one. Treating the source dataset as a point cloud in Euclidean space, Assumption 1 essentially implies that there are no extreme outliers as $n$ tends to infinity. Each observation must have a negligible contribution to the total variance along each principal component axis.

\subsection{Sketching estimators}
The central limit theorem for the sketched data suggests that the results about $\vect{\beta}_{S}$ and $\vect{\beta}_{P}$ for the Gaussian sketch will also approximately hold for the Hadamard and Clarkson-Woodruff sketches for large $n$. In order to establish convergence of the estimators it helps to adopt an extra assumption on the sequence of source datasets. 
\begin{assumption}
	\label{assump:limiting}
 \begin{align*}
	\underset{n \to \infty}{\lim } n^{-1}\begin{bmatrix}
	\vect{y}^{\T}_{(n)}\vect{y}_{(n)} & \vect{y}^{\T}_{(n)}\mat{X}_{(n)} \\
	\mat{X}^{\T}_{(n)}\vect{y}_{(n)} & \mat{X}^{\T}_{(n)}\mat{X}_{(n)}
	\end{bmatrix} = \mat{Q} \qquad \text{for some positive-definite matrix $\mat{Q}$. }
	\end{align*}
\end{assumption}
It is worth discussing the significance of the limiting matrix $\mat{Q}$. A useful comparison can be made to asymptotic theory for regression models, where a common assumption is that the design matrix satisfies the limit condition $n^{-1}
\mat{X}^{\T}_{(n)}\mat{X}_{(n)} \to \mat{B}$, where $\mat{B}$ is some positive definite matrix \citep{white1984asymptotic, greene1997econometric}. The development of asymptotic results is often eased by treating the covariates as a random sample, although this requires positing a realistic probability model for the covariates,  which may be difficult. Treating the covariates as an arbitrary fixed sequence relaxes this assumption and covers more general scenarios. Although it is possible to establish asymptotic results when $n^{-1}\mat{X}^{\T}_{(n)}\mat{X}_{(n)}$ is not required to converge to any fixed matrix, proofs can become very technical \citep[Appendix A.2]{fahrmeir1994multivariate}. Imposing a limiting value for $n^{-1}\mat{X}^{\T}_{(n)}\mat{X}_{(n)}$ simplifies arguments and can be seen as a compromise between making strong and weak assumptions about the covariates \citep[p.46]{fahrmeir1994multivariate}. There is an analogous motivation for Assumption \ref{assump:limiting}, the limiting matrix $\mat{Q}$ is present to avoid specifying a probability model for the source dataset, without overcomplicating the mathematical analysis. 

Setting up a limit theorem requires a little extra care with notation. As we have a sequence of datasets $\mat{A}_{(n)}$, there is a corresponding sequence of optimal least squares coefficients $\vect{\beta}_{F}^{(n)}$. Similarly, there is a sequence of squared residual errors $RSS_{F}^{(n)}$ and model sum of squares $MSS_{F}^{(n)}$. As the sequence of datasets are fixed, $\vect{\beta}_{F}^{(n)}$ , $RSS_{F}^{(n)}$ and $MSS_{F}^{(n)}$ are a deterministic sequence. 

Under Assumptions \ref{assump:leverage} and \ref{assump:limiting}, it is possible to establish an asymptotic result for $\vect{\beta}_{S}$ and $\vect{\beta}_{P}$.
\begin{theorem}
\label{thm:random_projection_beta_S}
Suppose that Assumptions 1 and 2 hold, $k \ge p $, and $\vect{\beta}_{S}$ is computed using a Hadamard or Clarkson-Woodruff sketch. Let $(\widetilde{\mat{X}}^{\T}\widetilde{\mat{X}})^{+}$ denote the Moore-Penrose pseudo-inverse of $(\widetilde{\mat{X}}^{\T}\widetilde{\mat{X}})$. Let
\begin{align*}
\widetilde{\mat{V}}_{(n)}  = \dfrac{RSS^{(n)}_{F}}{k}\left(\widetilde{\mat{X}}^{\T}\widetilde{\mat{X}}\right)^{+}  \ \mathrm{and } \
 {\mat{V}}_{(n)} = \dfrac{RSS^{(n)}_{F}}{k-p+1}\left({\mat{X}}^{\T}_{(n)}{\mat{X}_{(n)}}\right)^{-1}.
\end{align*}
Then as $n \to \infty$, convergence in distribution holds for
\begin{align*}
(i) [\mat{V}^{-1/2}_{(n)}(\vect{\beta}_{S} - \vect{\beta}_{F}^{(n)})\mid \vect{A}_{(n)}]  &\to \textnormal{Student}\left(\vect{0}, \mat{I}_{p},  \ k-p+1\right), \\
(ii) [\widetilde{\mat{V}}^{-1/2}_{(n)}(\vect{\beta}_{S} - \vect{\beta}_{F}^{(n)}) \ \mid \vect{A}_{(n)}] & \to N \left(\vect{0},  \mat{I}_{p} \right).
\end{align*}
\end{theorem}
For large $n$, we expect $\vect{\beta}_{S}$ to be approximately distributed as per Theorem \ref{thm:random_projection_beta_S} for both the Hadamard and Clarkson-Woodruff sketches.

It is harder to establish a comparable limit theorem for $\vect{\beta}_{P}^{*}$, due to the non-standard distribution of $\vect{\beta}_{P}^{*}$ when using a Gaussian sketch. There is no typical normalised distribution to target. Instead, we wish to show asymptotic equivalence in moments. The partially sketched estimator under the Hadamard and Clarkson-Woodruff sketches should have similar mean and variance properties to the Gaussian partially sketched estimator. An extra assumption has to be made to show convergence in moments. A sufficient condition is a stability condition on the singular values of the sketched data matrix. 
\begin{assumption}
\label{assump:moments}
Let $\mat{G}$ be the Gram matrix of the scaled sketched dataset, $\mat{G}=n^{-1}\widetilde{\mat{X}}^{\T}\widetilde{\mat{X}}$. Assume that the sequence of source datasets is such that $E_{S}\left( \dfrac{1}{\sigma^2_{\text{min}}(\mat{G})}\right)^2$ is finite for large enough $n$. 
\end{assumption}
This additional regularity condition enables a formal limit theorem regarding the moments of $\vect{\beta}_{P}^{*}$. 
\begin{theorem}
\label{thm:beta_p_asymptotics}
Suppose that Assumptions \ref{assump:leverage}, \ref{assump:limiting} and \ref{assump:moments} hold, $k > p +3 $, and $\vect{\beta}_{P}^{*}$ is computed using a Hadamard or Clarkson-Woodruff sketch. Let
\begin{align*}
 \mat{V}_{(n)} &= \dfrac{(k-p-1)}{(k-p)(k-p-3)}\left(MSS_{F}^{(n)}(\mat{X}^{\T}_{(n)}\mat{X}_{(n)})^{-1} + \dfrac{k-p+1}{k-p-1}\vect{\beta}_{F}^{(n)}\vect{\beta}_{F}^{(n)\T}\right).
\end{align*}
Then as $n \to \infty$, 
\begin{align*}
(i)\  E_{S}(\vect{\beta}_{P}^{*}-\vect{\beta}_{F}^{(n)} \mid \mat{A}_{(n)})  \ &{\to} \  \vect{0}.  \\
(ii) \ \mathrm{var}_{S}\left( {\mat{V}}_{(n)}^{-1/2}(\vect{\beta}_{P}^{*} - \vect{\beta}_{F}^{(n)}) \ \mid \vect{A}_{(n)} \right) \ & { \to} \ \mat{I}_{d}
\end{align*}
\end{theorem}
Once again, the heavy notation may obscure the essence of the result. The subscript $S$ is used to emphasise that the only source of randomness is the sketching matrix, and that the source dataset is fixed. The theorem suggests that the bias and variance of $\vect{\beta}_{P}^{*}$ under the Clarkson-Woodruff and Hadamard sketches should be approximately equal to that under the Gaussian sketch. Specifically,  we expect equations \eqref{eq:beta_p_variance} and \eqref{eq:beta_p_star_variance} to be good approximations for the variance of the sketched estimators using the Hadamard or Clarkson-Woodruff sketches.   

The results here are meant to be useful heuristics to assess the uncertainty attached to the output of the randomised approximation algorithm. There is a need to communicate and quantify the approximation error of sketching algorithms to end users \citep{lopes_2018_error, dobriban_2018_new}, and the asymptotic results developed in this section can be of use. 
\section{Unconditional results}
\label{sec:unconditional}
So far we have treated the source dataset as fixed to isolate the approximation error introduced by the random projection. When sketching is used for statistical inference, we can extend the hierarchical model of Section \ref{sec:gaussian_complete} to include a source of variation at the population level. We take the design matrix $\vect{X}$ as fixed and treat the response $\vect{y}$ as random.  We take the data generating process to be $
\vect{y} = \mat{X}\vect{\beta}_{0} +  \varepsilon$,
where $\varepsilon$ is a vector of $n$ independently and identically distributed random variables with mean zero and variance $\sigma^2$. Let $\gamma^2$ represent the average mean function sum of squares, so $\gamma^2=\lVert \mat{X}\vect{\beta}_{0}\rVert_{2}^{2}/n$. At the population level, the ordinary least squares estimator satisfies \citep{searle_linear_1997},
\begin{align*}
E_{y}(\vect{\beta}_{F}) &= \vect{\beta}_{0}, \\
\text{var}_{y}(\vect{\beta}_{F}) &= \sigma^2(\mat{X}^{\T}\mat{X})^{-1}, \\
E_{y}(RSS_{F}) &= (n-p)\sigma^2, \\
E_{y}(MSS_{F}) &= p\sigma^2 + n\gamma^2.
\end{align*}
Taking iterated expectations, we can see that the Gaussian sketch gives an unbiased estimator of the population parameter $\vect{\beta}_{0}$,
\begin{align*}
E_{y}(\vect{\beta}_{S})&= E_{y}\left\lbrace E_{S}(\vect{\beta}_{S}\mid\vect{y}, \mat{X})\right\rbrace \\
 &= E_{y}(\vect{\beta}_{F}) \\
 &= \vect{\beta}_{0}
\end{align*}
The unconditional variance of the Gaussian sketch can be obtained using the law of total variance,
\begin{align}
\text{var}_{y}(\vect{\beta}_{S}) &= E_{y}\left\lbrace \text{var}_{S}(\vect{\beta}_{S} \mid \vect{y}, \mat{X})\right\rbrace + \text{var}_{y}\lbrace E_{S}(\vect{\beta}_{S} \mid \vect{y}, \mat{X}) \rbrace \nonumber \\
&=  E_{y}(\dfrac{RSS_{F}}{(k-p+1)}(\mat{X}^{\T}\mat{X})^{-1}) + 0 \nonumber \\
&= \dfrac{(n-p)\sigma^2}{(k-p+1)}(\mat{X}^{\T}\mat{X})^{-1}.
\end{align}
We can also determine the unconditional properties of the partial sketch estimator $\vect{\beta}_{P}^{*}$. The estimator is also unbiased for $\vect{\beta}_{0}$,
\begin{align*}
E_{y}(\vect{\beta}_{P}^{*}) &= E_{y}\lbrace E_{S}(\vect{\beta}_{P}^{*} \mid \vect{y}, \mat{X}) \rbrace \\
&= E_{y}(\vect{\beta}_{F}) \\
&= \vect{\beta}_{0}.
\end{align*}
The unconditional variance of $\vect{\beta}_{P}^{*}$ is
\begin{align}
\text{var}_{y}(\vect{\beta}_{P}^{*}) &= 
E_{y}\left\lbrace \text{var}_{S}(\vect{\beta}_{P}^{*} \mid \vect{y}, \mat{X}) \right\rbrace  + \text{var}_{y}\left\lbrace E_{S}(\vect{\beta}_{P}^{*} \mid \vect{y}, \mat{X}) \right\rbrace \\
&= E_{y} \left\lbrace \dfrac{(k-p-1)}{(k-p)(k-p-3)}\left(MSS_{F}(\mat{X}^{\T}\mat{X})^{-1} + \dfrac{k-p+1}{k-p-1} \vect{\beta}_{F}\vect{\beta}_{F}^{\T}\right)\right\rbrace + 0 \\
&= \dfrac{(k-p-1)}{(k-p)(k-p-3)}\left\lbrace (p\sigma^2 + n\gamma^2)(\mat{X}^{\T}\mat{X})^{-1} + \left(\dfrac{k-p+1}{k-p-1}\sigma^2(\mat{X}^{\T}\mat{X})^{-1} + \dfrac{k-p+1}{k-p-1} \vect{\beta}_{0}\vect{\beta}_{0}^{\T} \right)\right\rbrace.
\end{align}
The most significant terms in the unconditional variance of $\vect{\beta}_{S}$ are $n\sigma^2$ and $(\mat{X}^{\T}\mat{X})^{-1}$. The dominating terms in the unconditional variance of $\vect{\beta}_{P}^{*}$ are $(\mat{X}^{\T}\mat{X})^{-1}$ and $n\gamma^2=\lVert \mat{X}\vect{\beta}_{0}\rVert_{2}^{2}$. We reach similar conclusions to the conditional analysis, in that we expect $\vect{\beta}_{S}$ to be more efficient when the signal to noise ratio is high, and $\vect{\beta}_{P}^{*}$ to be more efficient when the signal to noise ratio is low. Under Assumptions \ref{assump:leverage}, \ref{assump:limiting} and \ref{assump:moments}, the variance expression give asymptotic approximations for the Hadamard and Clarkson-Woodruff projections. These results can be extended to account for more complicated error models on $\varepsilon$ if it is still possible to determine $E_{y}(\vect{\beta}_{F})$, $\text{var}_{y}(\vect{\beta}_{F})$, $E_{y}(RSS_{F})$ and $E_{y}(MSS_{F})$. In independent work, \citet{chi_2018_randomized}, also study the error rates of sketched regression, and additionally consider cases where the sketched design matrix does not have same rank as the full design matrix.

\section{Data application}
\subsection{Human leukocyte antigen dataset}

We compared the performance of the sketching estimators on a real genetic dataset taken from the UK Biobank database.  We use a small extract from the data in \cite{astle_allelic_2016}. The selected response variable was mean red cell volume (MCV), taken from the full blood count assay and adjusted for various technical and environmental covariates. Genome-wide imputed genotype data in expected allele dose format were available on $n=132353$ study subjects \citep{howie_flexible_2009}. We consider 1000 genetic variants in the Human leukocyte antigen (HLA) region of chromosome 6, selected so that no pair of variants had Pearson correlation of allelic scores greater than 0.8. The region was chosen as many associations were discovered in a genome-wide scan using univariable models; these associations were with variants with different allele frequencies, suggesting multiple distinct causal variants in the region. The aim is to perform a multivariable regression analysis to obtain variant effect size estimates that are conditional on the other variants in the region.

\begin{figure}
\includegraphics[width=\textwidth]{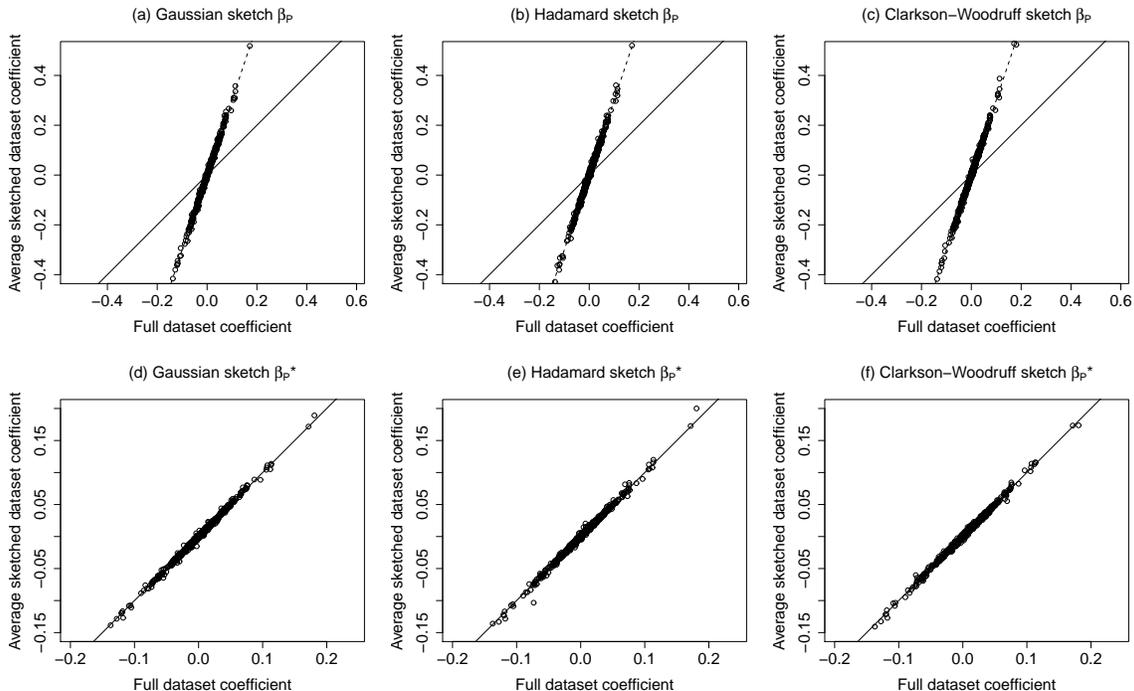}
\caption{Bias of sketching estimators on the HLA dataset. Mean estimates are plotted against true values. In this scenario $n=132353, p=1000, k=1500$. Solid line is the identity line and dashed line represents the theoretical bias factor.}
\label{fig:HLA_bias}
\end{figure}

An early theoretical finding was that the partial sketched estimator $\vect{\beta}_{P}$ was biased. One thousand sketches were taken to estimate the bias $E_{S}(\vect{\beta}_{P} - \vect{\beta}_{F})$ with $k=1500$. We also computed the bias corrected estimator $\vect{\beta}_{P}^{*}$ in each replication. Figure \ref{fig:HLA_bias} plots the average value of the estimators against the true value of the least squares coefficient using the full dataset. The top row (a)-(c) shows results for $\vect{\beta}_{P}$,  and the bottom row (d)-(f) shows results for $\vect{\beta}_{P}^{*}$. The first, second and third columns display the results for the Gaussian, Hadamard and Clarkson-Woodruff sketches respectively. The solid line in each panel is the identity line. The dashed line in the top row shows the theoretical bias, having slope $k/(k-p-1)$. 

The results in the top row show that $\vect{\beta}_{P}$ is biased for each of the random projections. The bias closely matches the theoretical factor. The bottom row shows that the adjusted estimator $\vect{\beta}_{P}^{*}$ appears to be unbiased, with the mean values falling closely along the identity line. 

We also compared the complete and partially sketched estimators on mean square error and the coverage of confidence intervals at $k=1500$ and $k=10000$. We also compared the data oblivious sketches to simple uniform subsampling with replacement. Simple random sampling is often referred to as the uniform sketch in the literature. We did not consider a combined estimator as the small $R^2_{F}$ value would mean give an optimal complete sketching weight of close to zero. Table \ref{tab:HLA_mse} reports the mean square error for each of the estimators. The signal to noise ratio is quite low for this dataset with $R^2_{F}=0.02$. We expect that partial sketching will be much more efficient than complete sketching on this dataset given the low signal to noise ratio. The simulation results support this idea, with $\vect{\beta}_{P}^{*}$ having a mean square error roughly sixty times smaller than $\vect{\beta}_{S}$ at both values of $k$. Results are very similar for each of the random projections, suggesting that the asymptotic approximations are reasonable for this dataset. For $k=1500$, the mean square error of $\vect{\beta}_{P}$ is approximately ten times that of $\vect{\beta}_{P}^{*}$. For  $k=10000$, there is less of a difference, as the ratio $k/(k-p-1)$ is closer to one. The bias adjusted estimator $\vect{\beta}_{P}^{*}$ has significant advantages over 
$\vect{\beta}_{P}$ when $k/(k-p-1)$ is larger than one.   

\begin{table}
\centering
\def~{\hphantom{0}}
\begin{tabular}{lccc|ccc}
 & \multicolumn{3}{c|}{$k=1500$} & \multicolumn{3}{c}{$k=10000$} \\ 
& $\vect{\beta}_{S}$ & $\vect{\beta}_{P}$ & $\vect{\beta}_{P}^{*}$ & $\vect{\beta}_{S}$ & $\vect{\beta}_{P}$ & $\vect{\beta}_{P}^{*}$ \\[5pt] \hline
Gaussian & 238 (3) & 39 (0.7) & 3.8 (0.08)& 13.3 (0.17) & 0.28 (0.004) & 0.21 (0.002) \\
Hadamard & 238 (4)& 39 (0.7) & 3.8 (0.07)  & 12.5 (0.16) & 0.26 (0.003) & 0.20 (0.002) \\
Clarkson-Woodruff & 241 (3) & 38 (0.8) & 4.0 (0.05)& 13.2 (0.16) & 0.28 (0.004) & 0.21 (0.002) \\
Uniform & 375 (15) & 105 (7.6) & 10.7 (0.55)& 13.8 (0.20) & 0.38 (0.007) & 0.29 (0.005) \\
\end{tabular}
\caption{Mean square error of sketched estimators on HLA dataset. Standard errors are in brackets.}
\label{tab:HLA_mse}
\end{table}
Table \ref{tab:HLA_ci} summarises the coverage of 95\% confidence intervals for the sketched estimators.  We report the overall proportion of intervals that contained the true value of the least squares estimate $\vect{\beta}_{F}$ over the two hundred and fifty sketches and $p=1000$ coefficients. The observed coverage is close the nominal level of 0.95 at both levels of $k$. The different random projections give very similar results, suggesting that the use of asymptotic approximations is again reasonable on this dataset.  The intervals for the Hadamard sketch appear to be slightly conservative at $k=10000$.

\begin{table}
\centering
\def~{\hphantom{0}}
\begin{tabular}{lcc|cc}
 \\
 & \multicolumn{2}{c|}{$k=1500$} & \multicolumn{2}{c}{$k=10000$} \\
& $\vect{\beta}_{S}$ & $\vect{\beta}_{P}^{*}$ & $\vect{\beta}_{S}$ & $\vect{\beta}_{P}^{*}$ \\[5pt] \hline
Gaussian & 0.950 & 0.953 & 0.950 & 0.951 \\
Hadamard & 0.949 &0.949 & 0.954 & 0.954 \\
Clarkson-Woodruff & 0.947 & 0.952 & 0.951 & 0.950 \\
\end{tabular}
\caption{Coverage of confidence intervals on the HLA dataset. The largest standard error is 0.002}
\label{tab:HLA_ci}
\end{table}

Table \ref{tab:HLA_timings} reports the average sketching time for the data oblivious sketches.  We computed ten sketches using each projection. The Gaussian sketch is an order of magnitude slower than the Hadamard projection and two orders of magnitude slower than the Clarkson-Wooduff sketch. The Gaussian sketch also scales more poorly as $k$ increases, as is expected from Table \ref{tab:run_time}. 

\begin{table}
\centering
\def~{\hphantom{0}}
\begin{tabular}{lccc}
& $k=1500$ & $k=10000$ \\[5pt] \hline
Gaussian & 522  & 3479  \\
Hadamard & 57  & 65  \\
Clarkson-Woodruff &5.3  & 5.4 \\
\end{tabular}
\caption{Timings for sketching the HLA dataset in seconds. We report the average time to compute the sketched dataset $\widetilde{\mat{A}} = \mat{S}\mat{A}$.}
\label{tab:HLA_timings}
\end{table}

\subsection{Flights dataset}
The sketching algorithms were also evaluated on the New York flights dataset available in the R package \texttt{nycflights13} \citep{wickham_nycflights13:_2014}. Arrival delay was taken as the response, and departure delay, distance,  departure time, origin and month and day were chosen to be the covariates. Rows of the dataset with missing data were omitted, leaving $n=327346$ and $d=47$. The goal was to compare the accuracy of the various sketches on real data rather than to build a statistical model for the flights dataset. We compared the mean square error of the estimators and the coverage of confidence intervals for $k=5000$. In contrast to the HLA dataset, the flights dataset has a very high $R^2_{F}$ value of $0.99$. We took five hundred sketches to compare complete and partial sketching. 

Table \ref{tab:flights_mse} reports the mean square error of $\vect{\beta}_{S}, \vect{\beta}_{P}$ and $\vect{\beta}_{P}^{*}$. As expected, complete sketching has a much smaller mean square error than partial sketching. Table \ref{tab:flights_ci} summarises the coverage rates of the 95\% confidence intervals. We report the overall proportion of intervals that contained the true value of the least squares estimate over the five hundred sketches and $p=46$ coefficients.

\begin{table}
\centering
\def~{\hphantom{0}}
\begin{tabular}{lccc}
& $\vect{\beta}_{S}$ & $\vect{\beta}_{P}$ & $\vect{\beta}_{P}^{*}$\\[5pt] \hline
Gaussian & 60 (2) &14900 (400) & 14900 (400) \\
Hadamard & 63 (2)& 14800 (500) & 13900 (400) \\
Clarkson-Woodruff &66 (2)& 15000 (500) & 13800 (400) \\
Uniform &64 (2)& 14600 (500) & 14600 (400) \\
\end{tabular}
\caption{Mean square error of sketched estimators on flights dataset with $k=5000$. Standard errors are in brackets.}
\label{tab:flights_mse}
\end{table}

\begin{table}
\centering
\def~{\hphantom{0}}
\begin{tabular}{lccc}
& $\vect{\beta}_{S}$ & $\vect{\beta}_{P}^{*}$ \\[5pt] \hline
Gaussian & 0.948 &0.951 \\
Hadamard & 0.950 & 0.948\\
Clarkson-Woodruff & 0.948& 0.947 \\
\end{tabular}
\caption{Coverage of 95\% confidence intervals on the flights dataset with $k=5000$.The largest standard error is 0.004}
\label{tab:flights_ci}
\end{table}

Table \ref{tab:flights_timings} reports the average sketching time for the data oblivious random projections. We generated ten sketches with each method. The Gaussian sketch is again considerably slower to apply than the Hadamard and Clarkson-Woodruff projections. 
\begin{table}
\def~{\hphantom{0}}
\begin{tabular}{lc}
& $k=5000$ \\[5pt] \hline
Gaussian  & 404  \\
Hadamard   & 5.8  \\
Clarkson-Woodruff & 0.2  \\
\end{tabular}
\caption{Timings for sketching the flights dataset in seconds. We report the average time to compute the sketched dataset $\widetilde{\mat{A}} = \mat{S}\mat{A}$.}
\label{tab:flights_timings}
\end{table}

\begin{figure}
\centering
\includegraphics[width=0.75\textwidth]{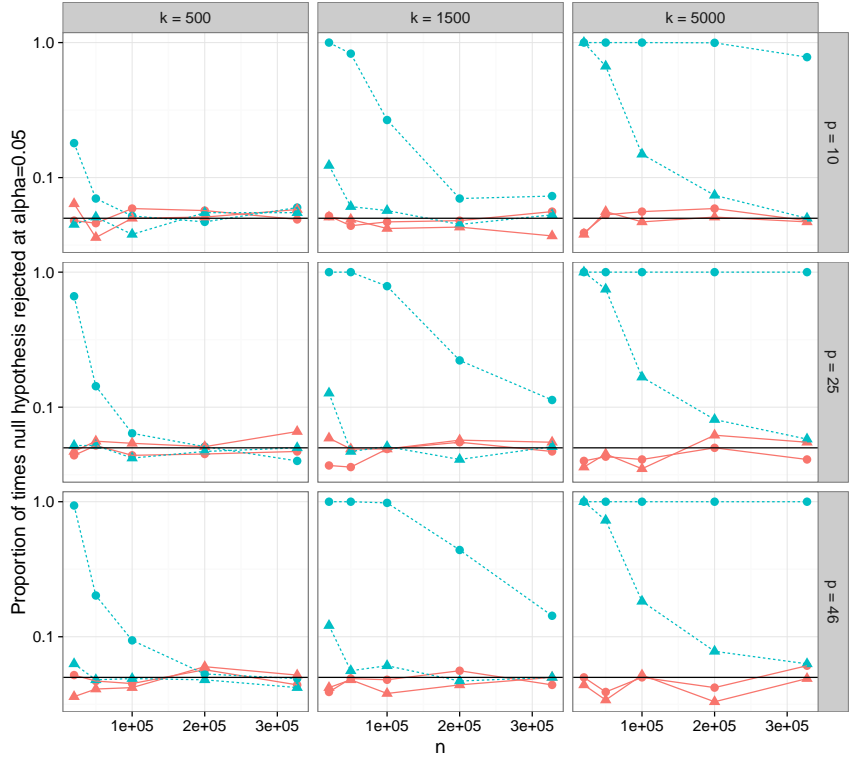}
\caption{Proportion of times null hypothesis of normality is rejected against size of the source dataset $(n)$ for the Hadamard (solid line) and Clarkson-Woodruff sketches (dashed line).  Results for tests of the sketched residual vector $\widetilde{\vect{e}} = \mat{S}(\vect{y} - \mat{X}\vect{\beta}_{F})$ are plotted as circles $(\circ)$, and results for tests of the entire sketched dataset $[\widetilde{\vect{y}}, \widetilde{\mat{X}}]$ are plotted as triangles $(\triangle)$. The horizontal line gives the type 1 error of 0.05. The $y$-axis is on a log scale.}
\label{fig:normality_tests}
\end{figure}

We also assessed the finite sample behaviour of the normal approximation in Theorem \ref{thm:sketching_clt} at different levels of $k$ and $p$. We dropped some predictors from the full flights dataset to give smaller datasets with $p=10$ and $p=25$ covariates. We then took subsamples of different sizes from each of the datasets. A single subsample was taken at each value of $n$, so the same subsampled dataset was being sketched each time. One thousand sketches were taken of each dataset at different values of $k$. We tested the joint multivariate normality of $[\vect{\widetilde{y}},\widetilde{\mat{X}}]$ and the normality of the sketched residual $\widetilde{\vect{e}} = \mat{S}(\vect{y} - \mat{X}\vect{\beta}_{F})$.  The squared Mahalanobis distance of the sketched observations was compared to the theoretical $\chi^2$-distribution. As $n$ increases the rejection rate is expected to fall to the type one error rate of 0.05. Figure \ref{fig:normality_tests} plots the proportion of times  the null hypothesis of normality is rejected against the size of the source dataset. 

The Hadamard sketch appears to have a much faster rate of convergence than the Clarkson-Woodruff sketch. When using a Hadamard sketch, each row in the sketched dataset is a linear combination of $n$ observations from the source dataset. When using a Clarkson-Woodruff sketch, each row in the sketched dataset is expected to be a combination of only $n/k$ observations from the source dataset. As such, $n/k$ must be large for the normal approximation to hold. As expected, the rejection rate for the Clarkson-Woodruff sketch increases with $k$, but remains stable for the Hadamard sketch. In Fig. \ref{fig:normality_tests} the rejection rate for the Clarkson-Woodruff sketch increases with $p$. The Hadamard sketch seems to be less sensitive to the number of covariates. The extra $\log k$ computation cost associated with the Hadamard sketch (Table \ref{tab:run_time}) appears to have the benefit of accelerated convergence to normality. Even though joint normality may not be holding for the Clarkson-Woodruff sketch for the flights dataset, the coverage of the confidence intervals is still very good. As $\widetilde{\vect{y}} = \widetilde{\mat{X}}\vect{\beta}_{F} +\widetilde{\vect{e}}$, normality of the sketched residual is perhaps sufficient in justifying the approximate confidence intervals using Theorem \ref{thm:random_projection_beta_S} (ii). The sketched residual converges much more quickly than the full sketched data matrix, which perhaps explains the good coverage properties of the confidence intervals for $\vect{\beta}_{S}$ in Table \ref{tab:flights_ci}.

\subsection{Synthetic data}
\label{subsec:synthetic_simulation}
\begin{figure}
\centering
\includegraphics[width=0.8\textwidth]{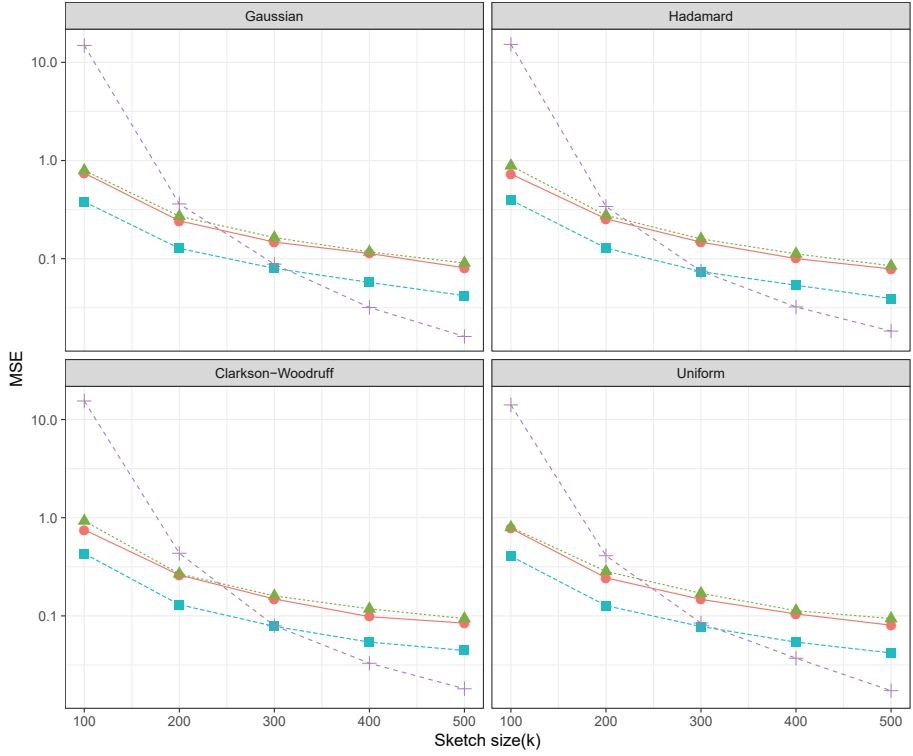}
\caption{Comparison of sketching estimators on synthetic dataset with $R^2_{F} \approx 0.5$. The $y$-axis is on a log scale. The average squared error for the sketching estimator is plotted against sketch size. Results are shown for $\vect{\beta}_{S}$ ($\circ$), $\vect{\beta}_{P}^{*}$ $(\triangle)$, the weighted combined estimator $\vect{\beta}_{C}$ $(\Box)$ and the one-step estimator $\vect{\beta}_{H}$ $(+)$.}
\label{fig:synthetic_results}
\end{figure}

We also generated a synthetic dataset with $n=10000$, $p=50$ and an $R^2_{F}$ of close to 0.5. The dataset consisted of responses $y_{i}$ and covariates $\vect{x}_{i}$, $(i=1, \ldots, n)$. Covariates $\vect{x}_{i}$ were drawn from a multivariate normal distribution with mean zero and covariance matrix $\Sigma$, with elements $\Sigma_{ij} = 0.5^{|i-j|}$. Responses were simulated independently using the standard linear model $y_{i} = \vect{x}_{i}^{\T}\vect{\beta}_{0} + \epsilon_{i}  \ (i=1, \ldots, n)$, where $\epsilon_{i}$ is a distributed as $N(0, 0.45)$. Each element of $\vect{\beta}_{0}$ was sampled independently from a $N(0, 0.01)$ distribution. We compared the single pass estimators $\vect{\beta}_{S}$, $\vect{\beta}_{P}^{*}$ to the combined estimator $\vect{\beta}_{C}$ with the optimal weight $\phi_{
\text{opt}}$, and the one-step estimator $\vect{\beta}_{H}$. We applied the Gaussian, Hadamard, Clarkson-Woodruff and uniform subsampling sketches.  We computed one hundred sketches at a range of sketch sizes $k$. We calculated the conditional sketching error $\lVert \widehat{\vect{\beta}} - \vect{\beta}_{F} \rVert_{2}^{2}$ for each sketched estimator $\widehat{\vect{\beta}}$ in each replicate. Figure \ref{fig:synthetic_results} plots the average error for the estimators $\vect{\beta}_{S}$, $\vect{\beta}_{P}^{*}$, $\vect{\beta}_{C}$ and $\vect{\beta}_{H}$ against the sketch size $k$. As expected, the combined estimator $\vect{\beta}_{C}$ has a mean square error that is roughly half that of $\vect{\beta}_{S}$ or $\vect{\beta}_{P}^{*}$ at all sketch sizes $k$. When $k/p$ is small, the one-step estimator $\vect{\beta}_{H}$ has a higher mean square error than the single pass estimator $\vect{\beta}_{S}$. As the ratio $k/p$ increases, the one-step estimator $\vect{\beta}_{H}$ becomes more efficient than the weighted estimator $\vect{\beta}_{C}$. This phenomenon can be studied in more detail using the moment results in \citet{letac_invariant_2004}. The results are similar for each of the data oblivious projections, suggesting that the asymptotic approximations are reasonable for this dataset. The uniform projection behaves similarly to the Gaussian projection, this is expected given that the covariates were simulated from a multivariate normal distribution.

\section{Discussion}
Sketching algorithms have emerged in the computer science community as a powerful device for the analysis of massive datasets \citep{mahoney_structural_2016}. Sketched regression algorithms use random projections to reduce the size of the original dataset, the sketched dataset is then used to estimate the optimal least squares coefficients. Most existing theory for sketched regression is from an algorithmic worst case perspective, and  connects with random matrix theory and computational geometry \citep{raskutti_statistical_2014, thanei_random_2017}. In this paper we have provided a complementary statistical perspective and derived new tools for assessing the uncertainty attached to sketched estimators, as well as guidelines for choosing between competing sketching algorithms.

The sketching central limit theorem was essential in establishing the asymptotic behaviour of the Clarkson-Woodruff and Hadamard projections. The regularity condition on the limiting leverage scores of the source dataset connects with both the existing computer science literature on sketching, and classic central limit theorems from the statistics literature. The field of randomised algorithms is clearly at the interface of computer science and statistics, and it is pleasing to see some overlap in the fundamental theory underpinning a Big Data algorithm. It is also possible to use other methods to develop uncertainty tools for randomised algorithms. \citep{lopes_2018_error} use the nonparametric bootstrap, and \citep{dobriban_2018_new} use asymptotic results in random matrix theory. Together, these provide a practical suite of tools for end users.

Iterative methods, in particular stochastic gradient descent, have not been mentioned so far. For large $n$ regression problems, stochastic gradient descent will produce iterates that converge to $\vect{\beta}_{F}$ under very mild conditions. Comparisons between single pass sketching and stochastic gradient methods are difficult, as the two techniques are not formulated for the exact same purpose. Single pass sketching algorithms are designed to return an approximate solution in finite time with probabilistically controlled error, whereas stochastic gradient methods are designed to converge to the exact solution asymptotically. It is perhaps more appropriate to compare stochastic gradient descent to iterative sketching methods, as iterative sketching algorithms also come with convergence guarantees to $\vect{\beta}_{F}$ \citep{pilanci_iterative_2016, gower_randomized_2015}. Iterative sketching methods make use of approximate second order information that can lead to a potential improvement compared to first order stochastic gradient methods \citep{roosta_subsampled_2016}. Our focus has been on characterising the approximation error attached to single pass sketching estimators.

There has been recent work in adapting sketching methods for statistical inference in large datasets, building from the worst case bounds in the computer science literature. \cite{geppert_random_2017} and \cite{bardenet_note_2015} investigate sketching algorithms for Bayesian regression, and derive bounds on the difference between the sketched posterior distribution and the full data posterior distribution. \cite{yang_fast_2015} consider sketched penalised regression, and give bounds between the sketched solution and the full data solution similar to the results in Section \ref{subsec:worst_case}. Only complete sketching is considered in the aforementioned work. The results on the advantages of partial sketching in this paper could motivate adaptations that make use of the exact marginal associations $\mat{X}^{\T}\vect{y}$.

Sketching ideas have been used to develop methods for approximate non-linear regression \citep{avron_subspace_2014, banerjee_efficient_2013}. A related branch of work uses random projections to reduce the number of predictors in regression and classification problems \citep{shah_min-wise_2013, cannings_random_2015, guhaniyogi_bayesian_2015}. 

\section*{Acknowledgement}
This work has been conducted using the UK Biobank resource under applications number 13745. Many thanks to Rajen Shah for helpful discussions.

\bibliographystyle{rss}
\bibliography{sketching}

\appendix 
\section*{Supplementary Information}
\setcounter{equation}{0}
\renewcommand\theequation{S.\arabic{equation}}
\section{Sketching examples}
As examples, we demonstrate the construction of a Hadamard sketch and a Clarkson-Woodruff sketch, for $k=3$, $n=4$. 

The Hadamard sketch matrix is formed as $\mat{S} = \Phi\mat{H}\mat{D}/\sqrt{k}$, where $\Phi$ is a $k \times n$ matrix and $\mat{H}$ and $\mat{D}$ are both $n \times n$ matrices. The fixed matrix $\mat{H}$ is a Hadamard matrix of order $n$. The random matrix $\mat{D}$ is a diagonal matrix where each nonzero element is an independent Rademacher random variable. The random matrix $\Phi$ subsamples $k$  rows of $\mat{H}$ with replacement. The display below shows an example of the random projection. The first matrix in the display represents $\mat{\Phi}\mat{H}$, a subsample of three rows from a $4 \times 4$ Hadamard matrix. In step 2, the diagonal matrix $\mat{D}$ is generated, with random Rademacher random variables along the diagonal. The diagonal elements are shown above the matrix. In step 3 the matrix multiplication $\mat{\Phi}\mat{H}\mat{D}$ is performed. This outputs the sketching matrix $\mat{S}$. 

\resizebox{\linewidth}{!}{
	\begin{minipage}{\linewidth}
		\begin{align*}
		\begin{blockarray}{cccc}
		\ & \ &\ & \ \\
		\ \ & \ \ & \ \ & \ \ \\
		\begin{block}{(cccc)}
		1 & -1 & \ \ 1 & -1 \ \\
		1 & -1 & -1 & \ \ 1 \ \\
		1 & -1 & \ \ 1 & -1 \ \\
		\end{block}
		\end{blockarray}
		\quad  \underset{\textrm{step 2}}{\to}  \quad 
		\begin{blockarray}{cccc}
		+1 & -1 & +1 & +1\\
		D_{11}  & D_{22} & D_{33}  & D_{44}  \\
		\begin{block}{(cccc)}
		1 & -1 & \ \ 1 & -1 \ \\
		1 & -1 & -1 & \ \ 1 \ \\
		1 & -1 & \ \ 1 & -1 \ \\
		\end{block}
		\end{blockarray}
		\quad  \underset{\textrm{step 3}}{\to}  \quad 
		\begin{blockarray}{cccc}
		+1 & -1 & +1 & +1\\
		\ \times & \ \times & \ \times & \ \times \\
		\begin{block}{(cccc)}
		1 & -1 & \ \ 1 & -1 \ \\
		1 & -1 & -1 & \ \ 1 \ \\
		1 & -1 & \ \ 1 & -1 \ \\
		\end{block}
		\end{blockarray}
		\quad  \underset{\textrm{output}}{\to}  \quad 
		\begin{blockarray}{cccc}
		\ & \ &\ & \ \\
		\ \ & \ \ & \ \ & \ \ \\
		\begin{block}{(cccc)}
		1  &\ \ 1 & \ \ 1 & -1 \ \\
		1 & \ \ 1 & -1 & \ \ 1 \ \\
		1 & \ \ 1 & \ \ 1 & -1 \ \\
		\end{block}
		\end{blockarray} \\
		\end{align*}
	\end{minipage}
}
The Clarkson-Woodruff sketch is a sparse random matrix. The  projection can be represented as the product of two independent random matrices, $\mat{S} = \mat{\Gamma}\mat{D}$, where $\mat{\Gamma}$ is a random $k \times n$ matrix and $\mat{D}$ is a random $n \times n$ matrix. The matrix $\mat{\Gamma}$ is formed by choosing one element in each column independently and setting the entry to $+1$. The matrix $\mat{D}$ is a diagonal matrix where each nonzero element is an independent Rademacher random variable. This results in a sparse $\mat{S}$, where there is only one nonzero entry per column. The display below shows an example of the random projection. The first matrix in the display represents $\mat{\Gamma}$, a random matrix where a single element in each column is set to one.  In step 2, the diagonal matrix $\mat{D}$ is generated, with random Rademacher random variables along the diagonal. The diagonal elements are shown above the matrix. In step 3 the matrix multiplication $\mat{\Gamma}\mat{D}$ is performed. This outputs the sketching matrix $\mat{S}$.

\resizebox{\linewidth}{!}{
	\begin{minipage}{\linewidth}
		\begin{align*}
		\begin{blockarray}{cccc}
		\ & \ &\ & \ \\
		\ \ & \ \ & \ \ & \ \ \\
		\begin{block}{(cccc)}
		1 & 0 & 0 & 0 \ \\
		0 & 0 &0 & 1 \ \\
		0 & 1 & 1 & 0 \ \\
		\end{block}
		\end{blockarray}
		\quad  \underset{\textrm{step 2}}{\to}  \quad 
		\begin{blockarray}{cccc}
		-1 & +1 & -1 & +1\\
		D_{11}  & D_{22} & D_{33}  & D_{44}  \\
		\begin{block}{(cccc)}
		1 & 0 & 0 &0 \\
		0 & 0 & 0 &1 \\
		0 & 1 & 1 &0  \\
		\end{block}
		\end{blockarray}
		\quad  \underset{\textrm{step 3}}{\to}  \quad 
		\begin{blockarray}{cccc}
		+1 & -1 & +1 & +1\\
		\ \times & \ \times & \ \times & \ \times \\
		\begin{block}{(cccc)}
		1 & 0 & 0 &0 \\
		0 & 0 & 0 &1 \\
		0 & 1 & 1 &0  \\
		\end{block}
		\end{blockarray}
		\quad  \underset{\textrm{output}}{\to}  \quad 
		\begin{blockarray}{cccc}
		\ & \ &\ & \ \\
		\ \ & \ \ & \ \ & \ \ \\
		\begin{block}{(cccc)}
		1 & \ \ 0 & \ \ 0 & \ \ 0 \\
		0 & \ \ 0 & \ \ 0 & -1 \\
		0 & -1 & \ \ 1 &\ \ 0  \\
		\end{block}
		\end{blockarray} 
		\end{align*}
	\end{minipage}
}

\section{Proof of Theorem \ref{thm:partial_sketching_worst_case}}
\begin{theorem}
Suppose that $\widetilde{\mat{X}}$ is an $\epsilon$-subspace embedding of $\mat{X}$ with $0<\epsilon < 0.5$. Then the following bound holds, 
\begin{align*}
\lVert \vect{\beta}_{P}- \vect{\beta}_{F}\rVert_{2}^{2} &\le \dfrac{4\epsilon^2}{\sigma_{\mathrm{min}}^2(\mat{X})}MSS_{F}. 
\end{align*}
\end{theorem}
Let the singular value decomposition of $\mat{X}$ be given by $\mat{X}=\mat{U}\mat{D}\mat{V}^{\mathsf{T}}$. The singular value decomposition will help to simplify expressions in later working. If the sketching matrix $\mat{S}$ is an $\epsilon$-subspace embedding for the source dataset with $0<\epsilon < 1$, then $\mat{U}^{\mathsf{T}}\mat{S}^{\mathsf{T}}\mat{S}\mat{U}$ is necessarily invertible. The expression for $\vect{\beta}_{P}$ can then be simplified to 
\begin{align*}
\vect{\beta}_{P} &= \mat{V}\mat{D}^{-1}(\mat{U}^{\mathsf{T}}\mat{S}^{\mathsf{T}}\mat{S}\mat{U})^{-1}\mat{D}^{-1}\mat{V}^{\mathsf{T}}\mat{X}^{\mathsf{T}}\vect{y} \\
&=  \mat{V}\mat{D}^{-1}(\mat{U}^{\mathsf{T}}\mat{S}^{\mathsf{T}}\mat{S}\mat{U})^{-1}\mat{D}^{-1}\mat{V}^{\mathsf{T}}\mat{V}\mat{D}\mat{U}^{\mathsf{T}}\vect{y}\\ 
&=  \mat{V}\mat{D}^{-1}(\mat{U}^{\mathsf{T}}\mat{S}^{\mathsf{T}}\mat{S}\mat{U})^{-1}\mat{U}^{\mathsf{T}}\vect{y}. 
\end{align*}
Similarly, $\vect{\beta}_{F}$ can be written as $\vect{\beta}_{F} = \mat{V}\mat{D}^{-1}\mat{U}^{\mathsf{T}}\vect{y}$. The Euclidean norm of the approximation error can thus be expressed as
\begin{align*}
\lVert \vect{\beta}_{P} - \vect{\beta}_{F}\rVert_{2} &= \lVert \mat{V}\mat{D}^{-1}(\mat{U}^{\mathsf{T}}\mat{S}^{\mathsf{T}}\mat{S}\mat{U})^{-1}\mat{U}^{\mathsf{T}}\vect{y} - \mat{V}\mat{D}^{-1}\mat{U}^{\mathsf{T}}\vect{y}\rVert_{2} \nonumber \\
&= \lVert  \left\lbrace \mat{V}\mat{D}^{-1}(\mat{U}^{\mathsf{T}}\mat{S}^{\mathsf{T}}\mat{S}\mat{U})^{-1} - \mat{V}\mat{D}^{-1} \right\rbrace \mat{U}^{\mathsf{T}}\vect{y}\rVert_{2}\nonumber \\
&= \lVert  \left\lbrace \mat{V}\mat{D}^{-1}[(\mat{U}^{\mathsf{T}}\mat{S}^{\mathsf{T}}\mat{S}\mat{U})^{-1} - \mat{I}_{p}]\right\rbrace \mat{U}^{\mathsf{T}}\vect{y}\rVert_{2}.
\end{align*}
The model sum of squares can be written as
\begin{align}
MSS_{F} &= \lVert \mat{X}\vect{\beta}_{F}\rVert_{2}^{2} \nonumber \\
&= \lVert \mat{X}\mat{V}\mat{D}^{-1}\mat{U}^{\mathsf{T}}\vect{y}  \rVert_{2}^{2}   \nonumber \\
&=  \lVert \mat{U}\mat{D}\mat{V}^{\mathsf{T}}\mat{V}\mat{D}^{-1}\mat{U}^{\mathsf{T}}\vect{y}  \rVert_{2}^{2} \nonumber \\
&=  \lVert \mat{U}\mat{U}^{\mathsf{T}}\vect{y}  \rVert_{2}^{2} \nonumber  \\
&= \lVert \mat{U}^{\mathsf{T}}\vect{y}  \rVert_{2}^{2}. \label{eq:MSS_norm_result}
\end{align}
The final line uses the fact that $\mat{U}^{\mathsf{T}}\mat{U} = \mat{I}_{p}$. Using the matrix norm induced by the Euclidean norm and the usual Euclidean norm for vectors we can form an upper bound on the error.  
\begin{align}
\lVert \vect{\beta}_{P} - \vect{\beta}_{F}\rVert_{2} &\le \lVert  \mat{V}\mat{D}^{-1}\left\lbrace (\mat{U}^{\mathsf{T}}\mat{S}^{\mathsf{T}}\mat{S}\mat{U})^{-1} - \mat{I}_{p}\right\rbrace\rVert_{2} \lVert \mat{U}^{\mathsf{T}}\vect{y}\rVert_{2} \nonumber \\
&\le \lVert  \mat{V}\mat{D}^{-1} \rVert_{2} \lVert \mat{U}^{\mathsf{T}}\vect{y}\rVert_{2}\lVert (\mat{U}^{\mathsf{T}}\mat{S}^{\mathsf{T}}\mat{S}\mat{U})^{-1} - \mat{I}_{p} \rVert_{2} \nonumber \\
&= \dfrac{MSS_{F}^{1/2}}{\sigma_{\text{min}}(\mat{X})}\lVert (\mat{U}^{\mathsf{T}}\mat{S}^{\mathsf{T}}\mat{S}\mat{U})^{-1} - \mat{I}_{p} \rVert_{2}. \label{eq:euclidean_bound}
\end{align}
It remains to upper bound the maximum singular value of the matrix $(\mat{U}^{\mathsf{T}}\mat{S}^{\mathsf{T}}\mat{S}\mat{U})^{-1} - \mat{I}_{p}$. Let $\mat{M}=\mat{U}^{\mathsf{T}}\mat{S}^{\mathsf{T}}\mat{S}\mat{U}$. The maximum absolute value of the singular values of $(\mat{U}^{\mathsf{T}}\mat{S}^{\mathsf{T}}\mat{S}\mat{U})^{-1} - \mat{I}_{p}$ will be given by 
$\text{max}(|1/\sigma_{\text{min}}(\mat{M}) - 1|,|1/\sigma_{\text{max}}(\mat{M}) - 1| )$, where $\sigma_{\text{min}}(\mat{M})$ is the minimum singular value of $\mat{M}$,  and $\sigma_{\text{max}}(\mat{M})$ is the maximum singular value of $\mat{M}$. If $\mat{S}$ is an $\epsilon$-subspace embedding for the source covariate matrix $\mat{X}$ then it must hold that $\sigma_{\text{min}}(\mat{M}) \ge 1-\epsilon$, and 
$\sigma_{\text{max}}(\mat{M}) \le 1+\epsilon$ \citep[p.11]{woodruff_sketching_2014}. As such, $\text{max}(|1/\sigma_{\text{min}}(\mat{M}) - 1|,|1/\sigma_{\text{max}}(\mat{M}) - 1| ) \le |1/(1-\epsilon)-1|$. It is simple to show that over the interval  $0 \le \epsilon \le 0.5$,  $|1/(1-\epsilon)-1| \le 2\epsilon$. This results in an upper bound on the singular value of interest,
\begin{align*}
\lVert (\mat{U}^{\mathsf{T}}\mat{S}^{\mathsf{T}}\mat{S}\mat{U})^{-1} - \mat{I}_{p} \rVert_{2} &\le |1/(1-\epsilon)-1| \\
&\le 2\epsilon.
\end{align*}
Substituting this back into \eqref{eq:euclidean_bound} gives that under the condition that $\epsilon < 0.5$
\begin{align*}
\lVert \vect{\beta}_{P} - \vect{\beta}_{F}\rVert_{2} \le \dfrac{MSS_{F}^{1/2}}{\sigma_{\text{min}}(\mat{X})} \times 2\epsilon.
\end{align*}
Squaring both sides gives the final result, that if $\epsilon < 0.5$
\begin{align*}
\lVert \vect{\beta}_{P} - \vect{\beta}_{F}\rVert_{2}^{2} \le \dfrac{4\epsilon^2}{\sigma^2_{\text{min}}(\mat{X})} MSS_{F}.
\end{align*}

\section{Proof of Theorem \ref{thm:gaussian_exact_distribution} (Hierarchical model for the Gaussian sketch)}
We use the following lemma about the Normal Inverse-Wishart distribution in many of our results \citep[p.73]{gelman_bayesian_2014}.

\begin{lemma}
 Suppose that $\mat{\Sigma}$ is a random $d \times d$ matrix and $\vect{y}$ is a $d$-dimensional random vector from the following hierarchical model
\begin{align*}
\vect{y} | {\Sigma} & \sim N\left(\vect{\mu}, \Sigma/\kappa\right), \\
\Sigma & \sim \text{Inv-Wishart}(\Lambda, \nu),
\end{align*}
where $\Lambda$ is a $d \times d$ scale matrix, $\nu$ is a scalar giving  degrees of freedom, and $\kappa$ is a scaling constant. Then marginally,
\begin{align*}
\vect{y} & \sim \text{Student}(\vect{\mu}, \Lambda/(\kappa(\nu-d+1)),\nu-d+1).
\end{align*}
\end{lemma}
Theorem \ref{thm:gaussian_exact_distribution} (ii) follows from setting $\vect{\mu}=\vect{\beta}_{F}$,  $\Sigma=(\widetilde{\mat{X}}^{\mathsf{T}}\widetilde{\mat{X}})^{-1}$, $\kappa=k/RSS_{F}$, $\Lambda = k(\mat{X}^{\mathsf{T}}\mat{X})^{-1}$, $\nu = k$ and $d=p$. Theorem \ref{thm:gaussian_exact_distribution}  $(i)$ follows from standard results on linear models, for example see \citet[Chapter 3]{searle_linear_1997}. 
\section{Variance for partial sketching}
Using a Gaussian sketch of size $k$ where $k>p+3$, the standard partial sketching estimator $\vect{\beta}_{P}$ has variance
\begin{align}
\text{var}(\vect{\beta}_{P}) &= \dfrac{k^2}{(k-p)(k-p-1)(k-p-3)}\left(MSS_{F}(\mat{X}^{\T}\mat{X})^{-1}+ \dfrac{(k-p+1)}{(k-p-1)}\vect{\beta}_{F}\vect{\beta}_{F}^{\mathsf{T}}  \right). \label{eq:beta_p_var_proof}
\end{align}
The bias corrected partial sketching estimator $\vect{\beta}_{P}^{*}$ has variance
\begin{align}
\text{var}(\vect{\beta}_{P}^{*}) &= \dfrac{(k-p-1)}{(k-p)(k-p-3)}\left(MSS_{F}(\mat{X}^{\mathsf{T}}\mat{X})^{-1} +\dfrac{(k-p+1)}{(k-p-1)} \vect{\beta}_{F}\vect{\beta}_{F}^{\mathsf{T}}\right).  \label{eq:beta_p_star_var_proof}
\end{align}
We now prove \eqref{eq:beta_p_var_proof} and \eqref{eq:beta_p_star_var_proof}.

Let the singular value decomposition of $\mat{X}$ be given by $\mat{X}=\mat{U}\mat{D}\mat{V}^{\mathsf{T}}$. The singular value decomposition will help to simplify expressions in later working. The sketched Gram matrix has the form $
\widetilde{\mat{X}}^{\mathsf{T}}\widetilde{\mat{X}} = \mat{V}\mat{D}\mat{U}^{\mathsf{T}}\mat{S}^{\mathsf{T}}\mat{S}\mat{U}\mat{D}\mat{V}^{\mathsf{T}}$. As $\mat{U}^{\mathsf{T}}\mat{S}^{\mathsf{T}}\mat{S}\mat{U} \sim \text{Wishart}(k, \mat{I}_{p}/k)$, the matrix $\mat{U}^{\mathsf{T}}\mat{S}^{\mathsf{T}}\mat{S}\mat{U}$ is almost surely invertible. The inverse Gram matrix can then be written as
\begin{align*}
(\widetilde{\mat{X}}^{\mathsf{T}}\widetilde{\mat{X}})^{-1} &=[\mat{D}\mat{V}^{\mathsf{T}}]^{-1} (\mat{U}^{\mathsf{T}}\mat{S}^{\mathsf{T}}\mat{S}\mat{U})^{-1}[\mat{V}\mat{D}]^{-1} \\
&=\mat{V}\mat{D}^{-1}(\mat{U}^{\mathsf{T}}\mat{S}^{\mathsf{T}}\mat{S}\mat{U})^{-1}\mat{D}^{-1}\mat{V}^{\mathsf{T}}.
\end{align*}
The expression for $\vect{\beta}_{P}$ can then be simplified to 
\begin{align*}
\vect{\beta}_{P} &= \mat{V}\mat{D}^{-1}(\mat{U}^{\mathsf{T}}\mat{S}^{\mathsf{T}}\mat{S}\mat{U})^{-1}\mat{D}^{-1}\mat{V}^{\mathsf{T}}\mat{X}^{\mathsf{T}}\vect{y} \\
&=  \mat{V}\mat{D}^{-1}(\mat{U}^{\mathsf{T}}\mat{S}^{\mathsf{T}}\mat{S}\mat{U})^{-1}\mat{D}^{-1}\mat{V}^{\mathsf{T}}\mat{V}\mat{D}\mat{U}^{\mathsf{T}}\vect{y}\\ 
&=  \mat{V}\mat{D}^{-1}(\mat{U}^{\mathsf{T}}\mat{S}^{\mathsf{T}}\mat{S}\mat{U})^{-1}\mat{U}^{\mathsf{T}}\vect{y}. 
\end{align*}
Let $\mat{M} = (\mat{U}^{\mathsf{T}}\mat{S}^{\mathsf{T}}\mat{S}\mat{U})^{-1}$. We know that  $\mat{M} \sim \text{Inverse-Wishart}(k, k\mat{I}_{p})$. Properties of the Inverse-Wishart distribution give that that for $i=1, \ldots, p$,
\begin{align}
\text{var}(M_{ii}) &= \dfrac{2k^2}{(k-p-1)^2(k-p-3)}. \label{eq:var_case_1}
\end{align}
Additionally,  for $i,j=1, \ldots, p$, where $j \ne i$
\begin{align}
\text{var}(M_{ij}) &= \dfrac{k^2(k-p-1)}{(k-p)(k-p-1)^2(k-p-3)}.  \label{eq:var_case_2}
\end{align}
Finally we have that for $i,j=1, \ldots, p$,  $i\ne j$,
\begin{align}
\text{cov}(M_{ij}, M_{ji}) &=  \dfrac{k^2(k-p-1)}{(k-p)(k-p-1)^2(k-p-3)}, \label{eq:cov_case_1} \\
\text{cov}(M_{ii}, M_{jj}) &=  \dfrac{2k^2}{(k-p)(k-p-1)^2(k-p-3)}. \label{eq:cov_case_2}
\end{align}
All other covariances $\text{cov}(M_{ij}, M_{br})$ are equal to zero unless they reduce to the cases in \eqref{eq:cov_case_1} or \eqref{eq:cov_case_2}. 
Let $\vect{z}= \mat{U}^{\mathsf{T}}\vect{y}$. Let $\mat{W} = \text{cov}\left(\mat{M}\mat{U}^{\mathsf{T}}\vect{y} \right) = \text{cov}(\mat{M}\vect{z})$. The elements of $\mat{W}$ can be determined using the properties in equations \eqref{eq:var_case_1} to \eqref{eq:cov_case_2}. Starting with the diagonal entries,
\begin{align*}
W_{ii} &= \text{var}\left(\sum_{j=1}^{p} M_{ij}z_{j}\right) \\
&= \sum_{j=1}^{p}z_{j}^2 \text{var}(M_{ij}) + \sum_{j=1}^{p}\sum_{w \ne j }^{p} z_{j}z_{w}\text{cov}(M_{ij}, M_{iw}).  
\end{align*}
As $\text{cov}(M_{ij}, M_{iw})$ is equal to zero for all $w \ne j$ this simplifies to 
\begin{align*}
W_{ii} &= \text{var}\left(\sum_{j=1}^{p} M_{ij}z_{j}\right) \\
&= \sum_{j=1}^{p}z_{j}^2 \text{ var}(M_{ij}).   
\end{align*}
It is helpful to split the sum into two pieces, a single term for $j=i$ and then a sum over the remaining indices. Grouping terms leads to an expression involving the model sum of squares $MSS_{F}$. 
\begin{align*}
W_{ii} &= z_{i}^2\dfrac{2k^2}{(k-p-1)^2(k-p-3)} +\sum_{j=1, j\ne i}^{p}z_{j}^2\dfrac{k^2(k-p-1)}{(k-p)(k-p-1)^2(k-p-3)} \\
&= z_{i}^2\dfrac{2k^2(k-p)}{(k-p)(k-p-1)^2(k-p-3)} +\sum_{j=1, j\ne i}^{p}z_{j}^2\dfrac{k^2(k-p-1)}{(k-p)(k-p-1)^2(k-p-3)}  \\
&= z_{i}^2\dfrac{2k^2(k-p-1)+2k^2}{(k-p)(k-p-1)^2(k-p-3)} +\sum_{j=1, j\ne i}^{p}z_{j}^2\dfrac{k^2(k-p-1)}{(k-p)(k-p-1)^2(k-p-3)}  \\
&= \dfrac{k^2(k-p-1)}{(k-p)(k-p-1)^2(k-p-3)}\sum_{j=1}^{p}z_{j}^2+\dfrac{k^2(k-p-1)+2k^2}{(k-p)(k-p-1)^2(k-p-3)}z_{i}^2 \\
&= \dfrac{k^2(k-p-1)}{(k-p)(k-p-1)^2(k-p-3)}MSS_{F}+\dfrac{k^2(k-p+1)}{(k-p)(k-p-1)^2(k-p-3)}z_{i}^2.
\end{align*}
In the second line the first term is modified to have the same denominator as the remainder sum. In the third line we add and subtract by $2k^2$ so that the numerator in the first term matches the numerator in the remainder sum. This allows the $z_{j}$ terms to be grouped into a sum over the full set of indexes $j=1, \ldots, p$ in the third line. The fourth line uses the fact that $\sum_{j=1}^{p}z_{j}^2 = \vect{z}^{\mathsf{T}}\vect{z} = MSS_{F}$. This was shown in the proof of Theorem 1 \eqref{eq:MSS_norm_result}. For the off diagonal entries $W_{ib}$ where $b \ne i$, 
\begin{align*}
W_{ib} &= \text{cov}\left( \sum_{j=1}^{p} M_{ij}z_{j}, \sum_{r=1}^{p} M_{br}z_{r}\right) \\
 &= \sum_{j=1}^{p}\sum_{r=1}^{p}z_{j}z_{r}\text{ cov}(M_{ij}, M_{br}).
\end{align*}
Now $\text{cov}(M_{ij}, M_{br})$ is only nonzero for $\text{cov}(M_{ib}, M_{bi})$ and $\text{cov}(M_{ii}, M_{bb})$. Using \eqref{eq:cov_case_1} and \eqref{eq:cov_case_2} we obtain
\begin{align*}
W_{ib} &= z_{i}z_{b}\text{ cov}(M_{ib}, M_{bi}) + z_{i}z_{b}\text{ cov}(M_{ii}, M_{bb})\\
&= \dfrac{k^2(k-p-1)}{(k-p)(k-p-1)^2(k-p-3)} z_{i}z_{b} + \dfrac{2k^2}{(k-p)(k-p-1)^2(k-p-3)} z_{i}z_{b} \\
&= \dfrac{k^2(k-p+1)}{(k-p)(k-p-1)^2(k-p-3)} z_{i}z_{b}.
\end{align*}
The entire covariance matrix $\vect{W}$ can therefore be written compactly as
\begin{align*}
\mat{W} &= \dfrac{k^2(k-p-1)}{(k-p)(k-p-1)^2(k-p-3)}\left(MSS_{F}\mat{I}_{p}\right) + \dfrac{k^2(k-p+1)}{(k-p)(k-p-1)^2(k-p-3)} \vect{z}\vect{z}^{T} \\
&=\dfrac{k^2(k-p-1)}{(k-p)(k-p-1)^2(k-p-3)}\left(MSS_{F}\mat{I}_{p} + \dfrac{(k-p+1)}{(k-p-1)}\vect{z}\vect{z}^{T}\right). \\
&= \dfrac{k^2}{(k-p)(k-p-1)(k-p-3)}\left(MSS_{F}\mat{I}_{p} + \dfrac{(k-p+1)}{(k-p-1)}\vect{z}\vect{z}^{T}\right).
\end{align*}
Now $\vect{\beta}_{P} = \mat{V}\mat{D}^{-1}\mat{M}\vect{z}$. Therefore $\text{var}(\vect{\beta}_{P})=\mat{V}\mat{D}^{-1}\text{var}(\mat{M}\vect{z})\mat{D}^{-1}\mat{V}^{\mathsf{T}}=\mat{V}\mat{D}^{-1}\mat{W}\mat{D}^{-1}\mat{V}^{\mathsf{T}}$. The variance of $\vect{\beta}_{P}$ is then a linear function of $\mat{W}$,
\begin{align}
\text{var}(\vect{\beta}_{P}) &= \mat{V}\mat{D}^{-1}\mat{W}\mat{D}^{-1}\mat{V}^{\mathsf{T}} \nonumber \\
&= \mat{V}\mat{D}^{-1}\dfrac{k^2}{(k-p)(k-p-1)(k-p-3)}\left(MSS_{F}\mat{I}_{p} + \dfrac{(k-p+1)}{(k-p-1)}\vect{z}\vect{z}^{T}\right)\mat{D}^{-1}\mat{V}^{\mathsf{T}} \nonumber \\
&= \dfrac{k^2}{(k-p)(k-p-1)(k-p-3)}MSS_{F}(\mat{V}\mat{D}^{-2}\mat{V}^{\mathsf{T}}) + \nonumber \\
& \qquad
\dfrac{k^2(k-p+1)}{(k-p)(k-p-1)^2(k-p-3)}\mat{V}\mat{D}^{-1}\vect{z}\vect{z}^\mathsf{T}\mat{D}^{-1}\mat{V}^{\mathsf{T}}. \label{eq:z_outer_prod}
\end{align}
Recall that $\vect{z}=\mat{U}^{\mathsf{T}}\vect{y}$ and
\begin{align}
\vect{\beta}_{F} &= (\mat{X}^{\mathsf{T}}\mat{X})^{-1}\mat{X}^{\mathsf{T}}\vect{y} \\
&= \mat{V}\mat{D}^{-1}\mat{U}^{\mathsf{T}}\vect{y} \\
&=  \mat{V}\mat{D}^{-1}\vect{z}. \label{eq:beta_f_z}
\end{align}
The term $\mat{V}\mat{D}^{-1}\vect{z}$ appears in \eqref{eq:z_outer_prod}. Substituting \eqref{eq:beta_f_z} into \eqref{eq:z_outer_prod} gives
\begin{align*}
\text{var}(\vect{\beta}_{P})&= \dfrac{k^2}{(k-p)(k-p-1)(k-p-3)}MSS_{F}(\mat{V}\mat{D}^{-2}\mat{V}^{\mathsf{T}}) + \\
& 
\quad \dfrac{k^2(k-p+1)}{(k-p)(k-p-1)^2(k-p-3)}\vect{\beta}_{F}\vect{\beta}_{F}^{\mathsf{T}}. \\
\end{align*}
A final simplification can be made by noting that $(\mat{X}^{\mathsf{T}}\mat{X})^{-1}=\mat{V}\mat{D}^{-2}\mat{V}^{\mathsf{T}}$ giving 
\begin{align*}
\text{var}(\vect{\beta}_{P})&= \dfrac{k^2}{(k-p)(k-p-1)(k-p-3)}MSS_{F}(\mat{X}^{\T}\mat{X})^{-1}+ \dfrac{k^2(k-p+1)}{(k-p)(k-p-1)^2(k-p-3)}\vect{\beta}_{F}\vect{\beta}_{F}^{\mathsf{T}} \\
&= \dfrac{k^2}{(k-p)(k-p-1)(k-p-3)}\left(MSS_{F}(\mat{X}^{\T}\mat{X})^{-1}+ \dfrac{(k-p+1)}{(k-p-1)}\vect{\beta}_{F}\vect{\beta}_{F}^{\mathsf{T}}  \right).
\end{align*}
The variance of $\vect{\beta}_{P}^{*}=[(k-p-1)/k]\vect{\beta}_{P}$ is then
\begin{align}
\text{var}(\vect{\beta}_{P}^{*})&= \left(\dfrac{k-p-1}{k}\right)^2\dfrac{k^2(k-p-1)}{(k-p)(k-p-1)^2(k-p-3)}\left(MSS_{F}(\mat{X}^{\mathsf{T}}\mat{X})^{-1} + \dfrac{(k-p+1)}{(k-p-1)}\vect{\beta}_{F}\vect{\beta}_{F}^{\mathsf{T}}\right)  \nonumber \\
&= \dfrac{(k-p-1)}{(k-p)(k-p-3)}\left(MSS_{F}(\mat{X}^{\mathsf{T}}\mat{X})^{-1} +\dfrac{(k-p+1)}{(k-p-1)} \vect{\beta}_{F}\vect{\beta}_{F}^{\mathsf{T}}\right).  \label{eq:beta_p_star_variance_supp}
\end{align}

\section{Combined estimator results}
We first show that $\vect{\beta}_{P}^{*}$ and $\vect{\beta}_{S}$ are uncorrelated. We again avoid explicitly conditioning on the source dataset $[\vect{y}, \mat{X}]$ in every step, it is always treated as fixed. The covariance between $\vect{\beta}_{P}^{*}$ and $\vect{\beta}_{S}$ computed from the same sketch can be shown to be zero. Using the definition of covariance, and taking iterated expectations
\begin{align*}
\text{cov}(\vect{\beta}_{P}^{*}, \vect{\beta}_{S}) &= \mathbb{E}_{S}\left\lbrace (\vect{\beta}_{P}^{*}-\vect{\beta}_{F})(\vect{\beta}_{S}-\vect{\beta}_{F})^{\mathsf{T}}\right\rbrace  \\
&= \mathbb{E}_{\widetilde{X}}\left[ \mathbb{E}_{\widetilde{y}}\left\lbrace (\vect{\beta}_{P}^{*}-\vect{\beta}_{F})(\vect{\beta}_{S}-\vect{\beta}_{F})^{\mathsf{T}} \mid \widetilde{\mat{X}} \right\rbrace \right].
\end{align*}
Recall the hierarchical model for complete sketching,
\begin{align*}
\widetilde{\vect{y}} \mid \widetilde{\mat{X}}  & \sim N\left(\widetilde{\vect{X}}\vect{\beta}_{F}, \dfrac{RSS_{F}}{k}\mat{I}_{k}\right).
\end{align*}
Equivalently,
\begin{align*}
\widetilde{\vect{y}} \mid \widetilde{\mat{X}} = \widetilde{\mat{X}}\vect{\beta}_{F} + \widetilde{\vect{e}},
\end{align*}
where $\widetilde{\vect{e}} \mid  \widetilde{\mat{X}}\sim N( \vect{0}, \dfrac{RSS_{F}}{k}\mat{I}_{k})$. So 
\begin{align*}
\vect{\beta}_{S}\mid \widetilde{\mat{X}}, \vect{y}, \mat{X} &= \vect{\beta}_{F} + (\widetilde{\mat{X}}^{\mathsf{T}}\widetilde{\mat{X}})^{-1}\widetilde{\mat{X}}^{\mathsf{T}}\widetilde{\vect{e}}.
\end{align*}
Substituting back into the expression for the covariance,
\begin{align*}
\text{cov}(\vect{\beta}_{P}^{*}, \vect{\beta}_{S}) 
&= \mathbb{E}_{\widetilde{X}}\left\lbrace \mathbb{E}_{\widetilde{e} \mid \widetilde{X}}\left[ (\vect{\beta}_{P}^{*}-\vect{\beta}_{F})(\vect{\beta}_{F} + (\widetilde{\mat{X}}^{\mathsf{T}}\widetilde{\mat{X}})^{-1}\widetilde{\mat{X}}^{\mathsf{T}}\widetilde{\vect{e}}-\vect{\beta}_{F})^{\mathsf{T}} \mid \widetilde{\mat{X}}\right]\right\rbrace \\
&= \mathbb{E}_{\widetilde{X}}\left\lbrace \left[ (\vect{\beta}_{P}^{*}-\vect{\beta}_{F})(\vect{\beta}_{F} + (\widetilde{\mat{X}}^{\mathsf{T}}\widetilde{\mat{X}})^{-1}\widetilde{\mat{X}}^{\mathsf{T}}\mathbb{E}_{\widetilde{e} \mid \widetilde{X}}[\widetilde{\vect{e}} \mid \widetilde{\mat{X}}]-\vect{\beta}_{F})^{\mathsf{T}} \mid \widetilde{\mat{X}}\right]\right\rbrace  \\
&= \mathbb{E}_{\widetilde{X}}\left\lbrace \left[ (\vect{\beta}_{P}^{*}-\vect{\beta}_{F})(\vect{\beta}_{F} -\vect{\beta}_{F})^{\mathsf{T}} \mid \widetilde{\mat{X}}\right]\right\rbrace \\
&= \mathbb{E}_{\widetilde{X}}\left\lbrace \left[ (\vect{\beta}_{P}^{*}-\vect{\beta}_{F})\vect{0}^{\mathsf{T}} \mid \widetilde{\mat{X}}\right]\right\rbrace \\
&= \mat{0}_{p \times p}. 
\end{align*}
Simple calculus shows that the value which minimises the expected mean square error
$\mathbb{E}_{S}(\lVert \vect{\beta}_{C} - \vect{\beta}_{F}\rVert_{2}^{2}\mid \vect{y}, \mat{X}) $ is 
\begin{align*}
\phi_{\text{opt}} &= \dfrac{\text{tr}(\text{var}(\vect{\beta}_{P}^{*}))}{\text{tr}(\text{var}(\vect{\beta}_{P}^{*})) + \text{tr}(\text{var}(\vect{\beta}_{S}))}. 
\end{align*}
\section{Proof of Theorem \ref{thm:bounded_clt} (central limit theorem under asymptotic negligibility condition)}
A triangular array of random variables is a useful structure for studying weak convergence. To establish a triangular array, define for every $n \in \mathbb{N}$ a collection of random variables  $Z_{n1}, Z_{n2}, \ldots, Z_{nr_{n}}$. There are $r_{n}$ random variables in row $n$ of the array. Suppose that $r_{n}=n$. Visually we can represent the first three rows of the array as
\begin{align*}
&Z_{11} \\
&Z_{21} \quad Z_{22} \\
&Z_{31} \quad Z_{32} \quad Z_{33} \\
\end{align*}

\begin{theorem*}[\citealp{billingsley_1995_probability}, Chapter 5, Section 27]
For each $n \in \mathbb{N}$, let $Z_{n1}, Z_{n2}, \ldots, Z_{nr_{n}}$ be a sequence of independent random variables with $\mathbb{E}(Z_{ni})=0$ and $\textnormal{var}(Z_{ni})=\sigma^2_{ni}$ for $i=1, \ldots, r_{n}$. Let $s_{n}^{2}=\sum_{i=1}^{r_n}\sigma^2_{ni}$ and assume that $r_{n} \to \infty$ as $n \to \infty$. Suppose that we can form a sequence of upper bounds $(K_{n})_{n \in \mathbb{N}}$ such that 
\begin{align*}
|Z_{ni}|  \le K_{n} \textnormal{ almost surely for $i=1, \ldots, r_n$}.    
\end{align*}
Then if $K_{n}/s_{n} \to 0$ as $n \to \infty$ we have the convergence in distribution
\begin{align*}
    \dfrac{1}{s_{n}}\sum_{i=1}^{r_{n}}Z_{ni} \overset{d}{\to} N(0,1)
\end{align*}
\end{theorem*}
Lindeberg's condition is a critical component in establishing asymptotic normality. We state Lindeberg's condition for triangular arrays of random variables. 
\begin{definition}[Lindeberg's condition]
\label{defn:lindeberg}
For each $n \in \mathbb{N}$, let $Z_{n1}, Z_{n2}, \ldots, Z_{nr_{n}}$ be a sequence of random variables with $\mathbb{E}(Z_{ni})=0$ and $\textnormal{var}(Z_{ni})=\sigma^2_{ni}$ for $i=1, \ldots, r_{n}$. 
Let $s_{n}^2 = \sum_{i=1}^{n}\sigma^2_{ni}$ and suppose that $r_{n} \to \infty$ as $n \to \infty$. The random variables are said to satisfy Lindeberg's condition if for all $\eta > 0$,
\begin{align}
\underset{n \to \infty}{\lim} \ \dfrac{1}{s_{n}^{2}}\sum_{i=1}^{r_n}\mathbb{E}(Z_{ni}^2 \mathbbm{1}_{\left\lbrace |Z_{ni}| > \eta s_{n} \right\rbrace}) = 0. \label{eq:lindeberg_condition}
\end{align}
\end{definition}
The triangular array of random variables does not have to have independent random variables in each row in order to satisfy the condition. The general form of the Lindeberg-Feller central limit theorem shows that a triangular array of independent random variables satisfying Lindeberg's condition is asymptotically normal after suitable scaling.
\begin{theorem}[Lindeberg-Feller]
\label{thm:lindeberg_feller_clt}
For each $n \in \mathbb{N}$, let $Z_{n1}, Z_{n2}, \ldots, Z_{nr_{n}}$ be a sequence of random variables with $\mathbb{E}(Z_{ni})=0$ and $\textnormal{var}(Z_{ni})=\sigma^2_{ni}$ for $i=1, \ldots, r_{n}$. 
Let $s_{n}^2 = \sum_{i=1}^{r_n}\sigma^2_{ni}$ and suppose that $r_{n} \to \infty$ as $n \to \infty$. Suppose the triangular array of random variables satisfies Lindeberg's condition (Definition \ref{defn:lindeberg}). Then 
	\begin{align*}
	\dfrac{1}{s_{n}}\sum_{i=1}^{r_n}Z_{ni} \overset{d}{\to} N(0, 1)
	\end{align*}
\end{theorem}
\noindent 
For a proof see \cite{loeve_probability_1977}. It can be difficult to show Lindeberg's condition directly. A stronger condition that implies the Lindeberg condition is the Lyapunov condition. \begin{definition}[Lyapunov's condition]
For each $n \in \mathbb{N}$, let $Z_{n1}, Z_{n2}, \ldots, Z_{nr_{n}}$ be a sequence of random variables with $\mathbb{E}(Z_{ni})=0$ and $\textnormal{var}(Z_{ni})=\sigma^2_{ni}$ for $i=1, \ldots, r_{n}$. Let $s_{n}^{2}=\sum_{i=1}^{r_n}\sigma^2_{ni}$ and suppose that $r_{n} \to \infty$ as $n \to \infty$. The triangular array of random variables is said to satisfy Lyapunov's condition if there exists a $\delta > 0$ such that
\begin{align}
    \underset{n \to \infty}{\lim}  \ \dfrac{1}{s_{n}^{2+\delta}}\sum_{i=1}^{r_n}\mathbb{E}(|Z_{ni}|^{2+\delta}) &= 0.
\end{align}
\end{definition}
The Lyapunov condition implies the Lindeberg condition. We state this in a Lemma for later reference.
\begin{lemma}
\label{lem:lyapunov_linderberg}
The Lyapunov condition implies the Lindeberg condition. 
\end{lemma}
To see this assume the Lyapunov condition is satisfied and fix $\eta > 0$. Now $|Z_{ni}| \ge \eta s_{n}$ implies that $1 \le |Z_{ni}/(\eta s_{n})|^{\delta} $. We can then form an upper bound on the sequence of partial sums that appear in Lindeberg's condition. 
\begin{align*}
\dfrac{1}{s_{n}^{2}}\sum_{i=1}^{r_n}\mathbb{E}(Z_{ni}^2 \mathbbm{1}_{\left\lbrace |Z_{ni}| > \eta s_{n} \right\rbrace}) &\le \dfrac{1}{s_{n}^{2}}\sum_{i=1}^{r_n}\mathbb{E}(Z_{ni}^2|Z_{ni}/(\eta s_{n})|^{\delta} \mathbbm{1}_{\left\lbrace |Z_{ni}| > \eta s_{n} \right\rbrace})   \\
&= \dfrac{1}{s_{n}^{2}}\sum_{i=1}^{r_n}\mathbb{E}(|Z_{ni}|^2|Z_{ni}/(\eta s_{n})|^{\delta} \mathbbm{1}_{\left\lbrace |Z_{ni}| > \eta s_{n} \right\rbrace})    \\
&= \dfrac{1}{s_{n}^{2}}\dfrac{1}{(\eta s_{n})^{\delta}}\sum_{i=1}^{r_n}\mathbb{E}(|Z_{ni}|^{2+\delta} \mathbbm{1}_{\left\lbrace |Z_{ni}| > \eta s_{n} \right\rbrace}) 
\\
&= \dfrac{1}{\eta^{\delta}}\dfrac{1}{s_{n}^{2+\delta}}\sum_{i=1}^{r_n}\mathbb{E}(|Z_{ni}|^{2+\delta}).
\end{align*}
Assuming that Lyapunov's condition holds we can establish zero as an upper bound
\begin{align*}
   \underset{n \to \infty}{\lim}  \  \dfrac{1}{s_{n}^{2}}\sum_{i=1}^{r_n}\mathbb{E}(Z_{ni}^2 \mathbbm{1}_{\left\lbrace |Z_{ni}| > \eta s_{n} \right\rbrace}) &\le  \underset{n \to \infty}{\lim}  \ \dfrac{1}{\eta^{\delta}}\dfrac{1}{s_{n}^{2+\delta}}\sum_{i=1}^{r_{n}}\mathbb{E}(|Z_{ni}|^{2+\delta}) \\
   &= \dfrac{1}{\eta^{\delta}}\underset{n \to \infty}{\lim} \ \dfrac{1}{s_{n}^{2+\delta}}\sum_{i=1}^{r_{n}}\mathbb{E}(|Z_{ni}|^{2+\delta}) \\
   &= 0.
\end{align*}
As the partial sums are lower bounded by zero, the Lyapunov condition implies the Lindeberg condition.

We now present a useful Lemma for showing the Lyapunov condition. The result is from \cite{billingsley_1995_probability} and applies to triangular arrays of uniformly bounded random variables. 
\begin{lemma}[\citealp{billingsley_1995_probability}]
\label{lem:bounded_lyapunov}
For each $n \in \mathbb{N}$, let $Z_{n1}, Z_{n2}, \ldots, Z_{nr_{n}}$ be a sequence of random variables with $\mathbb{E}(Z_{ni})=0$ and $\textnormal{var}(Z_{ni})=\sigma^2_{ni}$ for $i=1, \ldots, r_{n}$. Let $s_{n}^{2}=\sum_{i=1}^{r_n}\sigma^2_{ni}$ and suppose that $r_{n} \to \infty$ as $n \to \infty$. Suppose that we can form a sequence of upper bounds $(K_{n})_{n \in \mathbb{N}}$ such that 
\begin{align*}
|Z_{ni}|  \le K_{n} \textnormal{ almost surely for $i=1, \ldots, r_n$}.    
\end{align*}
Then if $K_{n}/s_{n} \to 0$ as $n \to \infty$  the Lyapunov condition holds for the triangular array of random variables.
\end{lemma}
Lemma \ref{lem:bounded_lyapunov} is useful as it does not impose a constant uniform bound on the random variables. In the special case where $|Z_{ni}| \le M$ almost surely for some constant $M$ for all $n \in \mathbb{N}$ and all $i = 1, \ldots, r_{n}$ we have that Lyapunov's condition is satisfied providing that $s_{n} \to \infty$. Lemma  \ref{lem:bounded_lyapunov} allows for the bound $K_{n}$ to increase with $n$ as long as the rate of growth is slower than the rate of growth of $s_{n}$. Lyapunov's condition holds providing that $K_{n}=o(s_{n})$.

The proof of Lemma \ref{lem:bounded_lyapunov} is given below. Again fix some $\delta > 0$. If $|Z_{ni}| \le K_{n}$ almost surely for $i=1, \ldots, r_n$ it must hold that $|Z_{ni}|^{\delta} \le K_{n}^{\delta}$ as $|Z_{ni}|, K_{n}$ and $\delta$ are all positive. As such $|Z_{ni}|^{2+\delta} = |Z_{ni}|^{2}|Z_{ni}|^{\delta} \le |Z_{ni}|^{2}K_{n}^{\delta}$. We can then form an upper bound on the sequence of partial sums that appear in Lyapunov's condition.
\begin{align}
    \dfrac{1}{s_{n}^{2+\delta}}\sum_{i=1}^{r_n}\mathbb{E}(|Z_{ni}|^{2+\delta}) &\le   \dfrac{1}{s_{n}^{2+\delta}}\sum_{i=1}^{r_n}\mathbb{E}(|Z_{ni}|^2) K_{n}^{\delta} \nonumber \\
    &=\dfrac{K_{n}^{\delta}}{s_{n}^{2+\delta}} \sum_{i=1}^{r_n} \mathbb{E} |Z_{ni}|^{2}.\nonumber \\
    &= \dfrac{K_{n}^{\delta}}{s_{n}^{2+\delta}} s_{n}^{2}  \nonumber \\
    &= \left(\dfrac{K_{n}}{s_{n}}\right)^{\delta}.\label{eq:bounded_inequality}
\end{align}
Now assuming that $K_{n}=o(s_{n})$ we have that $K_{n}/s_{n} \to 0$ as $n \to \infty$. We then also have that  
\begin{align*}
  \underset{n \to \infty}{\lim}  \  \left(\dfrac{K_{n}}{s_{n}}\right)^{\delta} &= \left(\underset{n \to \infty}{\lim}  \  \dfrac{K_{n}}{s_{n}}\right)^{\delta} \\
  &= 0,
\end{align*}
as the exponentiation by $\delta > 0$ is a continuous function. Now taking limits on both sides of the inequality \eqref{eq:bounded_inequality}:
\begin{align}
     \underset{n \to \infty}{\lim}  \  \dfrac{1}{s_{n}^{2+\delta}}\sum_{i=1}^{r_n}\mathbb{E}(|Z_{ni}|^{2+\delta}) &\le     \underset{n \to \infty}{\lim}  \  \left(\dfrac{K_{n}}{s_{n}}\right)^{\delta} \\
     &= 0.
\end{align}
We also have the lower bound
\begin{align*}
0 \le \underset{n \to \infty}{\lim}  \  \dfrac{1}{s_{n}^{2+\delta}}\sum_{i=1}^{r_n}\mathbb{E}(|Z_{ni}|^{2+\delta}).
\end{align*}
By the squeeze theorem we then have that $K_{n}=o(s_{n})$ is sufficient for Lyapunov's condition to hold. 

The triangular array of independent random variables in Theorem \ref{thm:bounded_clt} satisfies Lyapunov's condition by Lemma \ref{lem:bounded_lyapunov}. As the Lyapunov condition implies the Lindeberg condition (Lemma \ref{lem:lyapunov_linderberg}) the general Lindeberg-Feller central limit theorem (Theorem \ref{thm:lindeberg_feller_clt}) gives asymptotic normality of the scaled row sums, thus proving Theorem \ref{thm:bounded_clt}.

\section{Proof of Theorem \ref{thm:sketching_clt} (Sketching central limit theorem)}
\begin{description}
	\item[Assumption 1] Let the singular value decomposition of the $n \times d$ source dataset be given by $\mat{A}_{(n)}=\mat{U}_{(n)}\mat{D}_{(n)}\mat{V}_{(n)}^{\mathsf{T}}$. Let $\vect{u}_{(n)i}^{\T}$ give the $i$th row in $\mat{U}_{(n)}$. Assume that the maximum leverage score tends to zero, that is
	\begin{align*}
	\lim_{n \to \infty} \underset{i=1, \ldots, n}{\text{max}} \lVert \vect{u}_{(n)i} \rVert_{2}^{2} = 0. 
	\end{align*}
\end{description}
Theorem \ref{thm:sketching_clt} gives the sketching central limit theorem. 
\begin{theorem*}
	Consider a fixed sequence of arbitrary $n \times d$ data matrices $\mat{A}_{(n)}$, where $d$ is fixed. Let $\mat{A}_{(n)}=\mat{U}_{(n)}\mat{D}_{(n)}\mat{V}_{(n)}^{\mathsf{T}}$ represent the singular value decomposition of $\mat{A}_{(n)}$. Let $\mat{S}$ be a $k \times n$ Hadamard or Clarkson-Woodruff sketching matrix where $k$ is also fixed. Suppose that Assumption 1 on the maximum leverage score is satisfied. 	Then as $n$ tends to infinity with $k$ and $d$ fixed,
	\begin{align*}
[\widetilde{\mat{A}}\mat{V}_{(n)}\mat{D}_{(n)}^{-1} \ | \ \mat{A}_{(n)}]\overset{d}{\to} \emph{MN}(\mat{0}, \mat{I}_{k}, \mat{I}_{d}/k).
	\end{align*}
\end{theorem*}
To prove the sketching central limit theorem it helps to restate Lemma \ref{lem:bounded_lyapunov}. This helps to show the importance of the leverage scores in establishing asymptotic normality. Lemma \ref{lem:bounded_lyapunov} provided a sufficient condition for showing that Lindeberg's condition holds. 
We can restate Lemma \ref{lem:bounded_lyapunov} in terms of a normalised triangular array. 
\begin{theorem}[\citealp{billingsley_1995_probability}]
\label{thm:triangular_array}
For each $n \in \mathbb{N}$ let $Z_{n1}, Z_{n2}, \ldots, Z_{nr_{n}}$ be a sequence of random variables with $\mathbb{E}[Z_{ni}]=0$ and $\text{var}(Z_{ni}^{2})=\sigma_{ni}^{2}$ for $i=1, \ldots, r_{n}$. Define $s_{n}^{2}=\sum_{i=1}^{r_{n}}\sigma_{ni}^{2}$ each $n$. Suppose that the rows of the triangular array are standardised such that $s_{n}^{2}=1$ for all $n$. Suppose that $r_{n} \to \infty$ as $n \to \infty$. Suppose we have a sequence of upper bounds $(Z_{n})$ suck that $|Z_{ni}| \le K_{n}$ almost surely for all $i=1, \ldots, r_{n}$. Then a sufficient condition for Lyapnuov's condition to hold is $K_{n} \to 0$ as $n \to \infty$.
\end{theorem}
The standardisation of the triangular array gives an intuitive condition for Lyapunov's and hence Lindeberg's condition to hold. We require that $K_{n} \to 0$ as $n \to \infty$. We require that the upper bound tends to zero. All the random variables in the row must converge almost surely to zero. Almost sure convergence is stronger than convergence in probability and rules out pathological cases where a single random variable in  a row can take a large value with small probability. Assumption 1 on the leverage scores in the sketching central limit theorem enforces a bounded growth condition that relates to Theorem \ref{thm:triangular_array}. 

Let $n \in \mathbb{N}$ index the sequence of source datasets of increasing size.  We assume that the source dataset consists of $r_{n}$ observations where $r_{n} \to \infty$ as $n \to \infty$. For now we can take take $r_{n}=n$ to ease interpretation. We take the singular value decomposition of each dataset $\mat{A}_{(n)}=\mat{U}_{(n)}\mat{D}_{(n)}\mat{V}_{(n)}^{\mathsf{T}}$. All results in this section treat the source dataset $\mat{A}_{(n)}$ as fixed, only the sketching matrix is random. We consider the sequence of whitened sketched datasets
\begin{align*}
\widetilde{\mat{A}}\mat{V}_{(n)}\mat{D}_{(n)}^{-1} &= (\mat{S}\mat{A})\mat{V}_{(n)}\mat{D}_{(n)}^{-1} \\
 &= \mat{S}\mat{U}_{(n)}\mat{D}_{(n)}\mat{V}_{(n)}^{\mathsf{T}}\mat{V}_{(n)}\mat{D}_{(n)}^{-1}\\
&= \mat{S}\mat{U}_{(n)}.
\end{align*}
The whitened sketched dataset $\widetilde{\mat{A}}\mat{V}_{(n)}\mat{D}_{(n)}^{-1}$ has a $MN(\vect{0}, \mat{I}_{k}, \mat{I}_{d}/k)$ distribution when $\mat{S}$ is a Gaussian sketch. We need to show that as $n$ tends to infinity, 
$\mat{S}\mat{U}_{(n)}$ convergences in distribution to a  $MN(\vect{0}, \mat{I}_{k}, \mat{I}_{d}/k)$ random matrix for both the Clarkson-Woodruff and Hadamard sketches. 

Let $\vect{u}_{(n)i}^{\T}$ denote row $i$ of the matrix of left singular vectors $\mat{U}_{(n)}$. We write $\vect{u}_{(n)i}^{\T}$ so that that we can form a triangular array of left singular vectors. Taking $r_{n}=n$, the first three rows of the triangular array can be written as
\begin{align*}
&\vect{u}_{(1)1} \\
&\vect{u}_{(2)1} \quad \vect{u}_{(2)2} \\
&\vect{u}_{(3)1} \quad \vect{u}_{(3)2} \quad \vect{u}_{(3)3}
\end{align*}
An important property is that for all $n$, the sum of the norms of the leverage scores always equals the number of variables in the source dataset $d$. 
\begin{align}
\sum_{i=1}^{r_n}\lVert \vect{u}_{(n)i}\rVert_{2}^{2} = d. \label{eq:leverage_standardisation}
\end{align}
As $n$ increases, the typical norm of each vector $\vect{u}_{(n)i}$, $i\in\left\lbrace1, \ldots, r_n\right\rbrace$ is expected to decrease. For completeness we restate Assumption 1 in terms of the triangular array formulation.
\begin{description}
	\item[Assumption 1] Let the singular value decomposition of the $r_n \times d$ source dataset be given by $\mat{A}_{(n)}=\mat{U}_{(n)}\mat{D}_{(n)}\mat{V}_{(n)}^{\mathsf{T}}$. Let $\vect{u}_{(n)i}^{\T}$ give the $i$th row in $\mat{U}_{(n)}$ for $i=1, \ldots, r_{n}$. Assume that the maximum leverage score tends to zero, that is
	\begin{align*}
	\lim_{n \to \infty} \underset{i=1, \ldots, r_{n}}{\text{max}} \lVert \vect{u}_{(n)i} \rVert_{2}^{2} = 0. 
	\end{align*}
\end{description}
This increasing collection of smaller quantities is similar to the behaviour of the triangular array of random variables in Theorem \ref{thm:triangular_array}. The standardisation property in equation \eqref{eq:leverage_standardisation}, namely that $\sum_{i=1}^{r_n}\lVert \vect{u}_{(n)i}\rVert_{2}^{2} = d$ for all $n$ is similar to the assumption that $s_{n}=1$ in each row of the triangular array of random variables in Theorem \ref{thm:triangular_array}. Assumption 1 on the leverage scores, where the maximum individual norm tends to zero is similar to the assumption that $K_{n} \to 0$ in Theorem \ref{thm:triangular_array}. This will be made more explicit in the proofs. Before moving on we make a note that assumption 1 also implies that the maximum square root of the leverage scores also tends to zero. As 
\begin{align}
 \underset{i=1, \ldots, r_{n}}{\text{max}} \lVert \vect{u}_{(n)i} \rVert_{2}  &= \left( \underset{i=1, \ldots, r_{n}}{\text{max}} \lVert \vect{u}_{(n)i} \rVert_{2}^{2}  \right)^{1/2}
\end{align}
We have that
\begin{align}
	\lim_{n \to \infty} \underset{i=1, \ldots, r_{n}}{\text{max}} \lVert \vect{u}_{(n)i} \rVert_{2} &= 	\lim_{n \to \infty}  \left( \underset{i=1, \ldots, r_{n}}{\text{max}} \lVert \vect{u}_{(n)i} \rVert_{2}^{2}  \right)^{1/2} \nonumber \\
	&= \left(\lim_{n \to \infty} \underset{i=1, \ldots, r_{n}}{\text{max}} \lVert \vect{u}_{(n)i} \rVert_{2}^{2}\right)^{1/2} \nonumber \\
	&= 0. \label{eq:assumption_norm}
\end{align}

To establish joint asymptotic normality of the sketched data matrix we use the  Cram\'{e}r-Wold device.
\begin{lemma}[Cram\'{e}r-Wold device]
	\label{lem:cramer_wold}
	Let $(\vect{Z}_{n})_{n \in \mathbb{N}}$ be a sequence of random vectors in $\mathbb{R}^{v}$. Let $\vect{Z}$ denote another random vector also in $\mathbb{R}^{v}$. The sequence of random vectors $(\vect{Z}_{n})$ converges in distribution to $\vect{Z}$ as $n$ tends to infinity if and only if the sequence of random variables $(\vect{\lambda}^{\T}\vect{Z}_{n})_{n \in \mathbb{N}}$ converges in distribution to $\vect{\lambda}^{\T}\vect{Z}$ for all unit vectors $\vect{\lambda} \in \mathbb{R}^{v}$. 
\end{lemma}
A proof is given in \citet[][Chapter 13, Section 3]{shorack_2000_probability}. Let $\vect{z}_{n}$ represent the $kd$ length vector formed by stacking transposed rows of the whitened sketched dataset $\widetilde{\mat{U}} = \mat{S}\mat{U}_{(n)}$. Let $\widetilde{\vect{u}}_{j}^{\mathsf{T}}$ give row $j$ in $\widetilde{\mat{U}}$ for $j=1, \ldots, k$.
 Formally, 
\begin{align}
\vect{z}_{n} &= \begin{bmatrix}
\widetilde{\vect{u}}_{1} \\
\widetilde{\vect{u}}_{2}  \\
\vdots \\
\widetilde{\vect{u}}_{k} 
\end{bmatrix}.
\label{eq:zstack}
\end{align}
Let us define the random matrix $k \times d$ random matrix $\mat{W}$ as having the matrix normal distribution
\begin{align*}
\mat{W} \sim MN(\vect{0}, \mat{I}_{k}, \mat{I}_{d}/k)
\end{align*}
Let $\vect{w}_{i}^{\T}$ refer to row $i$ in $\mat{W}$ for $i=1, \ldots, k$. Let $\vect{z}_{L}$ refer to the stacked transposed rows of $\mat{W}$, so
\begin{align}
\vect{z}_{L} &= \begin{bmatrix}
{\vect{w}}_{1} \\
{\vect{w}}_{2}  \\
\vdots \\
{\vect{w}}_{k} 
\end{bmatrix}.
\label{eq:zlstack}
\end{align}
Let $\vect{\lambda}$ be an arbitrary unit vector in $\mathbb{R}^{k \times d}$. It will be useful to also partition the vector $\vect{\lambda}$ into $k$ sub-vectors,
\begin{align}
\vect{\lambda} &= \begin{bmatrix}
\vect{\lambda}_{1} \\
\vect{\lambda}_{2} \\
\vdots \\
\vect{\lambda}_{k}
\end{bmatrix},
\label{eq:lambda_stack}
\end{align}
where $\vect{\lambda}_{j}$ is a $d$-dimensional vector for $j=1, \ldots, k$. For any unit vector $\vect{\lambda} \in \mathbb{R}^{k \times d}$,  $\vect{\lambda}^{\T}\vect{z}_{L}$ is distributed as $N(0, 1/k)$. We will aim to show that the distribution of the whitened sketched data $\mat{S}\mat{A}_{(n)} \mat{V}_{(n)}\mat{D}_{(n)}^{-1}$ converges to that of $\mat{W}$ through  the Cram\'{e}r-Wold device. We must show that for any fixed $k \times d$ length unit vector $\vect{\lambda}$, $\vect{\lambda}^{\mathsf{T}}\vect{z}_{n}$ converges in distribution to $N(0, 1/k)$ as $n \to \infty$. 

We will rely on a central limit theorem for jointly symmetric, pairwise independent random variables \citep{pruss_central_2000}. A collection of random variables $(Z_{1}, \ldots Z_{n})$ is said to be jointly symmetric if $(Z_{1}, \ldots Z_{n})$ has the same distribution as $(q_{1}Z_{1},\ldots, q_{n} Z_{n})$, where $q_{i} \in \left\lbrace +1, -1 \right\rbrace$ for $i=1, \ldots, n$. Given a set of random variables $Y_{1}, \ldots, Y_{n}$, a jointly symmetric collection $Z_{1}, \ldots, Z_{n}$ can be formed by sampling $n$ independent Rademacher random variables $h_{1}, \ldots, h_{n}$, and setting $Z_{i}=h_{i}Y_{i}$ \citep{pruss_central_2000}. It is possible to establish a central limit theorem for jointly symmetric, pairwise independent random variables.

\begin{theorem}[\cite{pruss_central_2000}{, Theorem 1, Corollary 2}]
\label{thm:pairwise_clt}
For each $n \in \mathbb{N}$, let $Z_{n1}, Z_{n2}, \ldots, Z_{nr_{n}}$ be a sequence of jointly symmetric pairwise independent random variables with $\mathbb{E}(Z_{ni})=0$ and $\textnormal{var}(Z_{ni})=\sigma^2_{ni}$ for $i=1, \ldots, r_{n}$. Let $s_{n}^{2}=\sum_{i=1}^{r_n}\sigma^2_{ni}$ and assume that $r_{n} \to \infty$ as $n \to \infty$. Suppose the triangular array of random variables satisfies Lindeberg's condition. Then as $n \to \infty$, $s_{n}^{-1}\sum_{i=1}^{r_n}Z_{ni}$ converges in distribution to $N(0, 1)$.
\end{theorem}
Not all triangular arrays with pairwise independent random variables in each row satisfy  a central limit theorem. The joint symmetry property is very important \citep{pruss_central_2000, svante_1988_some}. 

To use Theorem \ref{thm:pairwise_clt} we need to show that the triangular array of random variables satisfies Lindeberg's condition. As discussed this can be very difficult to establish directly. If the triagular array of random variables can be appropriately bounded, we can use Theorem \ref{thm:triangular_array} to show that Lyapunov's condition holds, and subsequently that Lindeberg's condition holds.

This is the approach we take in proving the sketching central limit theorem. The Cram\'{e}r-Wold device is used to reduce the study of multivariate convergence to univariate convergence. We can then form a triangular array of random variables such that elements in each row are jointly symmetric and pairwise independent. We then show that triangular array satisfies Lindeberg's condition using Theorem \ref{thm:triangular_array}. Assumption 1 on the maximum leverage score enforces the necessary cap on the rate of growth. Theorem \ref{thm:pairwise_clt} is then used to establish asymptotic normality. 

\subsection{Clarkson-Woodruff sketch}
The Clarkson-Woodruff sketch can be represented as the product of two independent random matrices, $\mat{S} = \mat{\Gamma}\mat{D}$, where $\mat{\Gamma}$ is a random $k \times n$ matrix and $\mat{D}$ is a random $n \times n$ matrix. The diagonal matrix $\mat{D}$ contains  $n$ independent Rademacher random variables on the diagonal. Let $h_{i} \in \left\lbrace +1, -1\right\rbrace$ be the random sign in element $D_{ii}$. The matrix $\mat{\Gamma}$ is formed by choosing one element in each column independently and setting the entry to $+1$.  Element $\Gamma_{ij}$ is equal to $+1$ if we add observation $i$ in the original dataset to sketched observation $j$. The signs in row $i$ are flipped if $h_{i}$ is equal to negative one. Each observation in the original dataset is assigned to one sketched observation as each column of $\mat{\Gamma}$ contains a single $+1$ entry. Using a Clarkson-Woodruff sketch row $j$ in the sketched data matrix can be represented as
\begin{align*}
\widetilde{\vect{u}}_{j}^{\T} &= \sum_{i=1}^{n}h_{i}\Gamma_{ij}\vect{u}_{(n)i}^{\T},
\end{align*}
where $h_{i}$ represents the random sign flip applied to row $i$ of the original data matrix, and $\Gamma_{ij}$ is the indicator variable which is equal to one if row $i$ of the original data is added to row $j$ of the sketched dataset.

Let us consider the linear combination $\vect{\lambda}^{\T}\vect{z}$, where $\vect{\lambda}$ and $\vect{z}$ are defined as in \eqref{eq:zstack} and \eqref{eq:lambda_stack} respectively. The sum over the $k$ rows in the sketched dataset can be rearranged into a sum over the $n$ rows in the source dataset,
\begin{align}
\vect{\lambda}^{\mathsf{T}}\vect{z}_{n} &= \sum_{j=1}^{k}\vect{\lambda}_{j}^{\mathsf{T}}\widetilde{\vect{u}}_{j} \nonumber \\
&= \sum_{j=1}^{k}\vect{\lambda}_{j}^{\mathsf{T}}\sum_{i=1}^{n}h_{i}\Gamma_{ij}\vect{u}_{(n)i}\nonumber \\
&= \sum_{i=1}^{n}h_{i}\sum_{j=1}^{k}\Gamma_{ij}\vect{\lambda}_{j}^{\mathsf{T}}\vect{u}_{(n)i}. \label{eq:CW_sum}
\end{align}
The scalar $\vect{\lambda}^{\mathsf{T}}\vect{z}_{n}$ is equal to the sum of $n$ independent random variables. Independence holds as the signs flips $h_{i}$ on each observation are independent, and each column of $\mat{\Gamma}$ is independent.

In the language of Theorem \ref{thm:triangular_array} we can form a triangular array of random variables setting 
\begin{align}
    Z_{ni} = h_{i}\sum_{j=1}^{k}\Gamma_{ij}\vect{\lambda}_{j}^{\mathsf{T}}\vect{u}_{(n)i}.
    \label{eq:cw_triangular_array}
\end{align}
for $i=1, \ldots, n$ and $n \in \mathbb{N}$. The linear combination in \eqref{eq:CW_sum} then be expressed as a row sum over the triangular array defined in \eqref{eq:cw_triangular_array}:
\begin{align}
\vect{\lambda}^{\mathsf{T}}\vect{z}_{n} &= \sum_{i=1}^{n}Z_{ni}. 
\label{eq:cw_cramer_triangular}
\end{align}
Our goal of showing that $\vect{\lambda}^{\mathsf{T}}\vect{z}_{n}$ converges in distribution to a $N(0, 1/k)$ random variable is achieved if we can show that $\sum_{i=1}^{n}Z_{ni}$ converges in distribution to a $N(0, 1/k)$ random variable.

It is worth making a connection to Theorem \ref{thm:pairwise_clt}, because of the random sign flips $h_{i}$ appearing in \eqref{eq:cw_triangular_array}, we have a sequence of mutually independent jointly symmetric random variables. Mutually independent random variables are also necessarily pairwise independent. Theorem \ref{thm:pairwise_clt} can be used to establish asymptotic normality of the sum in \eqref{eq:cw_cramer_triangular} and hence the linear combination $\vect{\lambda}^{\mathsf{T}}\vect{z}_{n}$. To show that the triangular array of random variables defined in \eqref{eq:cw_triangular_array} satisfies Lindeberg's condition we use Theorem \ref{thm:triangular_array}. Set $s_{n}^{2}= \sum_{i=1}^{n}\text{var}(Z_{ni})$. We first determine $s_{n}^{2}$. We then form the necessary sequence of upper bounds $K_{n}$ such that $|Z_{ni}| \le K_{n}$ almost surely for $i=1, \ldots, n$. The variance of a single term in the sum \eqref{eq:CW_sum} is
\begin{align}
\text{var}(Z_{ni}) &= \text{var}\left( h_{i}\sum_{j=1}^{k}\Gamma_{ij}\vect{\lambda}_{j}^{\mathsf{T}}\vect{u}_{(n)i}\right) \\
&= \sum_{j=1}^{k} \dfrac{1}{k}\vect{\lambda}_{j}^{\mathsf{T}}  \vect{u}_{(n)i}\vect{u}_{(n)i}^{\mathsf{T}}\vect{\lambda}_{j}. \label{eq:CW_term_variance}
\end{align}
The row-wise variance totals $s_{n}^{2}$ are then 
\begin{align*}
s_{n}^{2} &= \sum_{i=1}^{n}\text{var}(Z_{ni}) \\
&= \sum_{i=1}^{n}\text{var}\left( h_{i}\sum_{j=1}^{k}\Gamma_{ij}\vect{\lambda}_{j}^{\mathsf{T}}\vect{u}_{(n)i}\right) \\
&= \dfrac{1}{k}\sum_{i=1}^{n}\sum_{j=1}^{k} \vect{\lambda}_{j}^{\mathsf{T}} \vect{u}_{(n)i}\vect{u}_{(n)i}^{\mathsf{T}}\vect{\lambda}_{j} \\
&= \dfrac{1}{k}\sum_{j=1}^{k} \vect{\lambda}_{j}^{\mathsf{T}}\left( \sum_{i=1}^{n}\vect{u}_{(n)i}\vect{u}_{(n)i}^{\mathsf{T}} \right) \vect{\lambda}_{j}\\
&=  \dfrac{1}{k}\sum_{j=1}^{k}\vect{\lambda}_{j}^{\mathsf{T}}\mat{U}^{\mathsf{T}}_{(n)}\mat{U}_{(n)}\vect{\lambda}_{j} \\
&= \dfrac{1}{k}\sum_{j=1}^{k}\vect{\lambda}_{j}^{\mathsf{T}}\mat{I}_{d}\vect{\lambda}_{j}. \\
&= \dfrac{1}{k}\sum_{j=1}^{k}\vect{\lambda}_{j}^{\mathsf{T}}\vect{\lambda}_{j} \\
&= \dfrac{1}{k}.
\end{align*}
The fact that $\mat{U}^{\mathsf{T}}_{(n)}\mat{U}_{(n)}=\mat{I}_{d}$ for all $n$ serves as a useful normalisation to give stable limiting behaviour.
The step in the last line follows as we have taken $\vect{\lambda}$ to be a unit vector. We have $s_{n}=1/k$ for all $n$ in the triangular array. We now establish a sequence of upper bounds $(K_{n})$. As the random variables in the construction of  construction of the sketch are bounded, we can bound the random variables in the triangular array using the leverage scores of the sequence of source dataset. Now as the random sign $h_{i} \in \left\lbrace +1, -1 \right\rbrace$ 
\begin{align}
 |Z_{ni}| &= \lvert h_{i}\sum_{j=1}^{k}\Gamma_{ij}\vect{\lambda}_{j}^{\mathsf{T}}\vect{u}_{(n)i}\rvert \nonumber \\
 &= \lvert \left(\sum_{j=1}^{k}\Gamma_{ij}\vect{\lambda}_{j}^{\mathsf{T}} \right)\vect{u}_{(n)i}\rvert. \label{eq:cw_abs_bound}
\end{align} 
Now by the Cauchy-Schwarz inequality
\begin{align}
    \lvert \left( \sum_{j=1}^{k}\Gamma_{ij}\vect{\lambda}_{j}^{\mathsf{T}}\right)\vect{u}_{(n)i}\rvert \le \lVert \sum_{j=1}^{k}\Gamma_{ij}\vect{\lambda}_{j} \rVert_{2}\lVert \vect{u}_{(n)i}\rVert_{2} \label{eq:cw_cs_bound}
\end{align}
Now as $\Gamma_{ij}=1$ for a single $j \in \left\lbrace 1, \ldots, k\right\rbrace$ and is zero otherwise we have that 
\begin{align}
\lVert \sum_{j=1}^{k}\Gamma_{ij}\vect{\lambda}_{j} \rVert_{2} &\le \underset{\text{j=1, \ldots, k}}{\text{max}} \lVert \lambda_{j}\rVert_{2} \nonumber  \\
&\le 1.\label{eq:cw_gamma_bound}
\end{align}
The last line follows as we have taken $\vect{\lambda}$ to be a unit vector. Substituting \eqref{eq:cw_gamma_bound} and \eqref{eq:cw_cs_bound} into \eqref{eq:cw_abs_bound} we arrive at
\begin{align*}
|Z_{ni}| \le \lVert \vect{u}_{(n)i} \rVert_{2}.
\end{align*}
We can then form the sequence of upper bounds $K_{n}$,
\begin{align*}
    K_{n} &= \underset{i=1, \ldots, n}{\text{max}}\lVert \vect{u}_{(n)i} \rVert_{2}.
\end{align*}
We have that $|Z_{ni}| \le K_{n}$ almost surely for $i=1, \ldots, n$ and $n \in \mathbb{N}$. 
 Assumption 1 controls the limiting behaviour of $K_{n}=\underset{i=1, \ldots, n}{\text{max}}\lVert \vect{u}_{(n)i} \rVert_{2}$ (recall equation \eqref{eq:assumption_norm}). Taking limits and using Assumption 1 shows that $K_{n} \to 0$,
\begin{align*}
\underset{n \to \infty}{\lim}  \  K_{n} &= \underset{n \to \infty}{\lim}  \  \underset{i=1, \ldots, n}{\text{max}}\lVert \vect{u}_{(n)i} \rVert_{2} \\
&= 0.
\end{align*}
By theorem \ref{thm:triangular_array} we have that the triangular array of random variables in \eqref{eq:cw_triangular_array} satisfies Lindeberg's condition. As such the conditions of Theorem \ref{thm:pairwise_clt} are satisfied, giving that $\vect{\lambda}^{\mathsf{T}}\vect{z}_{n}$ converges in distribution to $N(0, 1/k)$. Finally, the  Cram\'{e}r-Wold  device gives that the whitened sketched dataset has a limiting matrix normal distribution, that is  $
\widetilde{\mat{A}} \mat{V}_{(n)}\mat{D}_{(n)}^{-1} $ converges in distribution to a $MN\left(\vect{0}_{k \times d}, \mat{I}_{k},\mat{I}_{d}/k \right)$ random matrix.
\subsection{Hadamard sketch}
Recall that the Hadamard sketch is defined through $\mat{S} = \mat{\Phi}\mat{H}\mat{D}/\sqrt{k}$. Here $\mat{H}$ is a Hadamard matrix. Hadamard matrices are square matrices with $2^n$ rows for some integer $n$. To take limits we have to define our sequence of source datasets $(\mat{A}_{(n)}=\mat{U}_{(n)}\mat{D}_{(n)}\mat{V}_{(n)}^{\T})$ as having $r_{n}=2^n$ rows for $n \in \mathbb{N}^{+}$. In practice when taking a Hadamard sketch we pad the original dataset with zeros if the original number of observations is not a power of two. To rigourously establish asymptotic normality for the Hadamard sketch we have to take $r_{n}=2^n$. The first three rows of the triangular array of left singular vectors now looks like
\begin{align*}
&\vect{u}_{(1)1} \\
&\vect{u}_{(2)1} \quad \vect{u}_{(2)2} \\
&\vect{u}_{(3)1} \quad \vect{u}_{(3)2} \quad \vect{u}_{(3)3} \quad \vect{u}_{(n)4}.
\end{align*}
The intuition is the same as with the Clarkson-Woodruff sketch, as we move down the rows $n$ we expect the norms of $\vect{u}_{(n)i}$, $i\in\left\lbrace1, \ldots, 2^n\right\rbrace$  to decrease. This follows from the implicit row-wise normalisation property
\begin{align*}
\sum_{i=1}^{r_n}\lVert \vect{u}_{(n)i}\rVert_{2}^{2} = d.
\end{align*}
The indexing change to $r_{n}=2^{n}$ instead of $r_{n}=n$ has very little impact on the underlying arguments. 

There are two independent sources of randomness in a Hadamard sketch, the $r_n=2^n$ independent random Rademacher variables in the diagonal matrix $\mat{D}$, and the random matrix $\mat{\Phi}$ which subsamples $k$ rows with replacement from the Hadamard matrix $\mat{H}$. Hadamard matrices have a number of properties that we will use \citep[section 3.2]{anderson_combinatorial_1997}.
\begin{itemize}
\item (P1) The first column contains all ones.
\item (P2) Every column other than the first contains an equal number of $+1$ and $-1$ entries.
\item (P3) Consider any two different columns $i$ and $s$, where $i,s \in \left\lbrace 2, \ldots, r_{n}
\right\rbrace$, $i \ne s$. Columns $i$ and $s$ will have $+1$ together in a quarter of the rows, and $-1$ together in a quarter of the rows. Furthermore, a quarter of the rows will have $+1$ in column $i$ and $-1$ in column $s$. Similarly, a quarter of the rows will have $-1$ in column $i$ and $+1$ in column $s$. 
\end{itemize}
Let $\mat{M}$ represent the random $k \times n$ matrix from the subsampling operation $\mat{M}=\Phi \mat{H}$. Let $m_{ji}$ refer to the element in row $j$ and column $i$ of $\mat{M}$. Each element in $\mat{M}$ is equal to $+1$ or $-1$.  Let $h_{i} \in \left\lbrace +1, -1\right\rbrace$ be the random sign in element $D_{ii}$. We now represent the Hadamard sketch as $\mat{S}=\mat{M}\mat{D}/\sqrt{k}$.

The structure  of the Hadamard matrix gives the random matrix $\mat{M}$ some useful properties. Consider an arbitrary row $j$ in $\mat{M}$. By (P1) listed above regarding the first column of $\mat{M}$, $m_{j1}=1$ with probability one. For the other columns,  $m_{ji}=1$ with probability half, and $m_{ji}=-1$ with probability half for $i = 2, \ldots, r_{n}$ by (P2). By (P3) listed above, we have pairwise independence between elements in row $j$ of $\mat{M}$, that is $p(m_{ji} | m_{js}) = p(m_{ji})$ for $i,s \in \left\lbrace 1, \ldots, r_{n}
\right\rbrace$, $i \ne s$. As rows of $\mat{M}$ are sampled independently, each column of $\mat{M}$ is pairwise independent.

Row $j$ in the sketched dataset is given by
\begin{align*}
\widetilde{\vect{u}}_{j}^{\T} &= \dfrac{1}{\sqrt{k}}\sum_{i=1}^{r_{n}}m_{ji}h_{i}\vect{u}_{(n)i}^{\T}.
\end{align*}
Let us again consider the linear combination $\vect{\lambda}^{\T}\vect{z}_{n}$, where $\vect{\lambda}$ and $\vect{z}_{n}$ are defined as in  \eqref{eq:zstack} and \eqref{eq:lambda_stack} respectively. The sum over the $k$ rows in the sketched dataset can be rearranged into a sum over the $r_n=2^n$ rows in the source dataset,
\begin{align}
\vect{\lambda}^{\mathsf{T}}\vect{z}_{n}&=  \sum_{j=1}^{k} \vect{\lambda}_{j}^{\mathsf{T}}\widetilde{\vect{u}}_{j} \nonumber \\
&= \dfrac{1}{\sqrt{k}} \sum_{j=1}^{k}\vect{\lambda}_{j}^{\mathsf{T}} \sum_{i=1}^{r_n}m_{ji} h_{i}\vect{u}_{(n)i} \nonumber \\
&= \dfrac{1}{\sqrt{k}} \sum_{i=1}^{r_n}h_{i}\left(\sum_{j=1}^{k}m_{ji}\vect{\lambda}_{j}^{\mathsf{T}}\right) \vect{u}_{(n)i}. \label{eq:SRHT_linear}
\end{align}

In the language of Theorem \ref{thm:triangular_array} we can form a triangular array of random variables setting 
\begin{align}
    Z_{ni} =\dfrac{1}{\sqrt{k}} h_{i}\left(\sum_{j=1}^{k}m_{ji}\vect{\lambda}_{j}^{\mathsf{T}}\right) \vect{u}_{(n)i}. 
    \label{eq:hadmard_triangular_array}
\end{align}
for $i=1, \ldots, r_n$ and $n \in \mathbb{N}$. The linear combination in \eqref{eq:SRHT_linear} can then be expressed as a row sum of the triangular array defined by \eqref{eq:hadmard_triangular_array}
\begin{align}
\vect{\lambda}^{\mathsf{T}}\vect{z}_{n} &= \sum_{i=1}^{r_n}Z_{ni}. 
\label{eq:hadmard_cramer_triangular}
\end{align}
Our goal of showing that $\vect{\lambda}^{\mathsf{T}}\vect{z}_{n}$ converges in distribution to a $N(0, 1/k)$ random variable is achieved if we can show that $\sum_{i=1}^{n}Z_{ni}$ converges in distribution to a $N(0, 1/k)$ random variable.

The sequence of random variables in each row of the triangular array $Z_{n1}, \ldots, Z_{nr_{n}}$ 
are not mutually independent over $i=1\ldots , r_n$. This is because the columns of $\mat{M}$ are not mutually independent. However, as the columns of $\mat{M}$ are pairwise independent, the random sums  $\sum_{j=1}^{k}m_{ji}\vect{\lambda}_{j}^{\mathsf{T}}$ appearing in \eqref{eq:hadmard_triangular_array} are also pairwise independent. Again making a connection to Theorem \ref{thm:pairwise_clt}, the independent sign flips $h_{i}$ appearing in \eqref{eq:hadmard_triangular_array} ensure that the random variables in each row of the triangular array are jointly symmetric and pairwise independent. 

Theorem \ref{thm:pairwise_clt} can be used to establish asymptotic normality of the sum in \eqref{eq:hadmard_cramer_triangular} and hence the linear combination $\vect{\lambda}^{\mathsf{T}}\vect{z}_{n}$. To show that the triangular array of random variables defined in \eqref{eq:hadmard_triangular_array} satisfies Lindeberg's condition we use Theorem \ref{thm:triangular_array}. 
Set $s_{n}^{2}= \sum_{i=1}^{n}\text{var}(Z_{ni})$. We first determine $s_{n}^{2}$. We then form the necessary sequence of upper bounds $K_{n}$ such that $|Z_{ni}| \le K_{n}$ almost surely for $i=1, \ldots, n$.

We start by considering the variance of a single term in the triangular array $\text{var}(Z_{ni})$. We have that
 \begin{align}
  \text{var}(Z_{ni}) &= \dfrac{1}{k} \text{var}\left(h_{i}\left(\sum_{j=1}^{k}m_{ji}\vect{\lambda}_{j}^{\mathsf{T}}\right) \vect{u}_{(n)i} \right)
 \end{align}
It is important to consider the covariance between the elements of the sum over $j=1, \ldots, k$. For $i\ne 1$ and $j,v\in \left\lbrace 1, \ldots, k \right\rbrace$, $j \ne v$ the covariance is zero
\begin{align*}
\text{cov}\left(h_{i}m_{ji}\vect{\lambda}_{j}^{\mathsf{T}}\vect{u}_{(n)i}, h_{i}m_{vi}\vect{\lambda}_{v}^{\mathsf{T}}\vect{u}_{(n)i} \right) &=   \mathbb{E}\left[h_{i}^{2}m_{ji}m_{vi}\vect{\lambda}_{j}^{\mathsf{T}}\vect{u}_{(n)i} \vect{\lambda}_{v}^\mathsf{T}\vect{u}_{(n)i} \right]  \\
&= \mathbb{E}\left[m_{ji}m_{vi}\right]\vect{\lambda}_{j}^{\mathsf{T}}\vect{u}_{(n)i}\vect{\lambda}_{v}^{\mathsf{T}}\vect{u}_{(n)i}
 \\
&= 0.
\end{align*}
We use (P2) to conclude that $\mathbb{E}\left[m_{ji}m_{vi}\right]=0$. Therefore for $i=2, \ldots, r_n$ 
\begin{align}
\text{var}(Z_{ni}) &= \dfrac{1}{k} 
\text{var}\left(\sum_{j=1}^{k}h_{i}m_{ji}\vect{\lambda}_{j}^{\mathsf{T}} \vect{u}_{(n)i} \right) \nonumber \\
&= \dfrac{1}{k}  \sum_{j=1}^{k}\text{var}\left( h_{i}m_{ji}\vect{\lambda}_{j}^{\mathsf{T}} \vect{u}_{(n)i}\right) \nonumber \\
&=\dfrac{1}{k}\sum_{j=1}^{k}  \vect{\lambda}_{j}^{\mathsf{T}} \vect{u}_{(n)i}\vect{u}_{(n)i}^{\mathsf{T}} \vect{\lambda}_{j}. \label{eq:zgen_variance}
\end{align}
Results are different for $i=1$ as the first column of the Hadamard matrix is all ones (P1). For $j,v \in \left\lbrace 1, \ldots, k \right\rbrace$, $j \ne v$ the covariance is
\begin{align*}
\text{cov}\left(h_{1}m_{j1}\vect{\lambda}_{j}^{\mathsf{T}}\vect{u}_{(n)1}, h_{1}m_{v1}\vect{\lambda}_{v}^{\mathsf{T}}\vect{u}_{(n)1} \right) &=   \mathbb{E}\left[h_{1}^{2}m_{j1}m_{v1}\vect{\lambda}_{j}^{\mathsf{T}}\vect{u}_{(n)1}\vect{\lambda}_{v}^{\mathsf{T}}\vect{u}_{(n)1} \right]  \\
&= \mathbb{E}\left[m_{j1}m_{v1}\right]\vect{\lambda}_{j}^{\mathsf{T}}\vect{u}_{(n)1}\vect{\lambda}_{v}^{\mathsf{T}}\vect{u}_{(n)1} \\ 
&= \vect{\lambda}_{j}^{\mathsf{T}}\vect{u}_{(n)1}\vect{\lambda}_{v}^{\mathsf{T}}\vect{u}_{(n)1}.
\end{align*}
From (P1) $m_{j1}=m_{v1}=1$. Now using the Cauchy-Schwarz inequality,
\begin{align*}
\lvert \text{cov}\left(h_{1}m_{j1}\vect{\lambda}_{j}^{\mathsf{T}}\vect{u}_{(n)1}, h_{1}m_{v1}\vect{\lambda}_{v}^{\mathsf{T}}\vect{u}_{(n)1} \right) \rvert &= |\vect{\lambda}_{j}^{\mathsf{T}}\vect{u}_{(n)1}||\vect{\lambda}_{v}^{\mathsf{T}}\vect{u}_{(n)1}| \\
&\le \lVert \vect{\lambda}_{j}\rVert_{2}\lVert \vect{u}_{(n)1}\rVert_{2}\lVert \vect{\lambda}_{v}\rVert_{2}\lVert \vect{u}_{(n)1}\rVert_{2} \\
&\le \lVert \vect{u}_{(n)1}\rVert_{2}\lVert \vect{u}_{(n)1}\rVert_{2} \\
&= \lVert \vect{u}_{(n)1}\rVert_{2}^{2}
\end{align*}
The second last last uses the fact that $\vect{\lambda}$ is a unit vector and we must have $\lVert \vect{\lambda}_{j}\rVert_{2}  \le 1$, $\lVert \vect{\lambda}_{j}\rVert_{2} \le 1$ for any $j,k$. From assumption 1, the right hand side of the previous inequality tends to zero as $n$ tends to infinity. As such we conclude that $ \lvert \text{cov}\left(h_{1}m_{j1}\vect{\lambda}_{j}^{\mathsf{T}}\vect{u}_{(n)1}, h_{1}m_{v1}\vect{\lambda}_{v}^{\mathsf{T}}\vect{u}_{(n)1} \right) \rvert$ is $o(1)$. Some covariance terms appear in the expression for $\text{var}(Z_{n1})$
\begin{align}
\text{var}(Z_{n1}) &= \dfrac{1}{k}
\text{var}\left(\sum_{j=1}^{k}h_{1}m_{j1}\vect{\lambda}_{j}^{\mathsf{T}} \vect{u}_{(n)1} \right) \nonumber \\
&=  \dfrac{1}{k}  \sum_{j=1}^{k}\text{var}\left( h_{i}m_{ji}\vect{\lambda}_{j}^{\mathsf{T}} \vect{u}_{(n)1}\right) + \dfrac{1}{k} 2\sum_{j=1}^{k-1}\sum_{v=j+1}^{k}\text{cov}\left(h_{1}m_{j1}\vect{\lambda}_{j}^{\mathsf{T}}\vect{u}_{(n)1}, h_{1}m_{v1}\vect{\lambda}_{v}^{\mathsf{T}}\vect{u}_{(n)1} \right) \nonumber \\
 &=  \dfrac{1}{k} \sum_{j=1}^{k}\text{var}\left( h_{1}m_{j1}\vect{\lambda}_{j}^{\mathsf{T}} \vect{u}_{(n)1}\right) + \dfrac{1}{k}2\sum_{j=1}^{k-1}\sum_{v=j+1}^{k}  \vect{\lambda}_{j}^{\mathsf{T}}\vect{u}_{(n)1}\vect{\lambda}_{v}^{\mathsf{T}}\vect{u}_{(n)1} \nonumber \\
&= \dfrac{1}{k}\sum_{j=1}^{k}\text{var}\left( h_{1}m_{j1}\vect{\lambda}_{j}^{\mathsf{T}} \vect{u}_{(n)1}\right)  + o(1) \nonumber \\
&= \dfrac{1}{k} \sum_{j=1}^{k}  \vect{\lambda}_{j}^{\mathsf{T}} \vect{u}_{(n)1}\vect{u}_{(n)1}^{\mathsf{T}} \vect{\lambda}_{j}+ o(1) \label{eq:z1_variance}
\end{align}
The trailing term can be grouped into an $o(1)$ term as the sketch size $k$ is fixed in our analysis. Using \eqref{eq:zgen_variance} and \eqref{eq:z1_variance} we can then determine the row-wise variance totals $s_{n}^{2}$:
\begin{align*}
s_{n}^{2} &=\dfrac{1}{k} \sum_{i=1}^{r_{n}}\text{var}(Z_{ni}) \\
&= 
\dfrac{1}{k}\sum_{i=1}^{r_n}\sum_{j=1}^{k}\vect{\lambda}_{j}^{\mathsf{T}} \vect{u}_{(n)i}\vect{u}_{(n)i}^{\mathsf{T}} \vect{\lambda}_{j} + o(1) \\
&= \dfrac{1}{k}\sum_{j=1}^{k}\vect{\lambda}_{j}^{\mathsf{T}} \left( \sum_{i=1}^{r_n} \vect{u}_{(n)i}\vect{u}_{(n)i}^{\mathsf{T}} \right) \vect{\lambda}_{j} + o(1) \\
&= \dfrac{1}{k}\sum_{j=1}^{k}\vect{\lambda}_{j}^{\mathsf{T}} \mat{U}^{\mathsf{T}}_{(n)}\mat{U}_{(n)} \vect{\lambda}_{j} + o(1) \\
&= \dfrac{1}{k}\sum_{j=1}^{k}\vect{\lambda}_{j}^{\mathsf{T}} \mat{I}_{d} \vect{\lambda}_{j} + o(1) \\
&= \dfrac{1}{k}\sum_{j=1}^{k}\vect{\lambda}_{j}^{\mathsf{T}}\vect{\lambda}_{j} + o(1) 
\\
&= \dfrac{1}{k} + o(1).
\end{align*}
The step in the last line follows as we have taken $\vect{\lambda}$ to be a unit vector. The fact that $\mat{U}^{\mathsf{T}}_{(n)}\mat{U}_{(n)}=\mat{I}_{d}$ for all $n$ serves as a useful normalisation to give stable limiting behaviour. We are working with a triangular array where the rows are nearly standardised. Asymptotically in $n$,  $s_{n}^{2} \to 1/k$.

We now establish a sequence of upper bounds $(K_{n})$. As the random variables in the construction of  construction of the Hadamard sketch are bounded, we can bound the random variables in the triangular array \eqref{eq:hadmard_triangular_array} using the leverage scores of the sequence of source datasets. Now as the random sign $h_{i} \in \left\lbrace +1, -1 \right\rbrace$ we have that for all $i=1, \ldots, r_n$:
\begin{align*}
|Z_{ni}| &= \dfrac{1}{\sqrt{k}}
\lvert h_{i}\sum_{j=1}^{k}m_{ji}\vect{\lambda}_{j}^{\mathsf{T}} \vect{u}_{(n)i} \rvert \\ 
&= \dfrac{1}{\sqrt{k}} \lvert \sum_{j=1}^{k}m_{ji}\vect{\lambda}_{j}^{\mathsf{T}} \vect{u}_{(n)i} \rvert. 
\end{align*}
Now using the Cauchy-Schwarz inequality,
\begin{align}
    \dfrac{1}{\sqrt{k}}\lvert  \left(\sum_{j=1}^{k}m_{ji}\vect{\lambda}_{j}^{\mathsf{T}} \right)\vect{u}_{(n)i} \rvert &\le \dfrac{1}{\sqrt{k}} \lVert\left(\sum_{j=1}^{k}m_{ji}\vect{\lambda}_{j} \right) \rVert_{2}\lVert\vect{u}_{(n)i} \rVert_{2}. \label{eq:hadamard_cauchy}
\end{align}
 Using the triangle inequality,
 \begin{align}
 \lVert\left(\sum_{j=1}^{k}m_{ji}\vect{\lambda}_{j} \right) \rVert_{2} \le  \sum_{j=1}^{k}\lVert m_{ji}\vect{\lambda}_{j} \rVert_{2}. \label{eq:hadamard_triangle_1}
 \end{align}
Now as $m_{ji} \in \left\lbrace +1, -1\right\rbrace$ for all $j =1, \ldots, k$,
\begin{align}
    \sum_{j=1}^{k}\lVert m_{ji}\vect{\lambda}_{j} \rVert_{2} &= \sum_{j=1}^{k}\lVert \vect{\lambda}_{j} \rVert_{2}. \label{eq:hadamard_triangle_2}
\end{align}
As $\vect{\lambda}$ is a unit vector we can easily form the bound
\begin{align}
    \sum_{j=1}^{k}\lVert \vect{\lambda}_{j} \rVert_{2} &\le k. \label{eq:hadamard_weak_bound}
\end{align}
Substituting \eqref{eq:hadamard_triangle_2} and \eqref{eq:hadamard_weak_bound} into \eqref{eq:hadamard_cauchy} leads to the upper bound for $i=1, \ldots, r_{n}$:
\begin{align}
    |Z_{ni}| &\le \dfrac{1}{\sqrt{k}} k  \lVert\vect{u}_{(n)i} \rVert_{2} \nonumber \\
    &=  \sqrt{k}\lVert\vect{u}_{(n)i} \rVert_{2}. \label{eq:hadamard_weak_z_bound}
\end{align}
We can then form the sequence of upper bounds $K_{n}$:
\begin{align*}
K_{n} &= \sqrt{k}\underset{i=1, \ldots, r_{n}}{\text{max }} \lVert\vect{u}_{(n)i} \rVert_{2}.
\end{align*}
We have that $|Z_{ni}| \le K_{n}$ almost surely for $i=1, \ldots, r_{n}$ and $n \in \mathbb{N}$. Assumption 1 (recall equation \eqref{eq:assumption_norm}) gives the limiting behaviour of $K_{n}$. As the sketch size $k$ is fixed in our analysis,
\begin{align*}
\underset{n \to \infty}{\lim}  \  K_{n} &=  \sqrt{k}\underset{n \to \infty}{\lim}  \  \underset{i=1, \ldots, r_{n}}{\text{max }} \lVert\vect{u}_{(n)i} \rVert_{2} \\
&= 0.
\end{align*}
We have that $K_{n} \to 0$ as $n \to \infty$. As $s_{n}^{2}=1/k + o(1)$ we have an asymptotically standardised array, and $K_{n}/s_{n} \to 0$. We can use Theorem \ref{thm:triangular_array} to conclude that the triangular array of random variables defined in \eqref{eq:hadmard_triangular_array} satisfies Lindeberg's condition. As such, the conditions of Theorem \ref{thm:pairwise_clt} are satisfied. We conclude that the row sums in \eqref{eq:hadmard_cramer_triangular} converge in distribution to $N(0, 1/k)$. Finally, the Cram\'{e}r-Wold device gives that the whitened sketched dataset has a limiting matrix normal distribution. That is the sequence of random matrices  $
\widetilde{\mat{A}}\mat{V}_{(n)}\mat{D}_{(n)}^{-1}$ converges in distribution to a $MN\left(\vect{0}, \mat{I}_{k},\mat{I}_{d}/k \right)$ random matrix. 
\section{Proof of Theorem \ref{thm:random_projection_beta_S} (Complete sketching asymptotics)}
\begin{description}
	\item[Assumption 2:]  \begin{align*}
\underset{n \to \infty}{\lim } n^{-1}\begin{bmatrix}
\vect{y}^{\T}_{(n)}\vect{y}_{(n)} & \vect{y}^{\T}_{(n)}\mat{X}_{(n)} \\
\mat{X}^{\T}_{(n)}\vect{y}_{(n)} & \mat{X}^{\T}_{(n)}\mat{X}_{(n)}
\end{bmatrix} = \mat{Q} \qquad \text{for some positive-definite matrix $\mat{Q}$. }
\end{align*}
\end{description}
\begin{theorem*}
Suppose that Assumptions 1 and 2 hold, $k \ge p $, and $\vect{\beta}_{S}$ is computed using a Hadamard or Clarkson-Woodruff sketch. Let $(\widetilde{\mat{X}}^{\T}\widetilde{\mat{X}})^{+}$ denote the Moore-Penrose pseudo-inverse of $(\widetilde{\mat{X}}^{\T}\widetilde{\mat{X}})$. Let
\begin{align*}
\widetilde{\mat{H}}_{(n)}  = \dfrac{RSS^{(n)}_{F}}{k}\left(\widetilde{\mat{X}}^{\T}\widetilde{\mat{X}}\right)^{+}  \ \mathrm{and } \
 {\mat{H}}_{(n)} = \dfrac{RSS^{(n)}_{F}}{k-p+1}\left({\mat{X}}^{\T}_{(n)}{\mat{X}_{(n)}}\right)^{-1}.
\end{align*}
Then as $n \to \infty$, convergence in distribution holds for
\begin{align*}
(i) [\mat{H}^{-1/2}_{(n)}(\vect{\beta}_{S} - \vect{\beta}_{F}^{(n)})| \vect{A}_{(n)}]  &\to \textnormal{Student}\left(\vect{0}, \mat{I}_{p},  \ k-p+1\right), \\
(ii) [\widetilde{\mat{H}}^{-1/2}_{(n)}(\vect{\beta}_{S} - \vect{\beta}_{F}^{(n)}) \ | \vect{A}_{(n)}] & \to N \left(\vect{0},  \mat{I}_{p} \right).
\end{align*}
\end{theorem*}
Notation is slightly heavier in the proof compared to the main text for the sake of clarity. Again we do not explicitly condition on the source dataset $\mat{A}_{(n)}$, the source dataset is always fixed, and the only randomness is from the sketching matrix. The sketched data will be denoted $\widetilde{\vect{y}}_{(n)}$ and $\widetilde{\mat{X}}_{(n)}$ to denote the dependence on the $n \times d$ source dataset. So $\widetilde{\vect{y}}_{(n)} = \mat{S}\vect{y}_{(n)}$ and $\widetilde{\mat{X}}_{(n)} = \mat{S}\mat{X}_{(n)}$. The dimension of the sketched dataset does not change.  

Assumption 2 is of assistance in establishing the limit theorem.  Let
\begin{align*}
    \mat{Q}_{(n)} &= n^{-1}\begin{bmatrix}
\vect{y}^{\T}_{(n)}\vect{y}_{(n)} & \vect{y}^{\T}_{(n)}\mat{X}_{(n)} \\
\mat{X}^{\T}_{(n)}\vect{y}_{(n)} & \mat{X}^{\T}_{(n)}\mat{X}_{(n)}.
\end{bmatrix}
\end{align*}
The matrix $\mat{Q}_{(n)}$ contains the sufficient statistics needed to fit a Gaussian linear model, $\vect{y}^{\mathsf{T}}_{(n)}\vect{y}_{(n)}, \mat{X}^{\mathsf{T}}_{(n)}\vect{y}_{(n)}$ and $\mat{X}^{\mathsf{T}}_{(n)}\mat{X}_{(n)}$ given the source dataset $\mat{A}_{(n)}=[\vect{y}_{(n)}, \mat{X}_{(n)}]$. Assumption 2 states the averaged sufficient statistic matrix converges to a limiting matrix $\mat{Q}$. 
It will be helpful to partition the limiting matrix $\mat{Q}$ as
\begin{align}
\mat{Q} &= \underset{n \to \infty}{\lim } n^{-1}\begin{bmatrix}
\vect{y}^{\mathsf{T}}_{(n)}\vect{y}_{(n)} & \vect{y}^{\mathsf{T}}_{(n)}\mat{X}_{(n)} \\
\mat{X}^{\mathsf{T}}_{(n)}\vect{y}_{(n)} & \mat{X}^{\mathsf{T}}_{(n)}\mat{X}_{(n)}
\end{bmatrix} = \begin{bmatrix}
s & \vect{m}^{\mathsf{T}} \\
\vect{m} & \mat{G}
\end{bmatrix}, \label{eq:Q_partition}
\end{align}
where $s$ is a scalar, $\mat{G}$ is a $p \times p$ matrix and $\vect{b}$ is a $p$-length column vector. The matrix $\mat{G}$ is the limiting averaged Gram matrix of the predictors. The  vector $\vect{m}$ is the limit of the predictor response inner products $n^{-1}\mat{X}_{(n)}^{\mathsf{T}}\vect{y}_{(n)}$,  and the scalar $s$ is the limit of the mean total sum of squares $n^{-1}\vect{y}^{\mathsf{T}}_{(n)}\vect{y}_{(n)}$. 

As mentioned, the assumption of a sequence of source datasets also gives a sequence of optimal least squares coefficients and residual errors. Let $\sigma_{F}^{2(n)} = RSS_{F}/n$.  Define the limiting least squares coefficient estimate as $\vect{\beta} = \underset{n \to \infty}{\lim } \vect{\beta}_{F}^{(n)}$ and the limiting residual error as $\sigma^{2}= \underset{n \to \infty}{\lim } \sigma^{2(n)}_{F}$. Both $\vect{\beta}$ and $\sigma^2$ can be expressed as functions of the matrix $\mat{Q}$. Specifically,
\begin{align}
\vect{\beta} &= \mat{G}^{-1}\vect{m}, \label{eq:limiting_beta}\\
\sigma^2 &= s - \vect{m}^{\mathsf{T}}\mat{G}^{-1}\vect{m}.\label{eq:limiting_sigma2}
\end{align}
From Assumption 2, we have that $n^{-1}\mat{V}_{(n)}\mat{D}_{(n)}^{2}\mat{V}_{(n)}^{\mathsf{T}} \to \mat{Q}$. As such we have that $n^{-1/2}\mat{D}_{(n)}\mat{V}_{(n)}^{\mathsf{T}} \to \mat{Q}^{1/2}$
From the sketching central limit theorem the whitened sketched data converges to a matrix normal distribution
\begin{align*}
[\widetilde{\vect{y}}_{(n)}, \widetilde{\mat{X}}_{(n)}]\mat{V}_{(n)}\mat{D}_{(n)}^{-1} \overset{d}{\to} MN\left(\vect{0}_{k \times d},  \mat{I}_{k},\mat{I}_{d}/k\right)
\end{align*}
The benefit of adding Assumption 2 is that using Slutsky's theorem we have the additional convergence result
\begin{align*}
n^{-1/2}[\widetilde{\vect{y}}_{(n)}, \widetilde{\mat{X}}_{(n)}] \overset{d}{\to} MN\left(\vect{0},  \mat{I}_{k},\mat{Q}/k\right).
\end{align*}
To prove results $(i)$ and $(ii)$ we use the continuous mapping theorem \cite[p. 7]{van_der_vaart_asymptotic_1998} in conjunction with the previous convergence result. It will be helpful to define the random variables $\widetilde{\vect{y}}_{L}, \widetilde{\mat{X}}_{L}$ as having the above limiting matrix normal distribution
\begin{align*}
[\widetilde{\vect{y}}_{L}, \widetilde{\mat{X}}_{L}] \sim  MN\left(\vect{0}_{k \times d},  \mat{I}_{k},\mat{Q}/k\right). 
\end{align*}
This is so we can say that
\begin{align*}
n^{-1/2}[\widetilde{\vect{y}}_{(n)}, \widetilde{\mat{X}}_{(n)}] \overset{d}{\to} [\widetilde{\vect{y}}_{L}, \widetilde{\mat{X}}_{L}] .
\end{align*}

\begin{lemma}[Continuous Mapping Theorem]
\label{lem:continuous_mapping}
Let $(\vect{Z}_{n})_{n \in \mathbb{N}}$ indicate a sequence of random vectors in $\mathbb{R}^{d}$ and $\vect{Z}$ indicate another random vector in $\mathbb{R}^{d}$. Suppose the function $g: \mathbb{R}^{d} \to \mathbb{R}^{m}$ is continuous at every point of a set $C$ such that $P(\vect{Z} \in C) =1$. Then if $\vect{Z}_{n} \overset{d}{\to} \vect{Z}$ then $g(\vect{Z}_{n}) \overset{d}{\to} g(\vect{Z})$. 
\end{lemma}
In Lemma \ref{lem:continuous_mapping}, the function  $g: \mathbb{R}^{d} \to \mathbb{R}^{m}$ does not change with $n$, and the dimensions $d$ and $m$ are fixed when taking limits. The sketched estimator $\vect{\beta}_{S}$ can be defined as a function of the sketched data that is continuous over the set where $\widetilde{\mat{X}}_{(n)}$ is of full rank.  Formally we could say that $\vect{\beta}_{S}=g(n^{-1/2}\widetilde{\vect{y}}_{(n)},n^{-1/2}\widetilde{\vect{X}}_{(n)})$. As $\widetilde{\mat{X}}_{L}$ is of rank $p$ almost surely, and $\widetilde{\mat{X}}_{(n)} \overset{d}{\to} \widetilde{\mat{X}}_{L}$ we can apply the continuous mapping theorem to determine the limiting distribution of the $\vect{\beta}_{S}$. The random matrix $[\widetilde{\vect{y}}_{L}, \widetilde{\mat{X}}_{L}]$ can be described using  a hierarchical model completely analogous in structure to the hierarchical model established for the Gaussian sketch in Section \ref{sec:gaussian_complete} of the main text. Specifically, 
\begin{align*}
\widetilde{\vect{y}}_{L} \mid \widetilde{\mat{X}}_{L}  & \sim N\left(\widetilde{\mat{X}}_{L}\vect{\beta}, \dfrac{1}{k}\sigma^2\mat{I}_{k}\right), \\
\widetilde{\mat{X}}_{L}  &\sim MN\left(\vect{0}, \mat{I}_{k}, \dfrac{1}{k}\mat{G} \right). 
\end{align*}
From Theorem \ref{thm:gaussian_exact_distribution} in the main text, and recalling that the function $g$ outputs $\vect{\beta}_{S}$, we have that 
\begin{align*}
g(\widetilde{\vect{y}}_{L}, \widetilde{\mat{X}}_{L}) \sim \text{Student}\left(\vect{\beta}, \dfrac{\sigma^2}{k-p+1}\mat{G}^{-1}, k-p+1 \right).
\end{align*} 
As such, for the Hadamard and Clarkson-Woodruff sketches, 
\begin{align*}
[\vect{\beta}_{S} \mid \vect{y}_{(n)}, \mat{X}_{(n)}] & \overset{d}{\to} \text{Student}\left(\vect{\beta}, \dfrac{\sigma^2}{k-p+1}\mat{G}^{-1}, k-p+1 \right).
\end{align*}
Let 
\begin{align*}
\mat{H}_{(n)} &=  \sigma^{2(n)}_{F}/(k-p+1)\left(n^{-1}{\mat{X}}^{\mathsf{T}}_{(n)}{\mat{X}}_{(n)}\right)^{-1}. 
\end{align*}
Now as $n^{-1}{\mat{X}}^{\mathsf{T}}_{(n)}{\mat{X}}_{(n)} \to \mat{G}$, 
$\sigma^{2(n)}_{F} \to \sigma^2$, and 
$\vect{\beta}_{F}^{(n)}  \to \vect{\beta}$, Slutsky's theorem can be used to arrive at $(i)$, 
\begin{align*}
{\mat{H}}^{-1/2}_{(n)}(\vect{\beta}_{S} - \vect{\beta}_{F}) &\overset{d}{\to} \textnormal{Student}\left(\vect{0}, \mat{I}_{p},  \ k-p+1\right). 
\end{align*}
For result $(ii)$, let us define the function 
\begin{align*}
f(n^{-1/2}\widetilde{\vect{y}}_{(n)}, n^{-1/2}\widetilde{\mat{X}}_{(n)}) &= \left[n\left(\widetilde{\mat{X}}^{\mathsf{T}}\widetilde{\mat{X}}\right)^{+}\right]^{-1/2}
\left(\widetilde{\mat{X}}^{+}\widetilde{\vect{y}} - \vect{\beta}\right)  \\
&= \left[n\left(\widetilde{\mat{X}}^{\mathsf{T}}\widetilde{\mat{X}}\right)^{+}\right]^{-1/2}
\left(\vect{\beta}_{S} - \vect{\beta}\right).
\end{align*}
This function  transforms the $\vect{\beta}_{S}$ so that the output is uncorrelated. This function is also continuous over the set where $\widetilde{\mat{X}}_{(n)}$ is of rank $p$. Again using the fact that $\widetilde{\mat{X}}_{L}$ has rank $p$ almost surely, it follows from the continuous mapping theorem that $f(n^{-1/2}\widetilde{\vect{y}}_{(n)}, n^{-1/2}\widetilde{\mat{X}}_{(n)}) \overset{d}{\to} f(\widetilde{\vect{y}}_{L}, \widetilde{\mat{X}}_{L})$. Result (ii) in Theorem \ref{thm:gaussian_exact_distribution} also applies to the hierarchical model for $\widetilde{\vect{y}}_{L}, \widetilde{\mat{X}}_{L}$, and gives the  distribution of the transformed $\vect{\beta}_{S}$ under the Gaussian sketch. The distribution of $f(\widetilde{\vect{y}}_{L}, \widetilde{\mat{X}}_{L})$  will be 
\begin{align*}
f(\widetilde{\vect{y}}_{L}, \widetilde{\mat{X}}_{L})  \sim N\left( \vect{0}, \dfrac{\sigma^{2}}{k}\mat{I}_{p}\right).
\end{align*}
As such, for the Clarkson-Woodruff and Hadamard sketches, 
\begin{align*}
\left[n\left(\widetilde{\mat{X}}^{\mathsf{T}}\widetilde{\mat{X}}\right)^{+}\right]^{-1/2}
\left(\vect{\beta}_{S} - \vect{\beta}\right)  & \overset{d}{\to} \text{N}\left(\vect{0}, \dfrac{\sigma^2}{k}\mat{I}_{p}\right).
\end{align*}
Now let
\begin{align*}
\widetilde{\mat{H}}_{(n)} = n\sigma^{2(n)}_{F}/{k}\left(\widetilde{\mat{X}}^{\mathsf{T}}\widetilde{\mat{X}}\right)^{+}
\end{align*}
Now as $\sigma^{2(n)}_{F} \to \sigma^2$, and 
$\vect{\beta}_{F}^{(n)}  \to \vect{\beta}$, Slutsky's theorem can be used to arrive at $(ii)$
\begin{align*}
\widetilde{\mat{H}}^{-1/2}_{(n)}(\vect{\beta}_{S} - \vect{\beta}_{F}^{(n)})  & \overset{d}{\to} N \left(\vect{0},  \mat{I}_{p} \right).
\end{align*}
\section{Proof of Theorem \ref{thm:beta_p_asymptotics} (Partial sketching asymptotics)}
\begin{theorem*}
Suppose that Assumptions 1, 2 and 3 hold, $k > p +3 $, and $\vect{\beta}_{P}^{*}$ is computed using a Hadamard or Clarkson-Woodruff sketch. Let
\begin{align*}
 \mat{H}_{(n)} &= \dfrac{(k-p-1)}{(k-p)(k-p-3)}\left(MSS_{F}^{(n)}(\mat{X}^{\T}_{(n)}\mat{X}_{(n)})^{-1} + \dfrac{k-p+1}{k-p-1} \vect{\beta}_{F}^{(n)}\vect{\beta}_{F}^{(n)\T}\right).
\end{align*}
Then as $n \to \infty$, 
\begin{align*}
(i)\  E_{S}(\vect{\beta}_{P}^{*}-\vect{\beta}_{F}^{(n)} | \mat{A}_{(n)})  \ &{\to} \  \vect{0}.  \\
(ii) \ \mathrm{var}_{S}\left( {\mat{H}}_{(n)}^{-1/2}(\vect{\beta}_{P}^{*} - \vect{\beta}_{F}^{(n)}) \ | \vect{A}_{(n)} \right) \ & { \to} \ \mat{I}_{d}
\end{align*}
\end{theorem*}
Application of the continuous mapping theorem gives that the distribution of $\vect{\beta}_{S}$ and $\vect{\beta}_{P}^{*}$ under the Hadamard and Clarkson-Woodruff sketches converges to the distribution of the estimators under the Gaussian sketch. This does not necessarily guarantee convergence in moments. To establish a limit theorem for the bias and variance of the estimators, we need a uniform integrability condition on the sketched dataset. The sketched data will be denoted $\widetilde{\mat{X}}_{(n)}$ to denote the dependence on the $n \times p$ source covariate matrix. So  $\widetilde{\mat{X}}_{(n)} = \mat{S}\mat{X}_{(n)}$.
We again do not explicitly condition on the source dataset $\mat{A}_{(n)} =[\vect{y}_{(n)}, \mat{X}_{(n)}]$ in the following working. 

Let $\mat{G}_{(n)}=n^{-1}\widetilde{\mat{X}}^{\mathsf{T}}_{(n)}\widetilde{\mat{X}}_{(n)}$. From the continuous mapping theorem and Theorem \ref{thm:sketching_clt}, it is known that
\begin{align*}
\mat{G}^{-1}_{(n)} &\overset{d}{\to} \mat{W}, 
\end{align*}
where $\mat{W}$ has an Inverse-Wishart$(k, k\mat{Q}^{-1})$ distribution and $\mat{Q}$ is the limiting matrix from assumption 2. We would like to establish convergence in first and second moments, that is
\begin{align*}
\mathbb{E}(\mat{G}^{-1}_{(n)}) &\to  \mathbb{E}(\mat{W}), \\
\text{var}(\mat{G}^{-1}_{(n)}) &\to \text{var}(\mat{W}).
\end{align*}
If convergence in first and second moments occurs, then we can show that $(i)$ and $(ii)$ will hold. If $\mathbb{E}(\mat{G}^{-1}_{(n)}) \to  \mathbb{E}(\mat{W})$, we can say that
\begin{align*}
\mathbb{E}(\vect{\beta}_{P}^{*} - \vect{\beta}) \to \vect{0},
\end{align*}
where $\vect{\beta}$ is the limiting ordinary least squares estimator \eqref{eq:limiting_beta},  that is a function of the limiting matrix $\mat{Q}$ in Assumption 2. From here, using that $\underset{n \to \infty}{\text{lim}} \vect{\beta}_{F}^{(n)}$, Slutsky's theorem can be used to arrive at $(i)$
\begin{align*}
\mathbb{E}(\vect{\beta}_{P}^{*} - \vect{\beta}_{F}^{(n)}) \to \vect{0}.
\end{align*}
To show convergence of the variance of the sketched estimator $(ii)$, we define 
\begin{align*}
 {\mat{H}}_{(n)} &=\dfrac{(k-p-1)}{(k-p)(k-p-3)} \left( MSS_{F}^{(n)}\left({\mat{X}}^{\mathsf{T}}_{(n)}{\mat{X}_{(n)}}\right)^{-1} + \dfrac{(k-p+1)}{(k-p-1)} \vect{\beta}_{F}^{(n)}\vect{\beta}_{F}^{(n)\mathsf{T}}\right), \\
 \mat{H} &= \dfrac{(k-p-1)}{(k-p)(k-p-3)} \left( (s -\sigma^{2})\mat{G}^{-1} + \dfrac{(k-p+1)}{(k-p-1)}\vect{\beta}\vect{\beta}^{\mathsf{T}}\right).
\end{align*}
Where $s$, $\sigma^2$ and  and $\mat{G}$ are functions of the limiting matrix $\mat{Q}$, as in \eqref{eq:Q_partition}, \eqref{eq:limiting_beta} and \eqref{eq:limiting_sigma2}. If $\text{var}(\mat{G}^{-1}_{(n)}) \to \text{var}(\mat{W})$ it follows that
\begin{align*}
\mathrm{var}_{S}\left( \mat{H}^{-1/2}(\vect{\beta}_{P}^{*}-\vect{\beta})\right) \to \mat{I}_{d}. 
\end{align*}
As $\mat{H}_{(n)}$ converges to $\mat{H}$ and $\vect{\beta}_{F}^{(n)}$ converges to $\vect{\beta}$ asymptotically with $n$, an application of Slutsky's theorem gives $(ii)$,
\begin{align*}
\mathrm{var}_{S}\left( {\mat{H}}_{(n)}^{-1/2}(\vect{\beta}_{P}^{*} - \vect{\beta}_{F}^{(n)}) \right) \ & { \to} \ \mat{I}_{d}
\end{align*}
As such, if we can establish that $\text{var}(\mat{G}^{-1}_{(n)}) \to \text{var}(\mat{W})$ we have proved $(ii)$. The following theorem describes the necessary conditions for such convergence to occur.

\begin{theorem}{\citep[Theorem 5.4]{billingsley_convergence_1968}}
\label{thm:moment_convergence}
Let $\vect{X}_{1}, \ldots, \vect{X}_{n}$ be a sequence of random vectors. Suppose $\vect{X}_{n}$ converges in distribution to a random variable $\vect{Z}$ as $n$ tends to infinity. For the additional convergence of moments $\mathbb{E}[\vect{X}_{n}] \to \mathbb{E}[\vect{Z}]$ and $\text{var}[\vect{X}_{n}] \to \text{var}[\vect{Z}]$, it must hold that for all conformable constant vectors $\vect{\lambda}$ 
\begin{align*}
\lim_{M \to \infty} \underset{n \to \infty}{\limsup} \   |\vect{\lambda}^{\mathsf{T}}\vect{X}_{n}|^{2}\mathbbm{1}_{\left\lbrace|\vect{\lambda}^{\mathsf{T}}\vect{X}_{n}|^{2} \ge M\right\rbrace} = 0.
\end{align*}
\end{theorem}
The above condition can be difficult to verify directly. It can be shown that if asymptotically $|\vect{\lambda}^{\mathsf{T}}\vect{X}_{n}|$ has a bounded fourth moment, then the integrability condition is satisfied \citep[section 2.5]{van_der_vaart_asymptotic_1998}.

A linear combination of the elements of the random matrix $\mat{G}^{-1}_{(n)}$ can be written an $\text{trace}(\mat{\Lambda}\mat{G}^{-1}_{(n)})$ for a $p \times p$ matrix of constants $\mat{\Lambda}$. It is easier to work with this form rather than stacking the elements of the random matrix to form a random vector. From theorem \ref{thm:moment_convergence}, it is sufficient to show that show that the expected value of $|\text{trace}(\mat{\Lambda}\mat{G}^{-1}_{(n)})|^4$ is finite for large $n$ to show the desired convergence in moments.

As $\text{trace}(\mat{\Lambda}\mat{G}^{-1}_{(n)})$ equals the sum of the singular values of the matrix $\mat{\Lambda}\mat{G}^{-1}_{(n)}$, we can form an upper bound on the value,
\begin{align*}
\text{trace}(\mat{\Lambda}\mat{G}^{-1}_{(n)}) &\le  p \lVert\mat{\Lambda}\mat{G}^{-1}_{(n)} \rVert_{2} \\
&\le p \lVert\mat{\Lambda}\rVert_{2}\lVert\mat{G}^{-1}_{(n)} \rVert_{2} 
\end{align*}
Squaring both sides gives an upper bound on the quantity that must satisfy the uniform integrability condition,
\begin{align*}
|\text{trace}(\mat{\Lambda}\mat{G}^{-1}_{(n)})|^2 &\le  p^2 \lVert\mat{\Lambda}\rVert_{2}^{2}\lVert \mat{G}_{(n)}^{-1} \rVert_{2}^{2}.
\end{align*}
Squaring again gives an upper bound on the fourth moment of the linear combination of interest
\begin{align*}
|\text{trace}(\mat{\Lambda}\mat{G}_{(n)}^{-1})|^4 &\le p^{4}\lVert\mat{\Lambda}\rVert_{2}^{4}\lVert\mat{G}_{(n)}^{-1} \rVert_{2}^{4} \\
&= p^{4}\lVert\mat{\Lambda}\rVert_{2}^{4}\left( \dfrac{1}{\sigma^2_{\text{min}}(\mat{G}_{(n)})}\right)^2
\end{align*}
By Assumption 3, the expectation of the right hand side is finite. As such, the uniform integrability condition holds and we can conclude that
\begin{align*}
\mathbb{E}(\mat{G}^{-1}_{(n)}) &\to  \mathbb{E}(\mat{W}), \\
\text{var}(\mat{G}^{-1}_{(n)}) &\to \text{var}(\mat{W}).
\end{align*}
As discussed at the beginning of the proof this is sufficient to show that $(i)$ and $(ii)$ hold.

\end{document}